\documentclass[%
 reprint,
 superscriptaddress,
 amsmath,amssymb,
 aps,
 nofootinbib,
]{revtex4-2}

\usepackage{amsfonts}
\usepackage[english]{babel}
\usepackage[bbgreekl]{mathbbol}
\usepackage{mathtools}
\usepackage{graphicx}
\graphicspath{{./figures/}}
\usepackage{multirow}
\usepackage{bm}
\usepackage[
    usenames,
    dvipsnames,
    x11names,
]{xcolor}
\usepackage[
    colorlinks=True,
    linkcolor=NavyBlue,
    citecolor=NavyBlue,
    urlcolor=PaleGreen4,
]{hyperref}
\usepackage[capitalise]{cleveref}
\usepackage[
    range-units=single,
]{siunitx}
\usepackage{natbib}
\usepackage{enumitem}
\usepackage{microtype}

\usepackage{tikz}
\usepackage{pgfplots}
\usepackage{pgfmath}
\pgfplotsset{compat=1.18}
\usetikzlibrary{
    decorations.markings, decorations.pathreplacing, arrows.meta, bending,
    calc, patterns, intersections,
}
\usetikzlibrary{fillbetween}
\tikzset{
  semithick,
  midarrow/.style={decoration={markings, mark=at position 0.5 with {\arrow{Stealth[round,scale=1.35,flex,sep=-3pt]}}},postaction={decorate}},
  endarrow/.style={decoration={markings, mark=at position 1 with {\arrow{Stealth[round,scale=1.35,flex,sep=-3pt]}}},postaction={decorate}},
  dot/.style = {circle, fill, minimum size=#1, inner sep=0pt, outer sep=0pt},
  dot/.default = 4pt  
}

\newcommand{\rme}{\mathrm{e}}
\newcommand{\rmi}{\mathrm{i}}
\newcommand{\rmd}{\mathrm{d}}
\newcommand{\ta}{\text{a}}
\newcommand{\ts}{\text{s}}
\newcommand{\ti}{\text{i}}

\newcommand{\Fa}{\mathcal{F}_\ta}
\newcommand{\Fs}{\mathcal{F}_\ts}
\newcommand{\wa}{\gamma_\ta}
\newcommand{\wrse}{\gamma_\text{rse}}
\newcommand{\wcc}{\gamma_\text{cc}}
\newcommand{\wsr}{\gamma_\text{sr}}
\newcommand{\wsec}{\gamma_\ts}
\newcommand{\dwa}{\Omega - \delta\omega_\ta}
\newcommand{\wsql}{\Omega_\text{sql}}
\newcommand{\qfrac}{\frac{\hbar\omega_0}{P_\ta}}
\newcommand{\Pabs}{P_\text{rel}}
\newcommand{\Krp}{\mathcal{K}}
\newcommand{\dww}{\Delta w / w}

\newcommand{\rfr}{\mathfrak{r}}
\newcommand{\tfr}{\mathfrak{t}}
\newcommand{\mat}[1]{\mathbf{#1}}
\newcommand{\matQ}[1]{\boldsymbol{#1}}
\newcommand{\Ui}{\mat{U}_\text{i}}
\newcommand{\Ur}{\mat{U}_\text{r}}
\newcommand{\rrse}{\rfr_\text{rse}}
\DeclareMathOperator{\re}{Re}
\DeclareMathOperator{\im}{Im}
\DeclareMathOperator{\sgn}{sgn}

\newcommand{\Lsa}{\mat{L}_\text{sa}}
\newcommand{\Las}{\mat{L}_\text{as}}
\newcommand{\Msh}{\mat{S}_\text{sh}}
\newcommand{\Mhs}{\mat{S}_\text{hs}}
\newcommand{\Mhh}{\mat{S}_\text{hh}}
\newcommand{\Mss}{\mat{S}_\text{ss}}
\newcommand{\qar}{q_\text{ar}}
\newcommand{\qsub}{q_\text{sub}}
\newcommand{\qhr}{q_\text{hr}}

\newcommand{\braket}[3]{\big\langle#1\big|#2\big|#3\big\rangle}

\newcommand{\Asharp}{A\textsuperscript{$\sharp$}}

\DeclareSIUnit{\ppm}{ppm}
\DeclareSIUnit{\partspermillion}{ppm}
\DeclareSIUnit{\ppb}{ppb}
\DeclareSIUnit{\partsperbillion}{ppb}
\DeclareSIUnit{\rtHz}{\big/\sqrt{Hz}}
\DeclareSIUnit{\diopter}{D}

\DeclareSymbolFontAlphabet{\mathbb}{AMSb}
\DeclareSymbolFontAlphabet{\mathbbl}{bbold}

\newcommand{\Tmat}[1]{\ensuremath{\mathbbl{#1}}}

\newcommand{\Marrowbbl}[1]{\overset{\raisebox{-1pt}{\text{\tiny$\bm\Leftrightarrow$}}}{#1}}
\newcommand{\Dmat}[1]{\ensuremath{\Marrowbbl{\mathbf{#1}}}}
\newcommand{\Mrarrowbbl}[1]{\overset{\raisebox{-1pt}{\text{\tiny$\bm\Rightarrow$}}}{#1}}
\newcommand{\Dvec}[1]{\ensuremath{\Mrarrowbbl{\mathbf{#1}}}}

\begin{document}

\title{
  Squeezed state degradations due to mode mismatch and thermal
  aberrations in gravitational-wave detectors
}

\author{Kevin Kuns}
\affiliation{
  LIGO Laboratory, Department of Physics, Massachusetts Institute of
  Technology, Cambridge, MA 02139, USA
}
\author{Daniel Brown}
\affiliation{
  OzGrav, University of Adelaide, Adelaide, South Australia 5005, Australia
}
\date{25 August 2026}

\begin{abstract}

To date, frequency-dependent squeezed light has been used to reduce
quantum noise in interferometric gravitational-wave detectors by
\qty{6.1}{\dB} (a factor of two). Future upgrades and detectors aim to
both reduce quantum noise by \qty{10}{\dB} (a factor of three) and to
increase the circulating power in the interferometer arm cavities.
Achieving these goals will be extremely challenging due, in part, to
the degradations to the squeezed state caused by mode mismatch between
the internal interferometer optical cavities and between the auxiliary
external cavities. It is therefore imperative to gain a detailed
understanding of all sources of mismatch and to obtain experience in
mitigating their effects in the current detectors in order to improve
astrophysical sensitivity now and in the future. Two types of internal
mismatch are identified which are due to the thermal aberrations
generated when the test mass optics absorb a small fraction of the
circulating arm power. It is found that the dynamics responsible for
the degradations caused by the mismatch between the quadratic part of
the wavefront of two modes has a characteristic low-pass frequency
dependence while the dynamics of the mismatch due to all higher order
thermal aberrations has a high-pass behavior. As a consequence, the
two types of mismatch are predominantly responsible for different
squeezing degradations---some of which are significant for the current
detectors and some of which will only be important for future
detectors with longer arms. The behavior of these two types of
internal mismatch are described, and the implications for detector
design, operation, and characterization are discussed.


\end{abstract}

\maketitle


\section{Introduction}

The LIGO-Virgo-KAGRA network of gravitational-wave observatories have
detected hundreds of compact binary coalescences to date~\cite{GWTC5} and
have opened the door to multimessenger astronomy by enabling the
simultaneous observation of both gravitational and electromagnetic
radiation~\cite{GW170817}. These detectors are now limited by quantum
noise throughout large regions of the detection
band~\cite{Capote2024}, and it is therefore necessary to understand
and reduce all sources of quantum noise in order to continue to
advance the astrophysics, cosmology, and fundamental physics enabled
by the observation of gravitational waves.

Gravitational-wave detectors now make routine use of squeezed light to
reduce quantum
noise~\cite{Barsotti2019,Tse2019,Acernese2019,Grote2013,Lough2020,Ganapathy2023,McCuller2020,Zhao2020}.
Most significantly, LIGO has achieved \qty{6.1}{\dB} of broadband
quantum noise reduction---simultaneously reducing radiation pressure
and shot noise---through the use of frequency-dependent squeezed
vacuum states~\cite{Ganapathy2023,Capote2024}.
These squeezed states are fragile and are susceptible to
several degradation mechanisms which include the effects of the
mismatch between the spatial modes of the numerous optical cavities which make
up these detectors.  In particular, the arm cavities must be matched to
the signal extraction cavity used to tune the detector sensitivity.
The interferometer must then be matched to three external optical
cavities which together generate the squeezed light, provide the
frequency dependence needed for broadband quantum noise reduction, and
filter the signal exiting the interferometer before detection. In
addition to matching the interferometer to these three external
cavities, they must also all be matched to one another.

The mismatch between these external cavities has been studied in
detail theoretically~\cite{McCuller2021,Kwee2014}, and these models
have been successfully compared with measurements of the current
detectors~\cite{McCuller2021}. Several degradation mechanisms to the
squeezed states were identified in this analysis. Critically, it was
shown that these sources of mismatch make complex frequency-dependent
contributions to the squeezing degradations due, in part, to the
coherent nature in which the higher order modes (HOMs) excited by the
mismatch interfere with themselves and with the fundamental
mode. While Ref.~\cite{McCuller2021} outlined how the effects of one
type of internal mode mismatch can be included in the analysis of
quantum noise, no in-depth study of internal mismatch has been done to
date as far as we are aware.

In this work we identify two types of internal mode mismatch which are
mainly caused by the absorption of a small fraction of the power
circulating in the interferometer cavities by the test mass optics.
This induces both a thermorefractive lens in the test mass substrates
and a thermoelastic deformation of the optic
surfaces~\cite{Hello_Vinet_1990,Hello_Vinet_1990b,Vinet2009}. These
aberrations degrade the signals needed to control the interferometers
and increase technical noise couplings. They are also the source of
several squeezing degradations that are the focus of this work. The
thermal aberrations can be decomposed into the mismatch between the
quadratic part of the wavefront of two optical modes and the mismatch
due to all higher order thermal aberrations. Due to the differences in
how the HOMs excited by these two types of mismatch interfere, the
HOMs generated by quadratic mismatch experience dynamics with a
low-pass frequency dependence while the HOMs generated by higher order
aberrations experience high-pass dynamics. As a consequence, each type
of mismatch is predominantly responsible for different
frequency-dependent squeezing degradations.


Efforts to quantitatively understand quantum noise in the current
detectors have met with some
success~\cite{McCuller2021,Capote2024,Jia2024}. Generally, after the
known losses have been included in the accounting, the remaining
unknown loss is attributed to mode mismatch. In order to continue to
improve the detectors, it is important to understand this putative
mismatch in more detail, and to this end efforts are made to reconcile
various measurements with noise models.
The analysis of Refs.~\cite{Capote2024,Jia2024} included the effects
of internal quadratic mismatch as outlined in
Ref.~\cite{McCuller2021}.  Importantly, however, these studies did not
consider higher order aberrations and the potentially significant
frequency-dependent losses that they contribute due to their high-pass
dynamics. Attributing losses to external mismatch can incorrectly
account for the effects of internal mismatch up to a point, and this
can confuse efforts to characterize the detectors. Failing to properly
account for internal mismatch---especially the loss due to higher
order aberrations---will become increasingly insufficient as the arm
power is increased and thermal aberrations become more significant.

In practice, the thermal state of the interferometers drifts, which is
accompanied by changes both to the optical dynamics and to different
thermal aberrations generating different mode mismatch. This produces
numerous technical challenges and further complicates efforts to
characterize and improve the detectors. An understanding of the
effects of internal mode mismatch provides insight into the changing
behavior of the interferometers and can suggest measurements and
adjustments to make in order to diagnose and tune the state of the
detectors---both to decrease squeezing degradations and to improve
other technical difficulties.

Moving beyond the current detectors, an upgrade to the existing LIGO
facilities, known as LIGO \Asharp{}~\cite{PostO5}, and an upgrade to
the Virgo facility, known as Virgo\_nEXT~\cite{VirgoNEXT}, are being
planned with the goals of achieving \qty{10}{\dB} of broadband
frequency-dependent quantum noise reduction and \qty{1.5}{\MW} of
circulating arm power---roughly a factor of four more than the highest
power achieved to date. At the same time, the next generation of
gravitational wave observatories are being designed. In the United
States, Cosmic Explorer (CE) would be a \qty{40}{\km} long detector
with the same quantum noise reduction and arm power
targets~\cite{Evans2021}. In Europe, the Einstein Telescope (ET) would
include a \qty{10}{\km} long interferometer with \qty{3}{\MW} arm
power and the same quantum noise reduction~\cite{ETDesign2020}.
Achieving these ambitious goals will require significant advances in
the ability to characterize and understand the effects of mode
mismatch responsible for squeezing degradations and to control the
sources of the mismatch responsible for them. Furthermore, the success
of future detectors will benefit greatly from the simultaneous design,
from the start, of the optical layout and the thermal compensation
schemes informed by knowledge of these squeezing degradations and by
experience mitigating their effects in the current detectors.



The long arms of future detectors further introduce new challenges not
present in the current kilometer-scale detectors. Higher order modes
will become resonant in the arm cavities at frequencies
within the detection band. Internal mismatch---especially quadratic
mismatch---will then be responsible for significant squeezing
degradations at these frequencies where the HOMs that they excite
become resonant in the arms. Understanding the details of
these degradations is critical in designing the future detectors, but
such knowledge may also prove useful for characterizing and improving
the current detectors even though these degradations do not impact
their sensitivity.


In order to investigate these effects of internal mode mismatch, we study a three mirror
coupled cavity system which provides a good description of a gravitational-wave detector
for many purposes. The main numerical results of this paper, shown in all of the
figures, are thus obtained using a modal model of such a system, described in
\cref{sec:simulation}, which includes the exact couplings between many HOMs generated by
thermal aberrations. The discussion in the main body of the work, however, relies on a
simpler phenomenological model which includes a single HOM, along with simple
approximations of limited validity to this model, in order to better understand the
exact behavior. This is described in \cref{sec:cc-dynamics,sec:full-gwinc-model}.
\Cref{sec:thermal-aberrations} describes the two types of thermal aberrations and
connects their characteristics to the phenomenological parameters of this simpler model.
The resulting squeezing degradations are then discussed in \cref{sec:sqz-degradations},
and a summary of these effects and their implications for detector design choices is given
in \cref{subsec:detector-implications}. We conclude in \cref{sec:conclusion}.


We stress that we are not attempting to quantitatively account for the
quantum noise present in the current detectors and are not making
projections about the sensitivity that a future detector or upgrade
can reach. The quantitative results depend sensitively on the details
of the thermal aberrations---our treatment of these is
simplistic---and while briefly discussed, we have not included the
effects of external mode mismatch in our detailed analysis. The goal
of this work is simply to identify two types of internal mismatch in a
gravitational-wave detector and to explain the phenomenology of their
impact to the degradations to the squeezed states in order to inform
efforts to understand and improve the current detectors and to design
the future ones.

\section{Coupled Cavity Dynamics}
\label{sec:cc-dynamics}

A significant part of understanding quantum noise is understanding how
quantum vacuum propagates throughout an optomechanical system. The
ultimate goal of this section is thus to obtain the transfer functions
necessary to calculate the squeezing degradations in
\cref{sec:sqz-degradations} and to understand how the low-pass
dynamics of quadratic mismatch and the high-pass dynamics of higher
order aberrations arise.


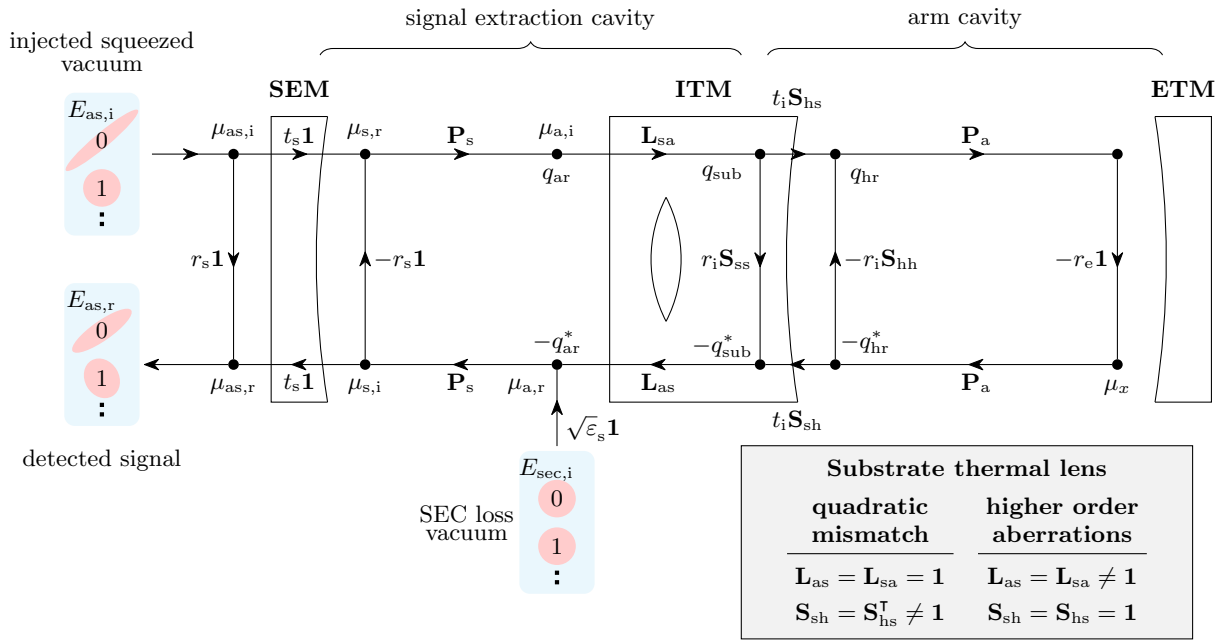
\begin{figure*}
    \centering
    \begin{tikzpicture}

        \def\ITMx{3.75}
        \def\ETMx{11}
        \def\SEMx{0}
        \def\L{1.65}  
        \def\M{3.8}  
        \def\dnx{0.5}  
        \def\dny{0.5}  
        \def\sag{0.15}  
        \def\ITMt{2.5}  
        \def\ETMt{0.75}  
        \def\dAL{0.75}  

        \newcommand{\drawArc}[4]{%
            \pgfmathsetmacro{\s}{#1}
            \pgfmathsetmacro{\y}{#2}
            \pgfmathsetmacro{\r}{\s / 2 + (\y * \y) / (8 * \s)}
            \pgfmathsetmacro{\theta}{asin(\y / (2 * \r))}
            \ifnum#4=1
                \draw #3 arc [start angle=-\theta, end angle=\theta, radius=\r];
            \else
                \draw #3 arc [start angle=180+\theta, end angle=180-\theta, radius=\r];
            \fi

        }

        \newcommand{\drawVac}[5]{%
            \def\H{1.9}  
            \def\W{1}  
            \def\vac{0.25}  
            \pgfmathsetmacro{\SQZf}{#1}
            \pgfmathsetmacro{\angf}{#2}
            \pgfmathsetmacro{\dQf}{\vac * \SQZf}
            \pgfmathsetmacro{\dPf}{\vac / \SQZf}
            \pgfmathsetmacro{\SQZh}{#4}
            \pgfmathsetmacro{\angh}{#5}
            \pgfmathsetmacro{\dQh}{\vac * \SQZh}
            \pgfmathsetmacro{\dPh}{\vac / \SQZh}
            \node (fund) at (0, \H/6) {};
            \node (hom) at (0, -\H/6) {};
            \fill[cyan!7, rounded corners=4, fill opacity=1]
                (-\W/2, -\H/2) rectangle (\W/2, \H/2);
            \node at (-\W/2 + 0.36, \H/2 - 0.25) {#3};
            \fill[red!20,rotate around={\angf:(fund)}, fill opacity=1]
                (fund) ellipse ({\dQf * 1cm} and {\dPf * 1cm});
            \fill[red!20,rotate around={\angh:(hom)}, fill opacity=1]
                (hom) ellipse ({\dQh * 1cm} and {\dPh * 1cm});
            \node at (fund) {0};
            \node at (hom) {1};
            \draw[dotted,ultra thick] (hom) ++(0, -1.3*\vac) -- ++(0, -0.2);
        }

        \begin{scope}[shift={(\ITMx, 0)}]
            \drawArc{0.2}{\L}{(\dAL, -\L/2)}{+1}
            \drawArc{0.2}{\L}{(\dAL, -\L/2)}{-1}
            \drawArc{\sag}{\M}{(\ITMt, -\M/2)}{-1}
            \node[dot] (hr_o) at (\ITMt + \dnx, \M/2 - \dny) {};
            \node[dot] (hr_i) at (\ITMt + \dnx, -\M/2 + \dny) {};
            \draw (\ITMt, \M/2) --node[above=1.5mm] {\textbf{ITM}} (0, \M/2)
                -- (0, -\M/2) -- (\ITMt, -\M/2);
            \node[dot] (sub_i) at (\ITMt - \dnx, \M/2 - \dny) {};
            \node[dot] (sub_o) at (\ITMt - \dnx, -\M/2 + \dny) {};
            \node[dot] (ar_i) at (-\dnx - 0.2, \M/2 - \dny) {};
            \node[dot] (ar_o) at (-\dnx - 0.2, -\M/2 + \dny) {};
        \end{scope}

        \begin{scope}[shift={(\ETMx, 0)}]
            \drawArc{\sag}{\M}{(0, -\M/2)}{+1}
            \draw (0, \M/2) --node[above=1.5mm] {\textbf{ETM}} (\ETMt, \M/2)
                -- (\ETMt, -\M/2) -- (0, -\M/2);
            \node[dot] (etm_i) at (-\dnx, \M/2 - \dny) {};
            \node[dot] (etm_o) at (-\dnx, -\M/2 + \dny) {};
        \end{scope}

        \begin{scope}[shift={(\SEMx, 0)}]
            \drawArc{\sag}{\M}{(0, -\M/2)}{-1}
            \draw (0, \M/2) --node[above=1.5mm] {\textbf{SEM}} (-\ETMt, \M/2)
                -- (-\ETMt, -\M/2) -- (0, -\M/2);
            \node[dot] (sem_fr_o) at (\dnx, \M/2 - \dny) {};
            \node[dot] (sem_fr_i) at (\dnx, -\M/2 + \dny) {};
            \node[dot] (sem_bk_i) at (-\ETMt - \dnx, \M/2 - \dny) {};
            \node[dot] (sem_bk_o) at (-\ETMt - \dnx, -\M/2 + \dny) {};
        \end{scope}

        \draw[midarrow] (sem_bk_i) --node[above] {$t_\ts\mat{1}$} (sem_fr_o);
        \draw[midarrow] (sem_fr_o) --node[above] {$\mat{P}_\ts$} (ar_i);
        \draw[midarrow] (ar_i) --node[above] {$\Lsa$} (sub_i);
        \draw[midarrow] (sub_i) --node[above=5mm] {$t_\text{i}\Mhs$} (hr_o);
        \draw[midarrow] (hr_o) --node[above] {$\mat{P}_\ta$} (etm_i);
        \draw[midarrow] (etm_i) --node[left] {$-r_\text{e}\mat{1}$} (etm_o);
        \draw[midarrow] (etm_o) --node[below] {$\mat{P}_\ta$} (hr_i);
        \draw[midarrow] (hr_i) --node[right] {$-r_\text{i}\Mhh$} (hr_o);
        \draw[midarrow] (hr_i) --node[below=5mm] {$t_\text{i}\Msh$} (sub_o);
        \draw[midarrow] (sub_i) --node[left] {$r_\text{i}\Mss$} (sub_o);
        \draw[midarrow] (sub_o) --node[below] {$\Las$} (ar_o);
        \draw[midarrow] (ar_o) --node[below] {$\mat{P}_\ts$} (sem_fr_i);
        \draw[midarrow] (sem_fr_i) --node[right] {$-r_\ts\mat{1}$} (sem_fr_o);
        \draw[midarrow] (sem_fr_i) --node[below] {$t_\ts\mat{1}$} (sem_bk_o);
        \draw[midarrow] (sem_bk_i) --node[left] {$r_\ts\mat{1}$} (sem_bk_o);

        \node[shift={(-0.5, 0.25)}] at (sub_o) {$-\qsub^*$};
        \node[shift={(0, 0.3)}] at (ar_o) {$-\qar^*$};
        \node[shift={(0.4, 0.3)}] at (hr_i) {$-\qhr^*$};
        \node[shift={(0, -0.3)}] at (ar_i) {$\qar$};
        \node[shift={(-0.5, -0.25)}] at (sub_i) {$\qsub$};
        \node[shift={(0.4, -0.3)}] at (hr_o) {$\qhr$};

        \node[shift={(0, 0.3)}] at (ar_i) {$\mu_\text{a,i}$};
        \node[shift={(-0.4, -0.3)}] at (ar_o) {$\mu_\text{a,r}$};
        \node[shift={(0, 0.3)}] at (sem_bk_i) {$\mu_\text{as,i}$};
        \node[shift={(0, -0.3)}] at (sem_bk_o) {$\mu_\text{as,r}$};
        \node[shift={(0, -0.3)}] at (etm_o) {$\mu_x$};
        \node[shift={(0, -0.3)}] at (sem_fr_i) {$\mu_\text{s,i}$};
        \node[shift={(0, 0.3)}] at (sem_fr_o) {$\mu_\text{s,r}$};

        \begin{scope}[shift={(\SEMx - 3, \M/3)}]
            \drawVac{2.5}{40}{$E_\text{as,i}$}{1}{120}
            \node at (0, 1.5) {\shortstack{injected squeezed\\vacuum}};
        \end{scope}

        \begin{scope}[shift={(\SEMx - 3, -\M/3)}]
            \drawVac{1.9}{35}{$E_\text{as,r}$}{1.1}{120}
            \node at (0, -1.4) {detected signal};
        \end{scope}

        \node[shift={(0, -1.2)}] (sec_in) at (ar_o) {};
        \begin{scope}[shift={($(ar_o) - (0, 2.1cm)$)}]
            \drawVac{1}{0}{$E_\text{sec,i}$}{1}{0}
            \node at (-1.2, 0) {\shortstack{SEC loss\\ vacuum}};
        \end{scope}

        \node[shift={(-1.2, 0)}] (as_inj) at (sem_bk_i) {};
        \node[shift={(-1.2, 0)}] (as_refl) at (sem_bk_o) {};
        \draw[midarrow] (as_inj) -- (sem_bk_i);
        \draw[endarrow] (sem_bk_o) -- (as_refl);
        \draw[midarrow] (sec_in) --node[right, yshift=-2ex]
            {$\sqrt{\varepsilon}_\ts\mat{1}$} (ar_o);

        \draw[decorate,decoration={brace,amplitude=5pt,raise=6ex}]
            (\SEMx - 0.1, \M/2) -- (\ITMx + \ITMt - 0.75, \M/2)
            node[midway,yshift=4em]{signal extraction cavity};
        \draw[decorate,decoration={brace,amplitude=5pt,raise=6ex}]
            (\ITMx + \ITMt - 0.4, \M/2) -- (\ETMx + 0.1, \M/2)
            node[midway,yshift=4em]{arm cavity};

        \begin{scope}[shift={(\ETMx - 2.5, -\M + 0.2)}]
            \filldraw[fill=gray!10] (-3, -1.45) rectangle (3, 1.12);
            \node at (0, 0) {%
                \begin{minipage}{10cm}
                    \begin{equation*}
                    \begin{array}{c@{\hspace{1em}}c}
                        \multicolumn{2}{c}{\textbf{Substrate thermal lens}} \\
                        \noalign{\vskip 4pt}
                        \begin{array}{c}
                            \textbf{quadratic} \\
                            \textbf{mismatch} \\
                            \noalign{\vskip 1pt}
                            \hline
                            \noalign{\vskip 4pt}
                            \begin{aligned}
                                \Las &= \Lsa = \mat{1} \\
                                \Msh &= \Mhs^\intercal \neq \mat{1}
                            \end{aligned}
                        \end{array}
                        &
                        \begin{array}{c}
                            \textbf{higher order} \\
                            \textbf{aberrations} \\
                            \noalign{\vskip 1pt}
                            \hline
                            \noalign{\vskip 4pt}
                            \begin{aligned}
                                \Las &= \Lsa \neq \mat{1} \\
                                \Msh &= \Mhs = \mat{1}
                            \end{aligned}
                        \end{array}
                    \end{array}
                    \end{equation*}

                \end{minipage}
            };
        \end{scope}
    \end{tikzpicture}
    \caption{Coupled cavity equivalent to the differential arm mode of a
      gravitational-wave detector. The ellipses and circles represent (potentially
      squeezed) quantum vacuum of the electromagnetic field at several locations. The 0
      modes represent the fundamental vacuum and the 1 and subsequent modes represent
      the HOM vacuum. Several nodes of the system are marked with the $\mu_{i}$ labels.
      The squeezed states are injected at the node $\mu_\text{as,i}$ after being
      reflected off of the filter cavity (not shown but included in the analysis). The
      signal is detected at the node $\mu_\text{as,r}$ after passing through the output
      mode cleaner which ensures that only the fundamental mode is measured. The power
      absorbed in the test mass coatings generates a thermorefractive lens in the ITM
      substrate, depicted as a thin lens in the figure, and deforms the HR surfaces of
      the ITM and ETM. The thermal aberrations can be decomposed into quadratic mismatch
      and higher order aberrations as shown in \cref{fig:thermal-lens}. As summarized in
      \cref{tab:aberration-wide}, the operators $\mat{L}_{ij}$ describe how a field
      propagates through the lens, i.e.\ the ITM substrate, and the $\mat{S}_{ij}$
      describe how a field transmits through or reflects from the HR surface. The
      Gaussian $q$ parameters at several nodes important to the analysis of
      \cref{sec:thermal-aberrations} are also shown. Reversed parameters $-q^*$ describe
      beams propagating in a coordinate system with an inversion in the tangential
      direction. The box summarizes the main result of \cref{sec:thermal-aberrations}
      for thermal aberrations generated by a substrate thermal lens. (While $\Las = \Lsa
      = \mat{1}$ for the surface deformations, $\Msh$ and $\Mhs$ are not always
      $\mat{1}$ for the substrate lens since the lens can change the mode of the SEC and
      thus the coupling between the fields on either side of the HR surface.)
    }
    \label{fig:coupled-cavity}
\end{figure*}

Modern gravitational-wave detectors employ the so-called dual-recycled
Fabry--Perot Michelson (DRFPMI) optical topology. In this
configuration, the arms of a Michelson interferometer are made into
two mirror Fabry--Perot cavities each made up of an end test mass (ETM)
and an input test mass (ITM) in order to increase the power stored in
the arm cavities and to increase the sensitivity to gravitational wave
signals. A signal extraction mirror (SEM) is placed between the
beamsplitter and the readout of the interferometer---thus forming an
optical cavity with the ITMs known as the signal extraction cavity
(SEC)---in order to tune the response of the detector to a
gravitational wave signal. The SEC can be operated in the resonant
sideband extraction (RSE) configuration which broadens the
bandwidth~\cite{Mizuno1993} or in the signal recycling configuration
(SR) which narrows the bandwidth~\cite{Meers1988}. To a good
approximation, the dynamics of the differential arm motion of a DRFPMI
can be described by a three mirror coupled cavity as illustrated in
\cref{fig:coupled-cavity}. A single ITM and ETM form an effective arm
cavity sensitive to differential arm motion, and the SEC is formed by
the addition of an SEM.


We review the dynamics of both RSE and SR with perfect mode matching
in \cref{subsec:rse-sr-matched} before delving into the details of RSE
in the presence of mode mismatch in \cref{subsec:rse-mismatched}. We
will ultimately be interested only in RSE, the case relevant to
gravitational-wave detectors, however the dynamics of the higher order
modes in an interferometer operating in the RSE configuration will
sometimes be those of the fundamental mode operating in the SR
configuration and it is therefore useful to have an understanding of
both.

As this work is primarily focused on the effects of mode mismatch, we
ignore optical loss in most of the discussion in order to simplify the
analytic expressions. However, all sources of loss listed in
\cref{tab:detector-parameters} are included in all of the numerical
results presented in the figures. Furthermore, the rational
approximations given to the exact expressions are valid in the limit
of large cavity finesses and for frequencies far below the free
spectral range (FSR). While some require corrections to account for
the relatively low SEC finesse of LIGO and the low FSR of CE, they
properly account for the gross dynamics of the fields and their
scalings with detector parameters in all cases.

\subsection{Resonant sideband extraction and signal recycling with perfect mode matching}
\label{subsec:rse-sr-matched}

\Cref{fig:coupled-cavity} depicts an operator graph of a general
coupled cavity system with many higher order modes (HOMs) in the
presence of mode mismatch and thermal aberrations which will be
analyzed in \cref{sec:thermal-aberrations}. In the simple case of a
single mode in a perfectly matched system considered here, the lens
and surface operators describing the aberrations are $\mat{L}_{ij} =
\mat{S}_{ij} \to 1$. If the arm cavity is detuned from resonance by a
frequency $\delta\omega_\ta$ and the SEC is detuned from resonance by
an angle $\phi_\ts$, the one-way propagation of a field through the
arm cavity $\mat{P}_\ta$ and through the SEC $\mat{P}_\ts$ at Fourier frequency $\Omega$ are
\begin{equation}
  \mat{P}_\ta \to \rme^{-\rmi(\dwa)L_\ta / c}, \quad
  \mat{P}_\ts \to \rme^{-\rmi(\Omega L_\ts / c + \phi_\ts)}
\end{equation}
where $L_\ta$ and $L_\ts$ are the lengths of the arm cavity and SEC,
respectively. The arm cavity detuning could be due to introducing an
offset $\Delta L_\ta$ in the length of the cavity so that
$\delta\omega_\ta = -ck \Delta L_\ta/L_\ta$ for the fundamental
mode. If the field is a higher order mode, the detuning could be due
to the additional one-way Gouy phase $\psi_\ta$ that the HOM
accumulates relative to the fundamental which may itself be on
resonance with $\Delta L_\ta=0$. In this case $\delta\omega_\ta =
\omega_\text{fsr} \psi_\ta / \pi$ where $\omega_\text{fsr} = \pi
c/L_\ta$ is the free spectral range (FSR).

Consider first the dynamics of the arm alone, unmodified by the SEC,
without the presence of the SEM. A field in the cavity experiences a
round-trip gain of $r_\ti r_\text{e} \rme^{-2\rmi(\dwa)L_\ta/c}$. When
this open-loop gain is positive, the cavity enhances the dynamics of a
field in the cavity; when it is negative, the dynamics are
suppressed. These dynamics are described by the arm cavity optical loop
suppression $H_\ta(\Omega)=[1 - r_\ti r_\text{e} \rme^{-2\rmi(\dwa)L_\ta/c}]^{-1}$.

We will be chiefly interested in two quantities. The first is the
reflection of a field off of the cavity
\begin{equation}
  \rfr_\ta(\Omega) = \frac{E_\text{a,r}}{E_\text{a,i}} = r_\ti -
  \frac{t_\ti^2 r_\text{e}\rme^{-2\rmi(\dwa)L_\ta/c}}{1 - r_\ti
    r_\text{e}\rme^{-2\rmi(\dwa)L_\ta/c}}
  \label{eq:arm-reflection-planewave}
\end{equation}
which is made by the interference between the field which is promptly
reflected from the ITM and the field which enters the cavity,
experiences its closed-loop dynamics, and then leaks back out. The
second is the transmission of a signal through the cavity
\begin{equation}
  \\ \tfr_\ta(\Omega) = \frac{E_\text{a,r}}{E_x} =
  \frac{t_\ti\rme^{-\rmi(\dwa)L_\ta/c}}{1 - r_\ti
    r_\text{e}\rme^{-2\rmi(\dwa)L_\ta/c}}
  \label{eq:arm-transmission-planewave}
\end{equation}
where $E_x$ is the field reflecting off of the ETM surface at the node
$\mu_x$ in \cref{fig:coupled-cavity}.

Gravitational-wave detectors employ so-called high finesse overcoupled
cavities where the ITM is highly reflective yet still less
reflective than the ETM: $1 \gg t_\ti > t_\text{e} \approx
0$. For such an overcoupled cavity,
\cref{eq:arm-transmission-planewave,eq:arm-reflection-planewave} can
be written in a zero, pole, gain form as
\begin{align}
  \rfr_\ta(\Omega) &= - \frac{1 - \rmi(\dwa)/\wa}{1 +
    \rmi(\dwa)/\wa} \label{eq:arm-reflection-pade} \\
  \tfr_\ta(\Omega)
  &= \sqrt{\frac{2\Fa}{\pi}}\frac{1}{1 + \rmi(\dwa)/\wa}
  \label{eq:arm-transmission-pade}
\end{align}
where the half-bandwidth of the arm $\wa$, also known as the arm
cavity pole, and arm cavity finesse are defined as
\begin{subequations}
\label{eq:cav-finesse-pole}
\begin{align}
\Fa &= \frac{\pi}{1 - r_\ti},\quad \wa = \frac{c}{L_\ta}\frac{1 -
  r_\ti}{1 + r_\ti} = \frac{\pi c}{2\Fa L_\ta} \\ \Fs &= \frac{\pi}{1
  - r_\ts},\quad \wsec = \frac{c}{L_\ts}\frac{1 - r_\ts}{1 + r_\ts} =
\frac{\pi c}{2\Fs L_\ts}
\end{align}
\end{subequations}
For future use, we have also defined the SEC finesse $\Fs$ and SEC
pole $\wsec$. Of particular note, the cavity imparts a $\pi$ phase
shift on reflection for fields resonant in the cavity:
$\rfr_\ta(|\dwa|\ll\wa) = -1$, rather than $+r_\ti$; and imparts
no phase shift for fields far from resonance:
$\rfr_\ta(|\dwa|\gg\wa) = +1$. This sign change around the cavity
pole has important implications for the coupled cavity
dynamics.\footnote{It is sometimes said that a non-resonant field does
not ``see the cavity,'' but this is misleading as the field still
enters the cavity and the ensuing dynamics, while being suppressed by
the arm loop rather than being enhanced by it, still interferes with
and modifies the prompt reflection. In addition to changing the
magnitude of the reflection---even more so than
\cref{eq:arm-reflection-pade} suggests in the presence of optical
loss---it also imparts a small phase shift for a field which is not
exactly anti-resonant, i.e.\ $\dwa \neq N\omega_\text{fsr}$ for integer $N$.}

Next consider how the SEC, formed by the addition of the SEM, modifies
the dynamics of the arm cavity. If the SEC is tuned so that the
fundamental accumulates a one-way phase shift $\phi_\ts$ in the
cavity, then the reflection of a field off of the coupled cavity is
again made by the interference of the field which is promptly
reflected from the SEM and the field that enters the coupled cavity:
\begin{equation}
  \rfr_\text{cc}(\Omega) = \frac{E_\text{as,r}}{E_\text{as,i}}
   = r_\ts +
  \frac{t_\ts^2 \rfr_\ta(\Omega)\rme^{-2\rmi(\Omega L_\ts / c + \phi_\ts)}}
       {1 + r_\ts \rfr_\ta(\Omega) \rme^{-2\rmi(\Omega L_\ts / c + \phi_\ts)}},
       \label{eq:cc-reflection}
\end{equation}
and the transmission through the coupled cavity is
\begin{equation}
  \tfr_\text{cc}(\Omega) = \frac{E_\text{as,r}}{E_x}
   = \frac{t_\ts \tfr_\ta(\Omega) \rme^{-\rmi(\Omega L_\ts / c + \phi_\ts)}}
      {1 + r_\ts \rfr_\ta(\Omega) \rme^{-2\rmi(\Omega L_\ts / c + \phi_\ts)}}.
      \label{eq:cc-transmission}
\end{equation}
This is exactly like the single arm cavity with the arm cavity itself
serving as a compound mirror and taking the place of the ETM and with
the SEM taking the place of the ITM.

We will only consider in detail the extreme cases of signal recycling
(SR)~\cite{Meers1988}, where $\phi_\ts=0$, and resonant sideband
extraction (RSE)~\cite{Mizuno1993}, where $\phi_\ts=\pi/2$. In these
two cases, the SEC optical loop suppression, evident in
\cref{eq:cc-reflection,eq:cc-transmission}, is
\begin{align}
  H_\ts(\Omega) &= \frac{1}{1 + r_\ts
    \rfr_\ta(\Omega)\rme^{-2\rmi(\Omega L_\ts/c + \phi_\ts)}}
  \nonumber \\ &= \frac{1}{1 \pm r_\ts} \frac{1 + \rmi (\dwa) / \wa}{1
    + \rmi(\dwa)/\wcc}
  \label{eq:fundamental-sec-closed-loop}
\end{align}
where the coupled cavity pole is
\begin{equation}
  \wcc = \frac{1 \pm r_\ts}{1 \mp r_\ts}\wa =
  \begin{cases}
    \displaystyle \wrse = \frac{2\Fs}{\pi}\wa & \phi_\ts = \pi/2
    \\[1em] \displaystyle \wsr = \frac{\pi}{2\Fs}\wa & \phi_\ts = 0
  \end{cases}
  \label{eq:cc-pole}
\end{equation}
and where the upper sign is taken for RSE and the lower for SR. With
RSE, the field picks up a round-trip propagation phase of $\pi$. Since
the arm cavity reflection is negative for frequencies $|\dwa|\ll\wa$
within the arm cavity bandwidth, and since the HR surface of the SEM
is inside the cavity, the total round-trip phase is negative and the
SEC suppresses the dynamics by a factor of 2 within the arm cavity
bandwidth. Outside the arm cavity bandwidth, the reflection changes
sign, the round-trip phase is positive, and the SEC enhances the
dynamics by a factor of $\Fs/\pi$ for frequencies $|\dwa|\gg\wrse$
outside the RSE bandwidth. With SR, on the other hand, these phasings
differ by $\pi$ since a field in the cavity does not accumulate any
propagation phase at low frequencies. Therefore for SR, the SEC
enhances the dynamics by a factor of $\Fs/\pi$ for frequencies
$|\dwa|\ll\wsr$ within the SR bandwidth and suppresses them by a
factor of 2 outside the arm cavity bandwidth.

The reflection from the coupled cavity can then be written in a zero,
pole, gain form as
\begin{equation}
  \rfr_\text{cc}(\Omega) = \begin{cases}
    \displaystyle
    \rfr_\text{rse} = \frac{1 - \rmi(\dwa)/\wrse}{1 + \rmi(\dwa)/\wrse} \\[1em]
    \rfr_\text{sr} = -\displaystyle\frac{1 - \rmi(\dwa)/\wsr}{1 + \rmi(\dwa)/\wsr}
  \end{cases},
  \label{eq:cc-reflection-zpk}
\end{equation}
and the transmission through the coupled cavity can be written as
\begin{equation}
  \tfr_\text{cc}(\Omega) = \begin{cases}
    \displaystyle \tfr_\text{rse} =
    -\sqrt{\frac{\Fa}{\Fs}}\frac{\rmi}{1 + \rmi(\dwa)/\wrse} \\[1em]
    \displaystyle \tfr_\text{sr} =
    \sqrt{\frac{4\Fa\Fs}{\pi^2}}\frac{1}{1 + \rmi(\dwa)/\wsr}
  \end{cases}.
  \label{eq:cc-transmission-zpk}
\end{equation}
The coupled cavity therefore behaves like the simple arm cavity with three differences.
First, the bandwidth of the coupled cavity is broadened with RSE and narrowed with SR by
a factor of $2\Fs/\pi$ (\cref{eq:cc-pole}). Second, the DC gain of the cavity is
decreased with RSE and increased with SR by a factor of $\sqrt{2\Fs/\pi}$
(cf.\ \cref{eq:arm-transmission-pade,eq:cc-transmission-zpk}). And third, the RSE reflection
phase differs by that of the arm reflection by a factor of $\pi$, and the RSE
transmission is rotated by $\pi/2$ relative to the arm transmission.

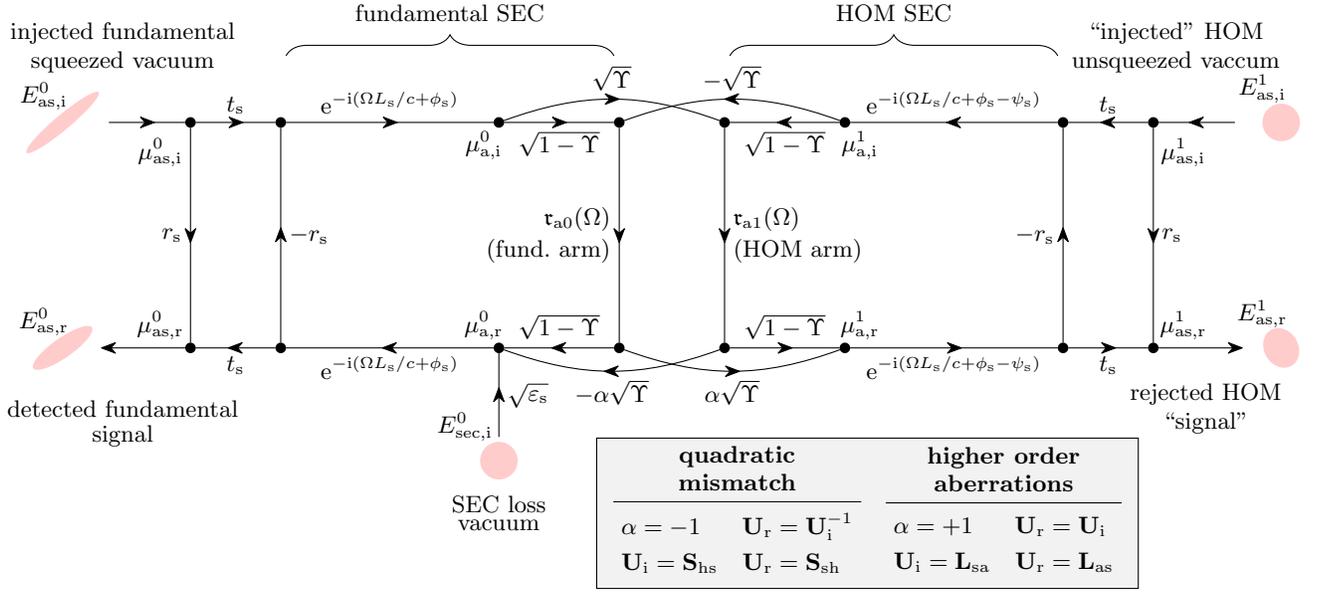
\begin{figure*}
    \centering
    \begin{tikzpicture}
        \def\H{1.5}  
        \def\Wsem{1.2}  
        \def\Ls{2.9}  
        \def\Wmm{1.6}  
        \def\dCav{5.2}  
        \pgfmathsetmacro{\Lcc}{\Wsem + \Ls + \Wmm + \dCav}

        \newcommand{\drawVac}[2]{%
            \def\vac{0.25}  
            \pgfmathsetmacro{\SQZ}{#1}
            \pgfmathsetmacro{\ang}{#2}
            \pgfmathsetmacro{\dQ}{\vac * \SQZ}
            \pgfmathsetmacro{\dP}{\vac / \SQZ}
            \node (fund) at (0, 0) {};
            \fill[red!20,rotate around={\ang:(fund)}, fill opacity=1]
                (fund) ellipse ({\dQ * 1cm} and {\dP * 1cm});
        }

        \begin{scope}[shift={(-\dCav, 0)}]
            \node[dot] (sem_fr_o_f) at (0, \H) {};
            \node[dot] (sem_fr_i_f) at (0, -\H) {};
            \node[dot] (sem_bk_o_f) at (-\Wsem, -\H) {};
            \node[dot] (sem_bk_i_f) at (-\Wsem, \H) {};
            \node[dot] (arm_i_f) at (\Ls, \H) {};
            \node[dot] (arm_o_f) at (\Ls, -\H) {};
            \node[dot] (mm_i_f) at (\Ls + \Wmm, \H) {};
            \node[dot] (mm_o_f) at (\Ls + \Wmm, -\H) {};

            \draw[midarrow] (sem_bk_i_f) --node[above] {$t_\ts$} (sem_fr_o_f);
            \draw[midarrow] (sem_fr_i_f) --node[right] {$-r_\ts$} (sem_fr_o_f);
            \draw[midarrow] (sem_fr_i_f) --node[below] {$t_\ts$} (sem_bk_o_f);
            \draw[midarrow] (sem_bk_i_f) --node[left] {$r_\ts$} (sem_bk_o_f);
            \draw[midarrow] (sem_fr_o_f)
                --node[above] {$\rme^{-\rmi (\Omega L_\ts/c + \phi_\ts)}$} (arm_i_f);
            \draw[midarrow] (arm_i_f)
                --node[below] {$\sqrt{1 - \Upsilon}$} (mm_i_f);
            \draw[midarrow] (mm_i_f)
                --node[left] {\shortstack[r]{$\rfr_{\ta0}(\Omega)$\\ (fund.\ arm)}} (mm_o_f);
            \draw[midarrow] (mm_o_f)
                --node[above] {$\sqrt{1 - \Upsilon}$} (arm_o_f);
            \draw[midarrow] (arm_o_f)
                --node[below] {$\rme^{-\rmi (\Omega L_\ts/c + \phi_\ts)}$} (sem_fr_i_f);

            \draw[decorate,decoration={brace,amplitude=8pt,raise=6.5ex}]
                (sem_fr_o_f) -- (mm_i_f)
                node[midway,yshift=4.5em]{fundamental SEC};
        \end{scope}

        \begin{scope}[shift={(+\dCav, 0)}]
            \node[dot] (sem_fr_o_h) at (0, \H) {};
            \node[dot] (sem_fr_i_h) at (0, -\H) {};
            \node[dot] (sem_bk_o_h) at (+\Wsem, -\H) {};
            \node[dot] (sem_bk_i_h) at (+\Wsem, +\H) {};
            \node[dot] (arm_i_h) at (-\Ls, \H) {};
            \node[dot] (arm_o_h) at (-\Ls, -\H) {};
            \node[dot] (mm_i_h) at (-\Ls - \Wmm, \H) {};
            \node[dot] (mm_o_h) at (-\Ls - \Wmm, -\H) {};

            \draw[midarrow] (sem_bk_i_h) --node[above] {$t_\ts$} (sem_fr_o_h);
            \draw[midarrow] (sem_fr_i_h) --node[left] {$-r_\ts$} (sem_fr_o_h);
            \draw[midarrow] (sem_fr_i_h) --node[below] {$t_\ts$} (sem_bk_o_h);
            \draw[midarrow] (sem_bk_i_h) --node[right] {$r_\ts$} (sem_bk_o_h);
            \draw[midarrow] (sem_fr_o_h)
                --node[above] {$\rme^{-\rmi (\Omega L_\ts/c + \phi_\ts - \psi_\ts)}$} (arm_i_h);
            \draw[midarrow] (arm_i_h)
                --node[below] {$\sqrt{1 - \Upsilon}$} (mm_i_h);
            \draw[midarrow] (mm_i_h)
                --node[right] {\shortstack[l]{$\rfr_{\ta1}(\Omega)$\\ (HOM arm)}} (mm_o_h);
            \draw[midarrow] (mm_o_h)
                --node[above] {$\sqrt{1 - \Upsilon}$} (arm_o_h);
            \draw[midarrow] (arm_o_h)
                --node[below] {$\rme^{-\rmi (\Omega L_\ts/c + \phi_\ts - \psi_\ts)}$} (sem_fr_i_h);

            \draw[decorate,decoration={brace,amplitude=8pt,raise=6.5ex}]
                (mm_i_h) -- (sem_fr_o_h)
                node[midway,yshift=4.5em]{HOM SEC};
        \end{scope}

        \draw[midarrow, bend left=20] (arm_i_f) to node[above] {$\sqrt{\Upsilon}$} (mm_i_h);
        \draw[midarrow, bend right=20] (arm_i_h) to node[above] {$-\sqrt{\Upsilon}$} (mm_i_f);
        \draw[midarrow, bend right=20] (mm_o_f) to node[below] {$\alpha\sqrt{\Upsilon}$} (arm_o_h);
        \draw[midarrow, bend left=20] (mm_o_h) to node[below] {$-\alpha\sqrt{\Upsilon}$} (arm_o_f);

        \node[shift={(-1.2, 0)}] (as_inj_f) at (sem_bk_i_f) {};
        \node[shift={(-1.2, 0)}] (as_refl_f) at (sem_bk_o_f) {};
        \draw[midarrow] (as_inj_f) -- (sem_bk_i_f);
        \draw[endarrow] (sem_bk_o_f) -- (as_refl_f);

        \node[shift={(1.2, 0)}] (as_inj_h) at (sem_bk_i_h) {};
        \node[shift={(1.2, 0)}] (as_refl_h) at (sem_bk_o_h) {};
        \draw[midarrow] (as_inj_h) -- (sem_bk_i_h);
        \draw[endarrow] (sem_bk_o_h) -- (as_refl_h);

        \node[shift={(-0.4, -0.4)}] at (sem_bk_i_f) {$\mu_\text{as,i}^0$};
        \node[shift={(-0.4, 0.3)}] at (sem_bk_o_f) {$\mu_\text{as,r}^0$};
        \node[shift={(0.4, -0.4)}] at (sem_bk_i_h) {$\mu_\text{as,i}^1$};
        \node[shift={(0.4, 0.3)}] at (sem_bk_o_h) {$\mu_\text{as,r}^1$};
        \node[shift={(-0.2, 0.3)}] at (arm_o_f) {$\mu_\text{a,r}^0$};
        \node[shift={(-0.2, -0.3)}] at (arm_i_f) {$\mu_\text{a,i}^0$};
        \node[shift={(0.2, 0.3)}] at (arm_o_h) {$\mu_\text{a,r}^1$};
        \node[shift={(0.2, -0.3)}] at (arm_i_h) {$\mu_\text{a,i}^1$};

        \begin{scope}[shift={($(as_inj_f) - (5mm, 0)$)}]
            \drawVac{2.5}{40}
            \node at (-0.25, 0.35) {$E_\text{as,i}^0$};
            \node at (0.8, 1) {\shortstack[c]{injected fundamental\\ squeezed vacuum}};
        \end{scope}

        \begin{scope}[shift={($(as_refl_f) - (5mm, 0)$)}]
            \drawVac{1.9}{35}
            \node at (-0.25, 0.35) {$E_\text{as,r}^0$};
            \node at (0.8, -1) {\shortstack[c]{detected fundamental\\ signal}};
        \end{scope}

        \begin{scope}[shift={($(as_inj_h) + (5mm, 0)$)}]
            \drawVac{1}{0}
            \node at (-0.25, 0.45) {$E_\text{as,i}^1$};
            \node at (-1.4, 1) {\shortstack{``injected'' HOM\\ unsqueezed vaccum}};
        \end{scope}

        \begin{scope}[shift={($(as_refl_h) + (5mm, 0)$)}]
            \drawVac{1.1}{120}
            \node at (-0.25, 0.45) {$E_\text{as,r}^1$};
            \node at (-1, -0.8) {\shortstack{rejected HOM\\ ``signal''}};
        \end{scope}

        \begin{scope}[shift={($(arm_o_f) - (0, 1.5)$)}]
            \drawVac{1}{0}
            \node (sec_loss) at (0, 0.2) {};
            \draw[midarrow] (sec_loss) to node[right] {$\sqrt{\varepsilon_\ts}$} (arm_o_f);
            \node at (-0.45, 0.45) {$E_\text{sec,i}^0$};
            \node at (0, -0.7) {\shortstack{SEC loss\\ vacuum}};
        \end{scope}

        \begin{scope}[shift={(\Ls - 0.3, -\H - 2)}]
            \filldraw[fill=gray!10] (-3.6, -1.2) rectangle (3.6, 0.8);
            \node at (0, 0) {%
                \begin{minipage}{10cm}
                    \begin{equation*}
                    \begin{array}{c@{\hspace{1em}}c}
                        \begin{array}{c}
                            \textbf{quadratic} \\
                            \textbf{mismatch} \\
                            \noalign{\vskip 1pt}
                            \hline
                            \noalign{\vskip 4pt}
                            \begin{alignedat}{2}
                                & \alpha = -1 \quad & \Ur &= \Ui^{-1} \\
                                & \Ui = \Mhs \quad & \Ur &= \Msh
                            \end{alignedat}
                        \end{array}
                        &
                        \begin{array}{c}
                            \textbf{higher order} \\
                            \textbf{aberrations} \\
                            \noalign{\vskip 1pt}
                            \hline
                            \noalign{\vskip 4pt}
                            \begin{alignedat}{2}
                                & \alpha = +1 \quad & \Ur &= \Ui \\
                                & \Ui = \Lsa \quad & \Ur &= \Las
                            \end{alignedat}
                        \end{array}
                    \end{array}
                    \end{equation*}
                \end{minipage}
            };
        \end{scope}
    \end{tikzpicture}
    \caption{Coupled cavity signal flow diagram for the model of the fundamental and a
      single HOM described in \cref{subsec:rse-mismatched}. This system can be thought
      of as two separate coupled cavities which are coupled by the mismatch between the
      SEC and arm cavity at the ITM. As explained by
      \cref{eq:arm-reflection-interference}, the HOM SEC is AC coupled with the
      fundamental SEC through higher order aberrations while being DC coupled through
      quadratic mismatch due to the different interference conditions (parametrized by
      $\alpha$) between the HOM and the fundamental. The vacuum and important nodes from
      \cref{fig:coupled-cavity} are reproduced here with the superscripts 0 and 1 to
      indicate the fundamental and HOM, respectively. Rather than drawing the arm
      cavities in full, the fundamental arm cavity $\rfr_{\ta 0}(\Omega)$ is given by
      \cref{eq:arm-reflection-pade} with $\delta\omega_\ta=0$ and the HOM arm cavity
      $\rfr_{\ta 1}(\Omega)$ is given by \cref{eq:arm-reflection-pade} with
      $\delta\omega_\ta = \omega_\text{fsr}\,\psi_\ta / \pi$. The one-way Gouy phases of
      the arm cavity and the SEC are $\psi_\ta$ and $\psi_\ts$, respectively, and the
      SEC detuning $\phi_\ts$ is $\pi/2$ for RSE. The field $E_\text{as,i}^0$ is the
      squeezed fundamental mode of the field injected into the interferometer;
      $E_\text{as,i}^1$ is the unsqueezed HOM of this injected field. The field
      $E_\text{as,r}^0$ is the fundamental mode of the signal field which is detected by
      photodetectors; $E_\text{as,r}^1$ is the HOM of this signal field which is
      rejected by the output mode cleaner. Note that only a single mode propagates along
      the edges here while the full vector of HOMs propagate along the edges shown in
      \cref{fig:coupled-cavity}. The box summarizes how this model is related to the
      operators describing the exact dynamics shown in \cref{fig:coupled-cavity} and
      described in detail in \cref{subsec:cc-aberrations,tab:aberration-wide}. }
    \label{fig:phenom-cc-sflow}
\end{figure*}

\Cref{eq:cc-transmission-zpk} describes the propagation of a field leaving the ETM
($E_x(\Omega)$) to a field measured in reflection of the coupled cavity
($E_\text{as,r}(\Omega)$). In the context of gravitational-wave detectors, we will be
interested in the transduction of ETM motion to the fundamental mode of the same optical
field measured in reflection of the coupled cavity. The modulation of the position of a
mirror $x(\Omega)$ produces a modulation $E_x(\Omega) = 2k\,x(\Omega)\sqrt{P_\ta}$ in
the phase of an electromagnetic field of power $P_\ta$ reflected from that mirror. These
fluctuations will then be rotated into the amplitude quadrature due to the extra $\pi/2$
rotation for RSE, and so the amplitude quadrature (with respect to the light generating
the phase signal) is generally measured.  Furthermore, in the main text we will
ultimately only be interested in the case where $\delta\omega_\ta=0$ for the fundamental
so that $\tfr_\text{rse}(+\Omega) = -\tfr_\text{rse}^*(-\Omega)$, and so the optical
response to mirror motion, known as the optomechanical plant, is given by
\begin{equation}
  C(\Omega) = \frac{1}{\sqrt{2}} \times \frac{E_x(\Omega)}{x(\Omega)}
    = \frac{1}{\sqrt{2}}\times 2k\sqrt{P_\ta}\, \tfr_\text{rse}(\Omega).
  \label{eq:optical-plant}
\end{equation}
The extra factor of $1/\sqrt{2}$ accounts for the presence of the beamsplitter when
mapping the dynamics of the coupled cavity onto those of an interferometric
gravitational-wave detector.

Finally, we will occasionally be interested in the
radiation-pressure-mediated-correlation of the upper and lower
sidebands which is responsible for quantum radiation pressure noise
and ponderomotive squeezing. This is best analyzed in the two-photon
formalism~\cite{Caves1985}, but for our purposes it is enough to note
that an amplitude quadrature fluctuation of a field incident on a
mirror is converted into a phase quadrature fluctuation of the field
on reflection of that mirror through the radiation pressure force
$F_\text{rp}(\Omega)$ of the laser light acting on the mirror. This
force produces a motion $x(\Omega) = \chi(\Omega) F_\text{rp}(\Omega)$
of the mirror where $\chi(\Omega)$ is the mechanical susceptibility,
or mechanical plant. For a field of power $P_\ta$ incident on a
perfectly reflecting free mass mirror with susceptibility
$\chi(\Omega) = -1/M\Omega^2$, this optomechanical coupling
is~\cite{Danilishin2012}
\begin{subequations}
\label{eq:free-mass-rp-coupling}
\begin{align}
  \Krp_\text{fm}(\Omega) &= \frac{8k\chi(\Omega) P_\ta}{c} = -\frac{8kP_\ta}{cM\Omega^2}
  = - \left(\frac{\wsql^{\text{fm}}}{\Omega}\right)^2 \\
  \left(\wsql^{\text{fm}}\right)^2 &= \frac{8kP_\ta}{Mc}
\end{align}
\end{subequations}
where $M$ is the mass of the mirror and where $\wsql^\text{fm}$ is
known as the standard quantum limit (SQL) frequency for a free mass.
In a coupled cavity with two identical mirrors, this coupling is
modifed as\footnote{As in Ref.~\cite{McCuller2021} and unlike in many
other references, we keep the optomechanical coupling complex so that
it describes all of the optical dynamics.}
\begin{equation}
  \Krp_\text{cc}(\Omega) = \tfr^2_\text{cc}(\Omega) \!\times\! 2\Krp_\text{fm}(\Omega)
  = - \left(\frac{\wsql^\text{cc}}{\Omega}\right)^2 \frac{1}{(1 + \rmi \Omega / \wcc)^2}
\end{equation}
where the extra factor of two accounts for the fact that there are two
mirrors in the arm cavity. One factor of $\tfr_\text{cc}$ accounts for
amplitude fluctuations entering the back of the SEM propagating to the
arms, and the second factor for the phase fluctuations propagating
back to the SEM. The coupled cavity SQL frequency is
\begin{align}
  \left(\wsql^{\text{cc}}\right)^2 &= \left|\tfr_\text{cc}(0)\right|^2 \!\times\! 2 \left(\wsql^{\text{fm}}\right)^2 \nonumber \\
  &=
    \begin{cases}
    \displaystyle \left(\wsql^{\text{rse}}\right)^2 = \frac{\Fa}{\Fs} \frac{16kP_\ta}{Mc} \\[1em]
    \displaystyle \left(\wsql^{\text{sr}}\right)^2 = \frac{4\Fa\Fs}{\pi^2} \frac{16kP_\ta}{Mc}
    \end{cases}
    \label{eq:cc-sql-frequencies}
\end{align}
The coupled cavity thus modifies the optomechanical coupling by adding
two poles to the response and by effectively changing the arm power by
a factor of $\Fa/\Fs$ for RSE and $4\Fa\Fs/\pi^2$ for SR.

\subsection{Resonant sideband extraction in the presence of mode mismatch}
\label{subsec:rse-mismatched}

For the remainder of the paper we focus on the case where the coupled cavity is operated
in the RSE configuration for the fundamental mode ($\phi_\ts=\pi/2$) but where there is a
mismatch between the mode of the arm cavity and the mode of the SEC. For small
mismatch, these dynamics can be described to a good approximation by a phenomenological
model similar to that of Ref.~\cite{McCuller2021} tailored to studying internal mismatch
where the fundamental mode mixes with a single higher order mode at the interface
between the arm cavity and the SEC. We will see that it is useful to think of this
system as two separate coupled cavities---one for the fundamental operating in RSE and
one for the HOM with an arbitrary SEC detuning---with the coupling between the two
occurring at this interface. This picture is illustrated in \cref{fig:phenom-cc-sflow}.
Since we focus on internal mismatch, we imagine that the active wavefront control keeps
the external cavities perfectly matched to the SEC mode as the internal mismatch
changes; see \cref{subsec:external-mismatch} for a brief discussion of the general case.

Mathematically, this is described by $2\times2$ matrices transforming
a vector representing the amplitudes of the fundamental and higher
order mode. The dynamics are the same as those described in
\cref{subsec:rse-sr-matched} except that the modes incident on the arm
cavity mix with the matrix $\Ui$ before encountering the dynamics of
the arm cavity described above, and then mix with the matrix $\Ur$ on
reflection of the arm cavity before reentering the SEC. Referring to
the nodes in \cref{fig:phenom-cc-sflow}, the matrix describing the
reflection of both fields off of the mismatched arm cavity is thus
\begin{align}
  \mat{R}_\ta &=
  \begin{bmatrix}
    E_\text{a,r}^0 / E_\text{a,i}^0 & E_\text{a,r}^0 / E_\text{a,i}^1 \\
    E_\text{a,r}^1 / E_\text{a,i}^0 & E_\text{a,r}^1 / E_\text{a,i}^1
  \end{bmatrix}
  \equiv
  \begin{bmatrix}
    \rfr_{00}(\Omega) & \tfr_{01}(\Omega) \\
    \tfr_{10}(\Omega) & \rfr_{11}(\Omega)
  \end{bmatrix} \nonumber \\
  &=
  \mat{U}_\text{r}
  \begin{bmatrix}
    \rfr_{\ta0}(\Omega) & 0 \\
    0 & \rfr_{\ta1}(\Omega)
  \end{bmatrix}
  \mat{U}_\text{i},
  \label{eq:approx-cc-reflection}
\end{align}
where $\rfr_{\ta0}(\Omega)$ is the arm reflection for the fundamental in the absence of
any mismatch (\cref{eq:arm-reflection-pade} with $\delta\omega_\ta=0$) and
$\rfr_{\ta1}(\Omega)$ is the arm reflection for the HOM in the absence of any mismatch
(\cref{eq:arm-reflection-pade} with $\delta\omega_\ta=\omega_\text{fsr}\,\psi_\ta/\pi$).
These matrices describing the coupling between the two modes due to a mismatch of magnitude
$\Upsilon$ are given by
\begin{equation}
  \Ui = \begin{bmatrix} \sqrt{1 - \Upsilon} & -\sqrt{\Upsilon}
    \\ \sqrt{\Upsilon} & \sqrt{1 - \Upsilon}
  \end{bmatrix}, \quad
  \Ur = \begin{bmatrix} \sqrt{1 - \Upsilon} & -\alpha\sqrt{\Upsilon}
    \\ \alpha\sqrt{\Upsilon} & \sqrt{1 - \Upsilon}
  \end{bmatrix}
  \label{eq:mismatch-matrices}
\end{equation}
where $\alpha=\pm1$.\footnote{More generally, $\Ui$ and $\Ur$ could be
arbitrary unitary matrices with a complex phase describing the phasing
of the mismatch between the two cavities.
However, we only discuss the mismatch between the arms and the SEC in
this work and this possibility results in different dynamics only if
there are multiple sources of mismatch. This is accounted for by the
model of \cref{sec:full-gwinc-model}. Importantly, while $\Ui$ and
$\Ur$ can be unitary in general, $\Ur=\Ui^\dag$ only for quadratic
mismatch and the restriction to $\alpha=\pm 1$ is a convenient way of
expressing the defining relationship of
\cref{eq:phenom-mismatch-mapping}.
}
As will be shown in \cref{sec:thermal-aberrations}, the exact arm
reflection given by \cref{eq:exact-cc-reflection} is equivalent to
\cref{eq:approx-cc-reflection,eq:mismatch-matrices} for small
mismatch, and these two possibilities for $\alpha$ correspond to the
two types of thermal aberrations which can occur inside a coupled
cavity:
\begin{equation}
  \Ur = \begin{cases}
    \Ui^{-1}, & \alpha=-1 ~(\text{quadratic mismatch}) \\
    \Ui, &\alpha=+1 ~(\text{higher order aberrations})
  \end{cases}
  \label{eq:phenom-mismatch-mapping}
\end{equation}
\textit{Whether} $\Ur$ \textit{is the same as} $\Ui$ \textit{or is its inverse is the
  fundamental difference between the two types of mismatch from which all of the
  differences in their dynamics follow.} The origin of this difference is explained by
  \cref{eq:cc-operators} and the surrounding discussion where the lens $\mat{L}_{ij}$
  and surface $\mat{S}_{ij}$ operators describing the exact dynamics in the presence of
  different thermal aberrations are related to these $\Ui$ and $\Ur$ matrices describing
  the mode scattering in this phenomenological two-mode model.

In order to understand the squeezing degradations in
\cref{sec:sqz-degradations}, it is necessary to understand how
fundamental and HOM fields couple into the fundamental mode of the
SEC, i.e.\ how they enter the fundamental SEC of
\cref{fig:phenom-cc-sflow} at the location $\mu_\text{a,r}^0$. It is
therefore instructive to analyze the reflection of fields in the SEC
incident on the arm (those at $\mu_\text{a,i}^0$ and $\mu_\text{a,i}^1$)
off of the mismatched arm cavity in some detail.
The $\rfr_{ii}(\Omega)$ and $\tfr_{ij}(\Omega)$ in
\cref{eq:approx-cc-reflection} describe the reflection of fields off
of the arm including the dynamics of the arm cavity alone. Two
quantities are of particular interest when additionally accounting for
the full dynamics of the HOM in the SEC.
In the picture suggested by \cref{fig:phenom-cc-sflow}, in addition to
the dynamics of its own cavity, the fundamental can be thought of as
interacting with the HOM's coupled cavity which itself behaves as an
effective cavity as described by
\cref{eq:arm-reflection-planewave,eq:arm-transmission-planewave}. Here,
the transmission into the HOM cavity is $\tfr_{10} = \sqrt{\Upsilon(1
  - \Upsilon)}(\rfr_{\ta1} + \alpha \rfr_{\ta0})$, rather than
$t_\ti$, the transmission out of the cavity is $\tfr_{01} =
-\sqrt{\Upsilon(1 - \Upsilon)}(\rfr_{\ta0} + \alpha\rfr_{\ta1})$,
rather than $t_\ti$,
and the HOM cavity SEC optical loop suppression is $[1 - r_\ts
  \rfr_{11}\rme^{-2\rmi(\Omega L_\ts/c - \psi_\ts)}]^{-1}$ where
$\rfr_{11} = \rfr_{\ta1} - \Upsilon (\rfr_{\ta1} + \alpha
\rfr_{\ta0})$.

The first quantity of interest is the reflection of the fundamental
mode off of the mismatched arm cavity,
i.e.\ $E_\text{a,r}^0/E_{\text{a,i}}^0$ including the full dynamics of
the HOM but without the presence of the fundamental's SEC, which is
\begin{equation}
  \tilde{\rfr}_\ta(\Omega) = \rfr_{00}(\Omega)
  + \frac{r_\ts\tfr_{01}(\Omega)\tfr_{10}(\Omega)
    \rme^{-2\rmi(\Omega L_\ts/c - \psi_\ts)}}{1 - r_\ts\rfr_{11}(\Omega)\rme^{-2\rmi(\Omega L_\ts / c - \psi_\ts)}}.
\end{equation}
To order $\Upsilon$, this is
\begin{multline}
  \tilde{\rfr}_\ta(\Omega) = (1 - \Upsilon) \rfr_{\ta0}
  - \alpha\Upsilon \rfr_{\ta1} \\
  - \Upsilon (\rfr_{\ta0} + \alpha \rfr_{\ta1}) (\rfr_{\ta 1} + \alpha \rfr_{\ta0})
  \frac{r_\ts \rme^{-2\rmi(\Omega L_\ts/c - \psi_\ts)}} {1 -
    r_\ts\rfr_{\ta1}\rme^{-2\rmi(\Omega L_\ts/c - \psi_\ts)}}.
  \label{eq:mismatched-arm-reflection}
\end{multline}
The mismatched reflection $\tilde{\rfr}_\ta(\Omega)$ given by
\cref{eq:mismatched-arm-reflection} should then be used in place of
the perfectly matched reflection $\rfr_\ta(\Omega)$ given by
\cref{eq:arm-reflection-planewave} in \cref{subsec:rse-sr-matched} in
order to compute the RSE reflection $\rfr_\text{rse}(\Omega)$
\cref{eq:cc-reflection} in the presence of mode mismatch.

The first term in this mismatched fundamental arm reflection
\cref{eq:mismatched-arm-reflection} is just the direct reflection of
the fundamental off of the arm cavity (reduced by the fraction
$\Upsilon$ of the fundamental scattered into the HOM). The second term
represents the process where the fundamental scatters into the HOM,
the HOM reflects directly off of the arm cavity, and then the HOM
scatters back into the fundamental. This is the analogue of the prompt
reflection off of a simple cavity, i.e.\ the first term in
\cref{eq:arm-reflection-planewave}. The final term describes the
process where the fundamental scatters into the HOM, the HOM
experiences the dynamics of its SEC, followed by the scattering of the
HOM back into the fundamental. This is the analogue of the second term
in \cref{eq:arm-reflection-planewave}.

The second quantity of interest is the reflection of the HOM off of
the mismatched arm and the subsequent scattering into the
fundamental, i.e.\ $E_\text{a,r}^0/E_\text{a,i}^1$ including the full
dynamics of the HOM but without the presence the fundamental's SEC,
which is
\begin{equation}
  \tfr_\text{a,h}(\Omega) =
  \frac{\tfr_{01}(\Omega)}{1 - r_\ts\rfr_{11}(\Omega)\rme^{-2\rmi(\Omega L_\ts / c - \psi_\ts)}}.
\end{equation}
To order $\sqrt{\Upsilon}$, which is sufficient for calculating the
squeezing degradations to order $\Upsilon$, this is
\begin{equation}
  \tfr_\text{a,h}(\Omega) =
  \frac{\sqrt{\Upsilon}(\rfr_{\ta0} + \alpha \rfr_{\ta1})}{
    1 - r_\ts\rfr_{\ta1}\rme^{-2\rmi(\Omega L_\ts/c - \psi_\ts)}}.
  \label{eq:mismatched-hom-arm-transmission}
\end{equation}
In the picture suggested by \cref{fig:phenom-cc-sflow}, this is the
transmission of a
HOM through the HOM coupled cavity, which is analogous to the
transmission \cref{eq:arm-transmission-planewave} through a simple
cavity.

The HOM does not experience the same optical system configured for RSE
that the fundamental experiences. In particular, the extra Gouy phase
$\psi_\ts$ that the HOM accumulates in the SEC means that the HOM SEC
is generically detuned even though the fundamental is not. The
different resonance conditions in the arm further change the dynamics
of the HOM in the SEC. There is a continuum of behavior between exact
HOM anti-resonance with $\psi_\ts=\pi/2$ and exact HOM resonance with
$\psi_\ts=0$, but we will often consider these two extreme cases in
detail in \cref{sec:sqz-degradations}. Since the HOM does not
experience the phase change on reflection of the arm cavities that the
fundamental does---except for frequencies where the HOM is resonant in
the arms as is discussed in \cref{subsec:hom-degradations}---their
dynamics are enhanced for $\psi_\ts=0$ when the round-trip phase is
positive, and are suppressed for $\psi_\ts=\pi/2$ when the round-trip
phase is negative. This enhancement is sometimes known as mode harming
and this suppression known as mode healing~\cite{Strain1991,Bochner2003}.


These dynamics will be analyzed in various limits in
\cref{sec:sqz-degradations} in order to calculate the squeezing
degradations. \textit{Crucially, the interference between the
  fundamental and HOM on reflection of the arm cavity determines the
  frequency dependence of the different types of internal mismatch
  within the coupled cavity.} Generally, either the fundamental or the
HOM will be near resonance of the arm cavity, and therefore either
$\rfr_{\ta0}(\Omega)$ or $\rfr_{\ta1}(\Omega)$ will be given by
\cref{eq:arm-reflection-pade} while the other will be $+1$. This
interference is therefore described by the following quantity,
featured prominently in
\cref{eq:mismatched-arm-reflection,eq:mismatched-hom-arm-transmission},\footnote{
All of the factors of $\rfr_{\ta i} + \alpha \rfr_{\ta j}$ are
proportional to $1 + \alpha \rfr_\ta$ since $\alpha=\pm 1$ and since
one of $\rfr_{\ta i}$ is $\rfr_\ta$ while the other is $+1$.
}
\begin{align}
  \frac{1 + \alpha \rfr_\ta(\Omega)}{2} &= \frac{1}{2}\frac{(1 -
    \alpha) + (1 + \alpha)\, \rmi (\dwa)/\wa}{1 + \rmi (\dwa)/\wa}
  \nonumber \\
  &=
  \begin{cases}
    \displaystyle\frac{1}{1 + \rmi(\dwa)/\wa},
    & \alpha=-1 ~(\text{quad}.) \\[1em]
    \displaystyle\frac{\rmi(\dwa)/\wa}{1 + \rmi(\dwa)/\wa},
    & \alpha=+1 ~ (\text{HOA})
  \end{cases}
  \label{eq:arm-reflection-interference}
\end{align}
With quadratic mismatch, the interference is constructive at low
frequencies and becomes destructive above the arm cavity pole when the
reflection of the resonant field off of the arm cavity changes
sign. The frequency dependence of quadratic mismatch thus has a
characteristic low-pass behavior. For higher order aberrations, on the
other hand, the interference is destructive at low frequencies and so
the frequency dependence has a characteristic high-pass
behavior. Since
\cref{eq:arm-reflection-interference} is proportional to
$\tfr_{ij}(\Omega)$, in the picture suggested by
\cref{fig:phenom-cc-sflow}, \cref{eq:arm-reflection-interference} is
the transmission between the two coupled cavities. The HOM SEC can
therefore be thought of as being AC coupled into the fundamental SEC
through higher order aberrations while being DC coupled through
quadratic mismatch.


\section{Quadratic and higher order thermal aberrations}
\label{sec:thermal-aberrations}



Thermal aberrations are the result of unwanted heating occurring in
optics that induce both thermorefractive and thermoelastic
effects. The primary issue in gravitational-wave detectors is the
thermal aberrations generated by heating in the arm cavity test mass
optics due to absorbing parts-per-million (ppm) of the high
intracavity power in their Bragg coatings~\cite{Granata2020}. In this
context, there are two significant sources of thermal
aberrations~\cite{Hello_Vinet_1990,Hello_Vinet_1990b,Vinet2009}. The
first and most significant for fused silica optics is the
thermorefractive lens in the input test mass (ITM) substrates through
which the optical fields must pass in order to enter and exit the arm
cavities. The second is the thermoelastic deformations which alter the
shape of the high reflective mirror surface of the test masses.
\Cref{subsec:hom-couplings} describes how these thermal
aberrations can be decomposed into quadratic and higher order effects
and the resulting higher order mode couplings that they
generate. \Cref{subsec:cc-aberrations} describes the effects of these
aberrations inside a coupled cavity, as in \cref{fig:coupled-cavity},
and justifies the simple model described in
\cref{subsec:rse-mismatched,fig:phenom-cc-sflow}.


\subsection{Higher order mode couplings due to quadratic and higher order thermal aberrations}
\label{subsec:hom-couplings}

\begin{figure}
  \includegraphics[width=\columnwidth]{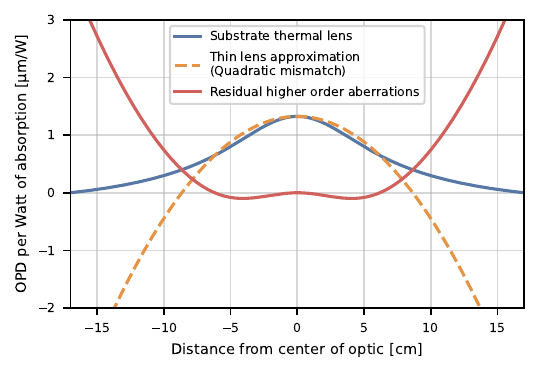}
  \caption{Optical path length differences resulting from the
    propagation through the substrate of a LIGO-like test mass due to
    thermal aberrations generated by the uniform absorption of a
    $2w=\qty{10.6}{\cm}$ diameter beam. Computed using the
    axisymmetric Hello-Vinet~\cite{Hello_Vinet_1990} model. The
    equivalent quadratic thin lens is shown along with the residual
    OPD relative to this lens. We refer to the aberrations generated
    by these distortions as quadratic mismatch and higher order
    aberrations (HOA), respectively. We parameterize the HOA mismatch
    by the equivalent OPD \textit{relative} to the perfect
    compensation of an optic which may have of order one watt
    absorbed in the coatings. This parameterization thus allows for negative
    relative absorbed power $\Pabs$ which corresponds to an imperfect
    compensation resulting in a negative thermal lens. The quadratic
    mismatch is parameterized by the fractional change $\dww$ between
    the beam sizes of the eigenmodes on either side of the ITM. See the
    last paragraph of \cref{subsec:hom-couplings} for details.}
  \label{fig:thermal-lens}
\end{figure}

The Hello-Vinet
model~\cite{Hello_Vinet_1990,Hello_Vinet_1990b,Vinet2009} provides a
reasonable approximation for the thermal lensing and thermoelastic
effects and is the basis for the analysis in this work. The model
computes the effective optical path difference (OPD) that an optical
field experiences across its wavefront due to thermal
aberrations. \Cref{fig:thermal-lens} shows the thermal lensing that
develops in the substrate of the test mass optic of a LIGO-like arm
cavity due to the heating of the laser incident on the high
reflective (HR) coating of the mirror.  It also depicts how we
separate the thermal aberrations into a quadratic and higher order
term by writing the OPD as
\begin{equation}
  Z(r, \phi) = ar^2 + z_\text{hoa}(r, \phi),
  \label{eq:opd-decomposition}
\end{equation}
where $r$ is the radial coordinate and $\phi$ is the azimuthal
coordinate of the plane perpendicular to the direction of travel of
the optical field.
When $Z(r,\phi)$ describes the OPD obtained by propagation through the
test mass substrate, the quadratic term is $a = -1/2f_\text{th}$ and
represents the equivalent thin lens with focal length $f_\text{th}$
that could be used to describe the thermal lensing effect and is shown
in orange in \cref{fig:thermal-lens}. When $Z(r,\phi)$ describes the
displacement due to elastic deformation of the surface of a mirror, the quadratic term is
$a= 1/2R_\text{th}$. If the radius of curvature of the cold mirror is
$R_\text{c}$, a mirror with radius of curvature $R = (1/R_\text{c} +
1/R_\text{th})^{-1}$ could be used to describe how the thermoelastic
effect changes the wavefront curvature.
More broad analysis of this equivalent thin lens and effective mirror
has been performed to quantify the impact of such aberrations on the
detectors~\cite{Strain_1994,Winkler1991}.


On top of this quadratic lensing there are higher order terms
$z_\text{hoa}(r,\phi)$ in the OPD that can be expanded into bases such
as Zernike or Seidel polynomials. The residual between the actual
thermal lensing and the quadratic approximation is shown in red in
\cref{fig:thermal-lens}. This residual is relatively small at the
center where the bulk of the optical intensity is found but quickly
rises towards the edges of the optic. We define this residual as the
higher order aberrations (HOA) which contains everything beyond the
quadratic term, such as spherical aberrations due to the non-parabolic
nature of the thermal lensing and elastic deformations.


In the absence of higher order modes, an optical field at a given
location is described by two parameters, which can be taken to be the beam
radius $w$ and the wavefront curvature, or defocus, $S$. These are
typically combined into a complex $q$ parameter as $1/q = S -
2\rmi/kw^2$ so that the spatial profile of the field is proportional
to the Gaussian $\rme^{-\rmi k r^2/2q}$.  The thin lensing is used to
propagate beams throughout the optical system using the standard ABCD
matrices for components, such as lenses and surface reflections and
transmissions~\cite{Siegman1986}. In particular, in the
case of the substrate thermal lens, the beam parameter $q$ before
passing through the substrate is related to the beam parameter
$\hat{q}$ after the substrate by
\begin{equation}
  \frac{1}{\hat{q}} =
  \frac{C + D/q}{A + B/q}
  = \frac{1}{q} - \frac{1}{f_\text{th}}.
  \label{eq:abcd-lens}
\end{equation}
The case of the thermoelastic deformation is similar, with the defocus
changing by something proportional to $2/R_\text{th}$, rather than
$1/f_\text{th}$, with the details depending on whether the field is
reflecting from or transmitting through the substrate and from which
direction~\cite{Siegman1986}. In general, we use hats on $q$ parameters to denote the $q$
parameter transformed by the ABCD matrix between two neighboring nodes
in \cref{fig:coupled-cavity}.

The total round-trip ABCD matrix is used to approximate the resonant spatial eigenmode
of a cavity including all quadratic effects for infinite sized optics as is discussed
further in \cref{subsec:cavity-eigenmodes}. For a geometrically stable cavity, a complex
$q$ parameter can be determined at every spatial point $\mu$. \Cref{fig:coupled-cavity}
shows several points of particular interest. Following the beam entering the arm cavity
from the SEC, $\qar$ in the SEC incident on the antireflective (AR) surface of the ITM,
$\qsub$ inside the substrate of the ITM incident on the HR surface, and $\qhr$ inside
the arm cavity transmitted through the ITM. The reversed beam parameters $-q^*$ describe
the beam propagating in the opposite direction leaving the arm cavity and entering the
SEC.

The above analysis only describes the fundamental mode and thus the
quadratic part of the wavefront characterized by the beam size and
defocus. When there is higher order spatial structure, optical fields
are typically defined in terms of orthogonal modes whose spatial
profile is usually described by Hermite- or Laguerre-Gauss modes with
shapes that are also determined by the complex $q$
parameters~\cite{Siegman1986}. For our purposes it is most convenient
to define the mode shapes in terms of the associated Laguerre
polynomials $L_p^\ell$ as
\begin{align}
  u&_{p\ell}(r,\phi;q) \nonumber \\
  &=
  \sqrt{\frac{2p!}{\pi(p + |\ell|)!}} \frac{1}{w}
  \left(\frac{\sqrt{2} r}{w}\right)^{|\ell|}\!\!
  L_p^{|\ell|}\!\left(\frac{2r^2}{w^2}\right)
  \rme^{\rmi\ell\phi}\,\rme^{-\rmi kr^2/2q} \nonumber \\
  &\equiv
  U_{p\ell}(r, \phi, w)\,\rme^{-\rmi kr^2/2q},
  \label{eq:LG-modes}
\end{align}
and then expand the field at a given point $\mu$ in terms of these
higher order modes (HOMs) as~\cite{Bond2016}
\begin{equation}
  E_\mu(r,\phi) = \sum_{p=0}^\infty\sum_{\ell=-\infty}^\infty c_{p\ell}(q_\mu)\, u_{p\ell}(r, \phi; q_\mu)\,
  \rme^{\rmi (2p + |\ell| + 1)\Xi_\mu}
  \label{eq:LG-expansion}
\end{equation}
where $\Xi_\mu = \arccos(\re q_\mu / \im q_\mu)$ is ``the'' Gouy phase at
the point $\mu$.\footnote{Most references include the exponential
$\rme^{\rmi(2p + |\ell| + 1)\Xi}$ in the definition of
$u_{p\ell}(r,\phi;q)$, however we write it like this because it makes
the concept of the Gouy phase of a cavity more clear and simplifies
the definition of the operators in
\cref{eq:exact-cc-reflection,eq:cc-operators}.}  See
\cref{subsec:cavity-eigenmodes} for a detailed discussion of the
various meanings of a Gouy phase. Different mode shapes,
i.e.\ different $q$ parameters, can then describe the same field with
different expansion coefficients $c_{p\ell}(q)$.  The coupling between
different modes described with different $q$ parameters is given by
the overlap
\begin{equation}
  \langle u_{sm}(q_2) | u_{p\ell}(q_1)\rangle =
  \int\! u_{sm}^*(r, \phi;q_2)\, u_{p\ell}(r, \phi; q_1)\,\rmd A
\end{equation}
where $\rmd A = r\,\rmd r\,\rmd\phi$ and the integral is taken over
the cross-sectional area of the beam.

The quadratic mode mismatch between two beams described by $q_1$ and $q_2$ is
represented in a two-dimensional space of beam size $w$ and defocus
$S$~\cite{Perreca2020,Anderson1984}. For two separate cavities connected by some
arbitrary telescope, they could be mismatched in either $w$ or $S$. However, the
steady-state boundary condition for a coupled cavity sharing an interface with standing
waves on either side is that the wavefront curvature of the resonant fields must match
at this interface. This has the consequence that any quadratic lensing change in the
extraction cavity, regardless of where it happens, results only in a beam size
difference between the arm and extraction cavity eigenmodes at the ITM HR surface. The
fractional mismatch between two modes with the same curvature but different beam sizes
$w_1$ and $w_2$ is
\begin{equation}
  1 - |\langle u_{00}(q_2) | u_{00}(q_1)\rangle|^2 = 1 - 4\left(\frac{w_1
    w_2}{w_1^2 + w_2^2}\right)^2 \approx \left(\frac{\Delta
    w}{w}\right)^2
\end{equation}
where $\Delta w = w_1 - w_2$ and $w = (w_1 + w_2) /2$.

Now suppose that an optical field described by $q_1$ passes through a
substrate thermal lens with an OPD given by
\cref{eq:opd-decomposition} and then interacts with another field
described by $q_2$. Since $a=-1/2f_\text{th}$ in this case, the
scattering of the HOMs between these two fields is
\begin{align}
  \braket{u_{sm}(&q_2)}{\rme^{-\rmi k Z(r,\phi)}}{u_{p\ell}(q_1)} \nonumber \\
  &= \begin{aligned}[t]
    \int\! U_{sm}^*(&r, \phi, w_2)\, U_{p\ell}(r, \phi, w_1)\, \rme^{-\rmi k z_\text{hoa}(r,\phi)} \\
    &\times\exp\!{\left\{\!
      \frac{\rmi k r^2}{2}\!
      \left[
        \frac{1}{q_2^*} - \left(\!\frac{1}{q_1} - \frac{1}{f_\text{th}}\!\right)
        \right]
      \!\right\}} \,\rmd A
    \end{aligned} \nonumber \\
  &= \braket{u_{sm}(q_2)}{\rme^{- \rmi k z_\text{hoa}(r,\phi)}}{u_{p\ell}(\hat{q}_1)}
  \label{eq:general-opd-coupling}
\end{align}
where the second line follows since the term in parenthesis is
$1/\hat{q}_1$ (cf.~\cref{eq:abcd-lens}) and since the lens does not
change the beam size $\hat{w}_1 = w_1$. In other words, the coupling
between the HOMs of a field directly before it encounters a thermal
lens with those of a field after it experiences those thermal
aberrations is described by two effects. First, the quadratic term
$ar^2$ transforms $q$, thus changing the defocus $S$. The remaining
coupling is then the scattering between the two fields with the new
defocus (in the new basis $\hat{q}_1$) due only to the higher order
aberrations. The situation is the same for a thermoelastic deformation
where $a=1/2R_\text{th}$, and \cref{eq:general-opd-coupling} holds for
any OPD or elastic deformation of the form
\cref{eq:opd-decomposition}.

\begin{table*}
  \footnotesize
  \setlength{\tabcolsep}{4pt}
  \begin{ruledtabular}
    \begin{tabular}{ccc}
      & \centering Quadratic mismatch & \centering\arraybackslash Higher order aberrations \\
      \hline
      \noalign{\vspace{2pt}}
      parametrized by
      & \parbox[t]{7.5cm}{\raggedright Fractional beam size error $\dww$}
      & \parbox[t]{7.5cm}{\raggedright Absorption $\Pabs$ relative to perfect compensation} \\[2pt]
      \hline
      \noalign{\vspace{2pt}}
      substrate lens
      & \parbox[c]{7.5cm}{\raggedright
          $\dww$ mismatch due to $f_\text{th}\neq\infty$ \\[4pt]
          $\Msh = \Mhs^\intercal \neq \mat{1}$ describes mode scattering due to $\Delta w\neq0$ \\[4pt]
          $\Las = \Lsa = \Mhh = \Mss = \mat{1}$\\[4pt]
          $\Ui=\Mhs$, $\Ur=\Msh$ $\quad\Rightarrow\quad \Ur=\Ui^{-1}$ to $O(\dww)$}
      & \parbox[c]{7.5cm}{\raggedright
          $\Delta w = 0$ since $f_\text{th} = R_\text{th} = \infty$ \\[4pt]
          $\Las = \Lsa \neq \mat{1}$ describes mode scattering due to $z_\text{l}\neq0$ \\[4pt]
          $\mat{S}_{ij} = \mat{1}$\\[4pt]
          $\Ui=\Lsa$, $\Ur=\Las$ $\quad\Rightarrow\quad \Ur=\Ui$} \\[2pt]
      \noalign{\vspace{3pt}}
      \hline
      \noalign{\vspace{2pt}}
      \shortstack{surface\\deformation}
      & \parbox[c]{7.5cm}{\raggedright
          $\dww$ mismatch due to  $R_\text{th}\neq\infty$ \\[4pt]
          $\Msh = \Mhs^\intercal \neq \mat{1}$ describes mode scattering due to $\Delta w\neq0$\\[4pt]
          $\Las = \Lsa = \Mhh = \Mss = \mat{1}$\\[4pt]
          $\Ui=\Mhs$, $\Ur=\Msh$ $\quad\Rightarrow\quad \Ur=\Ui^{-1}$ to $O(\dww)$}
      & \parbox[c]{7.5cm}{\raggedright
          $\Delta w = 0$ since $f_\text{th} = R_\text{th} = \infty$ \\[4pt]
          $\mat{S}_{ij}\neq \mat{1}$ describes mode scattering due to $z_\text{s}\neq0$ \\[4pt]
          $\Las = \Lsa = \mat{1}$\\[4pt]
          No corresponding $\Ui$ and $\Ur$} \\[2pt]
    \end{tabular}
  \end{ruledtabular}
  \caption{Summary of how the quadratic and higher order aberrations are parametrized,
  as described in the last paragraph of \cref{subsec:hom-couplings}, and how the lens
  and surface operators describe both kinds of mismatch in the case of a thermorefactive
  lens in the ITM substrate or a thermoelastic deformation of the optic surfaces. As
  depicted in \cref{fig:coupled-cavity}, the lens operators $\mat{L}_{ij}$ describe the
  propagation through the substrate, modeled here as a thin lens with focal length
  $f_\text{th}$, and the surface operators $\mat{S}_{ij}$ describe how a field transmits
  through or reflects from the HR surface, modeled here as a surface with radius of
  curvature $R_\ti = (1/R_{\ti0} + 1/R_\text{th})^{-1}$ where $R_{\ti0}$ is the radius
  of curvature of the perfectly compensated mirror. The higher order aberrations
  $z_\text{hoa}(r)$ due to the substrate lens are $z_\text{l}(r)$, and those due to the
  surface deformation are $z_\text{s}(r)$. The relationship of the lens and surface
  operators describing the exact dynamics to the matrices $\Ui$ and $\Ur$ describing the
  scattering in the phenomenological model of
  \cref{subsec:rse-mismatched,fig:phenom-cc-sflow} is also listed. Note that in practice
  the effects of higher order surface deformations are far weaker than those due to the
  other types of mismatch.}
  \label{tab:aberration-wide}
\end{table*}

For the rest of this work, we describe the thermal aberrations as the
deviation from a perfectly compensated state as follows. We imagine a
scenario where the thermal actuators are set to perfectly compensate
for of order one watt of absorbed power but where the actual absorbed
power differs from that by $\Pabs$. This difference can be positive in
which case the residual thermal aberrations result in a positive
thermal lens and elastic deformations with negative defocus; when
$\Pabs$ is negative, the signs are reversed.  We further imagine that
in this imperfectly compensated state the thermal actuators can
independently target the quadratic mismatch without changing the
higher order aberrations.  We then generate the appropriate OPD
\cref{eq:opd-decomposition} for a given $\Pabs$ relative to the
perfectly compensated state and parameterize the two types of
aberrations by
\begin{enumerate}[leftmargin=0.08\columnwidth,rightmargin=0.02\columnwidth]
\item Removing the quadratic term $ar^2$ from
  \cref{eq:opd-decomposition}. Rather than using the equivalent thin
  lens with focal length $f_\text{th}=-1/2a$, we directly set the
  focal length $f_\text{th}$ of the substrate lens to produce a given
  $\dww$ independent of $\Pabs$. \textbf{Quadratic mismatch is thus
    parameterized by $\dww$.}
\item The remaining terms $z_\text{hoa}(r)$ are added as the OPD of
  the substrate lens or the elastic deformation of the optic
  surface. \textbf{The higher order aberrations are thus parameterized
  directly by the absorption $\Pabs$ relative to the perfectly
  compensated state.}
\end{enumerate}
In other words, we use the red curve in \cref{fig:thermal-lens} to
define the higher order aberrations for a given power $\Pabs$ relative
to the perfectly compensated state and choose the orange curve
independently of the total blue curve to produce a given $\dww$.
For a sense of scale, the equivalent thin lens needed to produce
$\dww=\qty{5}{\%}$ is generated by of order \qty{30}{\mW} of uniform
absorption for both LIGO~\Asharp{} and CE. This parameterization is,
heuristically, a measure of how well a thermal actuator must correct
each type of aberration (expressed in terms of equivalent residual
beam-heating and beam size error) in order to reduce the squeezing
degradations discussed in \cref{sec:sqz-degradations} below some
target. However, this cannot be taken too far as the quantitative
squeezing degradations due to some thermal aberrations are
sensitive to the details of those aberrations and the way in which
they have been decomposed, and it would be difficult to derive
meaningful quantitative requirements without having a complete design
for a thermal actuator in hand.

\subsection{Thermal aberrations inside a coupled cavity}
\label{subsec:cc-aberrations}

Calculating squeezing degradations requires knowing how optical fields
propagate throughout the coupled cavity. As in
\cref{subsec:rse-mismatched}, it is therefore useful to analyze the
reflection of an optical field off of the arm cavity in the presence
of thermal aberrations. This is the reflection of a field at the node
$\mu_{\text{a,i}}$ (with beam parameter $\qar$) shown in
\cref{fig:coupled-cavity} to the field at the node $\mu_{\text{a,r}}$
(with beam parameter $-\qar^*$) without the SEM present. From
\cref{fig:coupled-cavity}, the exact operator for this reflection is
\begin{multline}
  \mat{R}_\ta = \Las\Big[ \\
      r_\text{i}\Mss
      -t_\text{i}^2 r_\text{e}\Msh
      \left(\mat{1} - r_\text{i} r_\text{e} \mat{P}_\ta^2 \Mhh\right)^{-1}
      \mat{P}_\ta^2\Mhs
      \Big] \Lsa.
  \label{eq:exact-cc-reflection}
\end{multline}
In the absence of any thermal aberrations all of the lens
$\mat{L}_{ij}$ and surface $\mat{S}_{ij}$ operators are $\mat{1}$.
The exact dynamics described by \cref{eq:exact-cc-reflection} are
equivalent to those described in \cref{subsec:rse-mismatched} if
\cref{eq:exact-cc-reflection} is equivalent to
\cref{eq:approx-cc-reflection}, i.e.\ if matrices $\Ui$ and $\Ur$
exist such that \cref{eq:exact-cc-reflection} can be written as
\begin{equation}
  \mat{R}_\ta =
  \Ur\Big[
      r_\text{i}\mat{1}
      - r_\text{e}t_\text{i}^2
      \left(\mat{1} - r_\text{i} r_\text{e} \mat{P}_\ta^2 \right)^{-1}
      \mat{P}_\ta^2
      \Big] \Ui
  \label{eq:cc-reflection-phenom-model}
\end{equation}
since the term in brackets is the arm reflection in the absence of any
aberrations.  This term is diagonal with each element corresponding to
the reflection of one of the HOMs off of the arm given by
\cref{eq:arm-reflection-planewave} with the cavity detuning
$\delta\omega_\ta = \omega_\text{fsr}\psi_\ta / \pi$ determined by the
one-way Gouy phase $\psi_\ta$ of that mode in the arm cavity.

First consider the quadratic mismatch alone which transforms the $q$
parameters and only affects the beam size and defocus. The round-trip
ABCD matrix of the arm cavity determines $\qhr$ and the round-trip
ABCD matrix of the SEC determines $\qar$ and $\qsub$. Therefore there
can be no quadratic mismatch between $\qar$ and $\qsub$,
i.e.\ $\hat{q}_\text{ar} = \qsub$. The only quadratic mismatch which
can occur is between the modes of the two cavities, and so
$\hat{q}_\text{sub}\neq \qhr$ in general. This also means that the
beam size and defocus of the SEC eigenmode are determined by both the
substrate thermal lens and the ITM thermoelastic deformation while the
arm eigenmode is only affected by the surface deformation. Following
the beam traveling from the SEC (at the node $\mu_\text{a,i}$) into the
arm cavity, the thermal lens first changes the defocus of the beam
$S_\text{sub} = S_\text{ar} - 1 / f_\text{th}$ but does not affect the
beam size $\hat{w}_\text{ar} = w_\text{ar} = w_\text{sub}$. The radius
of curvature of the ITM is $R_\ti = (1/R_{\ti0} + 1/R_\text{th})^{-1}$
where $R_{\ti0}$ is the radius of curvature of the perfectly
compensated mirror. Upon passing through the HR surface of the mirror,
the defocus of the beam is transformed to $\hat{S}_\text{sub} =
S_\text{sub} + (n-1)/R_\ti$, where $n$ is the index of refraction of the
substrate, but again does not change the beam size $\hat{w}_\text{sub}
= w_\text{sub}$. Due to the steady state boundary conditions at the
ITM HR surface discussed above, it must be the case that $S_\text{hr}
= \hat{S}_\text{sub}$ and so
\begin{align}
  S_\text{hr}&(R_{\ti0}, R_\text{th}) = \nonumber \\
  S&_\text{ar}(R_{\ti0}, R_\text{th}, f_\text{th})
  - \frac{1}{f_\text{th}}
  + (n - 1) \left(\frac{1}{R_{\ti0}} + \frac{1}{R_\text{th}}\right).
  \label{eq:itm-defocus}
\end{align}
It is not necessarily the case that $\hat{w}_\text{sub} =
w_\text{hr}$, though, and so the only question is whether
$w_\text{hr}$ is the same as $w_\text{ar}$.  Note that while the
defocus $S_\text{ar}$ is determined entirely by the ITM lens and
curvature, the beam size $w_\text{ar}$ is determined by the full
geometry of the SEC which, in the actual detectors, is additionally
controlled by other optics and thermal actuators which are not
necessarily located at the
ITM~\cite{Brooks2016,Rocchi2012,Jones2024}. Nevertheless,
\cref{eq:itm-defocus} is always true, and any lensing caused by any of
these actuators or optics can only change $w_\text{ar}$ regardless of
the Gouy phase in which they operate.


We next examine the exact form of the operators shown in
\cref{fig:coupled-cavity} and focus on the case of azimuthal
symmetry. From now on we will also denote the matrix of couplings
$\langle u_{sm}(q_2)|O(r)|u_{p\ell}(q_1)\rangle$ for some arbitrary
aberration $O(r)$ simply as $\langle q_2|O(r)|q_1\rangle$. The
coupling between any two of the nodes describing the aberrations in
the ITM shown in \cref{fig:coupled-cavity} is $\langle q_2|\rme^{-\rmi
  k Z(r)} | q_1 \rangle$ which, by \cref{eq:general-opd-coupling}, is
$\langle q_2|\rme^{-\rmi k z_\text{hoa}(r)} | \hat{q}_1 \rangle$ in
all cases. If $z_\text{l}(r)$ is the higher order OPD of the substrate thermal
lens and $z_\text{s}(r)$ is the higher order thermoelastic
deformations of the mirror HR surface, these operators are
therefore\footnote{
  Due to our simplification of azimuthal symmetry, we
  can ignore odd order modes. Therefore, two operations are simplified in the tangential
  plane:
  We exclude the parity operation
  for the backwards propagation through an OPD which would otherwise be
  necessary since a beam traveling right-to-left vs.\ left-to-right will
  see an inversion in this direction; and we do not explicitly include
  the parity operator for the coordinate system transformation on
  reflection which applies a $\pi$ phase shift to tangential odd order modes.
}
\begin{subequations}
  \label{eq:cc-operators}
\begin{alignat}{2}
  \Lsa\! &=\! \braket{\qsub}{\rme^{-\rmi k z_\text{l}}}{\hat{q}_\text{ar}} & \mspace{12mu}
  \Las\! &=\! \braket{\!\!-\!\qar^*}{\rme^{-\rmi k z_\text{l}}}{\!-\!\hat{q}^*_\text{sub}} \\
  \Mhh\! &=\! \braket{\qhr}{\rme^{2\rmi kz_\text{s}}}{\!-\!\hat{q}_\text{hr}^*} & \mspace{12mu}
  \Mss\! &=\! \braket{\!\!-\!\qsub^*}{\rme^{-2\rmi nkz_\text{s}}}{\hat{q}_\text{sub}}
  \label{eq:reflection-surface-operators} \\
  \Mhs\! &=\! \braket{\qhr}{\rme^{\rmi (1-n) kz_\text{s}}}{\hat{q}_\text{sub}} & \mspace{12mu}
  \Msh\! &=\! \braket{\!\!-\!\qsub^*}{\rme^{\rmi (1-n)kz_\text{s}}}{\!-\!\hat{q}_\text{hr}^*}
  \label{eq:transmission-surface-operators}
\end{alignat}
\end{subequations}
We now examine the thermal lens and thermoelastic deformation
separately to identify which operators in \cref{eq:cc-operators} are
the $\Ui$ and $\Ur$ in \cref{eq:cc-reflection-phenom-model} which
reproduce \cref{eq:exact-cc-reflection} in each of these cases, thus
justifying \cref{eq:approx-cc-reflection,eq:phenom-mismatch-mapping}
and the model of \cref{subsec:rse-mismatched,fig:phenom-cc-sflow}; the results are summarized in \cref{tab:aberration-wide}.

\paragraph{Substrate thermal lens}
First consider the case where there is only a substrate thermal lens,
which is the most significant effect for fused silica optics. In this
case, $z_\text{s}=0$ and so the surface operators are
\begin{equation}
  \Mhh = \Mss = \mat{1},\quad \Mhs = \langle \qhr |
  \hat{q}_\text{sub}\rangle = \Msh^\intercal = \langle
  -\hat{q}_\text{hr}^* | -\qsub^*\rangle.
  \label{eq:surface-operators-no-deformation}
\end{equation}
\begin{itemize}[leftmargin=0.05\columnwidth,rightmargin=0.02\columnwidth]
\item For quadratic lensing, $z_\text{l} = 0$ and so $\Las = \Lsa =
  \mat{1}$. In this case, the quadratic thermal lens $f_\text{th}$
  produces a beam size error $\dww$ between the modes of the two
  cavities so that $\hat{q}_\text{sub}\neq \qhr$. Therefore $\Mhs =
  \Msh^\intercal \neq 1$. \Cref{eq:cc-reflection-phenom-model} is thus
  \cref{eq:exact-cc-reflection} if $\Ui = \Mhs$ and $\Ur = \Msh$. For
  small mismatch, to order $\dww$, $\Msh^\intercal = \Msh^{-1}$, and
  so $\Ur=\Ui^{-1}$.
\item For higher order aberrations, on the other hand, $\Las = \Lsa
  \neq \mat{1}$ since $z_\text{l}\neq 0$, and $\Mhs = \Msh = \mat{1}$
  since there is no quadratic lensing ($\hat{q}_\text{sub} =
  \qhr$). \Cref{eq:cc-reflection-phenom-model} is thus
  \cref{eq:exact-cc-reflection} if $\Ui = \Lsa$ and $\Ur = \Las$, and
  so $\Ur = \Ui$.
\end{itemize}
Taken together, this reproduces \cref{eq:approx-cc-reflection} and the
defining relationship of \cref{eq:phenom-mismatch-mapping}.

\paragraph{Thermoelastic surface deformation}
Next consider the case where there is only a thermoelastic
deformation. In this case $z_\text{l}=0$ and so the lens operators are
$\Las = \Lsa = \mat{1}$.
\begin{itemize}[leftmargin=0.05\columnwidth,rightmargin=0.02\columnwidth]
\item For a quadratic deformation, $z_\text{s} = 0$ and so the surface
  operators are again given by
  \cref{eq:surface-operators-no-deformation} with $\Mhs =
  \Msh^\intercal \neq 1$ since the quadratic surface deformation
  $R_\text{th}$ produces a beam size error $\dww$. As with the
  quadratic thermal lens, $\Ui = \Mhs$ and $\Ur = \Msh$, and so
  $\Ur=\Ui^{-1}$ to order $\dww$.
\item The case for a higher order thermoelastic deformation with
  $z_\text{s}\neq 0$ is more complicated since the surface operators
  cannot be simplified from the general case given by
  \cref{eq:reflection-surface-operators,eq:transmission-surface-operators}. It
  is therefore not possible to find operators $\Ui$ and $\Ur$ such
  that \cref{eq:cc-reflection-phenom-model} is satisfied and the
  dynamics of this type of mismatch does not map cleanly onto the
  dynamics described in \cref{subsec:rse-mismatched}.
\end{itemize}
In practice, the effects of higher order surface deformations are far
weaker than those due to the other sources of mismatch and the
dynamics are still well described by
\cref{eq:cc-reflection-phenom-model} and the model of
\cref{subsec:rse-mismatched}.

Finally we note that, while we have focused on wavefront aberrations
generated by thermal distortions, any aberration can be broken up into
quadratic mismatch and higher order aberrations as in
\cref{eq:opd-decomposition} and all of the ensuing squeezing
degradations will behave in the same way.

\section{Squeezing degradations}
\label{sec:sqz-degradations}

Quantum noise in an interferometer used for gravitational wave
detection is caused by vacuum fluctuations in the fundamental mode of
the electromagnetic field which enter the anti-symmetric (AS)
port~\cite{Caves1981,Caves1980}. In the case of the effective coupled
cavity, these are the vacuum entering in the port $\mu_\text{as,i}$ of
\cref{fig:coupled-cavity}. Gravitational-wave detectors therefore
inject squeezed vacuum states, with reduced uncertainty in one optical
quadrature, into the AS port in order to reduce the quantum noise in
that
quadrature~\cite{Barsotti2019,Tse2019,Acernese2019,Grote2013,Lough2020,McCuller2020,Ganapathy2023,McCuller2020,Zhao2020}. Various
processes degrade the extent to which these squeezed vacuum states can
reduce the quantum noise, and in order to improve and design detectors
it is important to understand the frequency-dependent contribution of
each noise source to the total.

\begin{figure}
  \centering
  \includegraphics[width=\columnwidth]{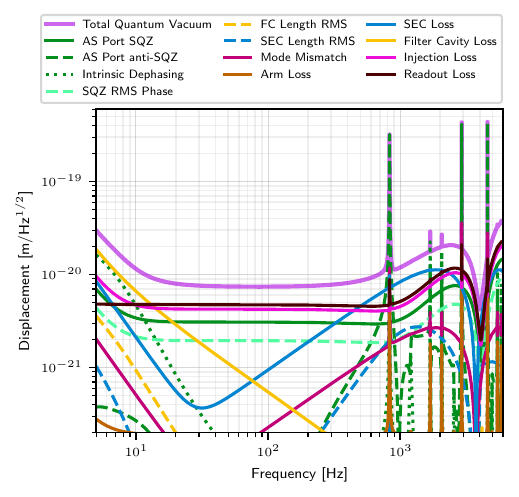}
  \caption{Quantum noise budget for Cosmic Explorer with $\dww=\qty{5}{\%}$ and
    $\Pabs=\qty{100}{\mW}$ using the parameters of \cref{tab:detector-parameters}. Note
    that this is a \textit{displacement}, rather than strain, budget. The peaks in the
    noise above \qty{800}{\Hz} are due to higher order mode resonances in the arm
    cavities which both rotate the squeezed state and induce up to
    \qty{520}{\milli\radian} of intrinsic phase noise; see
    \cref{subsec:hom-degradations}.}
  \label{fig:CE-quantum-budget}
\end{figure}

\Cref{fig:CE-quantum-budget} shows a budget of the quantum noises in
Cosmic Explorer using the baseline parameters of
\cref{tab:detector-parameters}. We will describe the squeezing
degradations by the McCuller squeezing
metrics~\cite{McCuller2021}. The details of using these four metrics
to make noise budgets such as \cref{fig:CE-quantum-budget} are given
in \cref{sec:qn-budget}. We first give a brief overview of these
metrics and how mode mismatch contributes to them before delving into
the details of how they quantify the effects of several sources of
squeezing degradations in the following sections. Note that while we
focus on mode mismatch, the HOMs excited by misalignment produce
squeezing degradations just as the HOMs responsible for mode mismatch
do, and so this discussion is applicable to both cases. The solid
green line labeled ``AS Port SQZ'' in \cref{fig:CE-quantum-budget}
represents the quantum noise due to the injected squeezed states in
the absence of any degradations (reduced by the fraction of squeezed
vacuum lost through either an optical loss or a mode mismatch). All of
the other traces represent one of the squeezing degradations which
limit the extent to which this squeezed vacuum can reduce the quantum
noise as follows.


\paragraph{Loss or efficiency}
What is called ``loss'' in an optomechanical system can be thought of
as the quantum noise generated by the unsqueezed vacuum coupling into
the fundamental mode of the system through either an optical loss or a
mode mismatch. When the squeezed state encounters a source of optical
loss, some of the reduced uncertainty squeezed vacuum is lost from the
system and is replaced with unsqueezed vacuum, thereby increasing the
quantum noise. This will be a frequency-dependent noise because the
transmission $\tfr_\mu(\Omega)$ of a field entering some point $\mu$ in
the system
to the readout is, in general, frequency-dependent, and this frequency
dependence will be different for every source of loss. The total
quantum noise due to optical loss can be understood by incoherently
adding the unsqueezed vacuum from every loss
location~\cite{BnC2001,Miao2019}.


Mode mismatch induces a loss by a similar mechanism, except in this
case, rather than being lost from the system, the squeezed vacuum in the
fundamental mode is scattered into higher order modes and is replaced
with the unsqueezed HOM vacuum which scatters into the
fundamental. Unlike with optical loss, however, this process is
coherent and the HOMs interfere with themselves and with the
fundamental. This process of scattering squeezed vacuum between the
various optical modes as they propagate throughout the optical system
and encounter different sources of mismatch will then become a true
loss when only the fundamental mode is ultimately measured on a
photodetector. Rather than coherently adding the contributions from
each HOM, it is therefore easier
to calculate the loss as the fraction of the squeezed vacuum injected
into the system which is ultimately not detected.
However, conceptually mismatch loss is caused by the same mechanism as
optical loss and its frequency dependence could, in principle, be
understood by following the dynamics of the HOMs throughout the
optical system. The combined effects of these loss mechanisms could be
quantified by the efficiency $\eta(\Omega)$, which is the fraction of
the squeezed state injected into the AS port that is ultimately
detected.

Additionally, vacuum fluctuations produce quantum radiation
pressure noise (QRPN) when they enter the arm cavities and beat with
the carrier power circulating there. (Unsqueezed vacuum entering the system
through an optical loss or a mode mismatch in the output path between
the interferometer and photodetectors do not cause QRPN since they
never enter the arm cavities.)  Ref.~\cite{McCuller2021} therefore
introduces the quantum noise gain $\Gamma(\Omega)$, which describes how
the ponderomotive squeezing of the interferometer produces QRPN by
defining the efficiency--gain product as
\begin{equation}
  \Gamma(\Omega)\eta(\Omega) = \frac{|\rrse(+\Omega)|^2 + |\rrse^*(-\Omega)|^2}{2}.
  \label{eq:sqz-metrics-efficiency}
\end{equation}
Likewise, $\Gamma(\Omega)\Lambda_\mu(\Omega)$ quantifies how much
unsqueezed vacuum coupling into the system through some source of
optical loss or mode mismatch is detected in the fundamental mode
including the effects of QRPN. In the case of optical loss, if
$\tfr_\mu(\Omega)$ is the transfer function of the field entering the
system at the location $\mu$ to the detected field, this loss is
\begin{equation}
  \Gamma(\Omega)\Lambda_\mu(\Omega)
  = \frac{|\tfr_\mu(+\Omega)|^2 + |\tfr_\mu^*(-\Omega)|^2}{2}.
  \label{eq:sqz-metrics-loss}
\end{equation}
Since the loss due to mode mismatch is coherent, all sources of
mismatch must be analyzed together and this loss cannot be calculated
as simply in general. In the absence of QRPN, $\Gamma(\Omega) = 1$ and
the total loss $\Lambda(\Omega)=\sum_\mu\Lambda_\mu(\Omega)$ is related to
the efficiency by $\Lambda(\Omega) = 1 - \eta(\Omega)$. It
is important to stress that $\Gamma(\Omega)\eta(\Omega)$ and
$\Gamma(\Omega)\Lambda_\mu(\Omega)$ are \textit{dimensionless couplings}
of how quantum vacuum (squeezed or unsqueezed) entering the system
through some mechanism is detected at the readout and do not quantify
how quantum noise affects a gravitational wave strain signal itself,
as is further discussed in \cref{subsec:strain-referred-loss}. Several
sources of loss are shown in \cref{fig:CE-quantum-budget} as the
solid non-green colored traces.

The remaining two degradation mechanisms are due to the squeezed state
when it encounters dynamics that the upper and lower sidebands
experience differently. They are not caused by the unsqueezed vacuum,
but the sources of optical loss and mode mismatch which couple
unsqueezed vacuum into the fundamental mode add to the dynamics
generating these imbalances---either in phase or in magnitude. Detuned
optical cavities are a common cause of imbalanced sidebands. Since a
HOM accumulates an extra phase shift relative to the fundamental mode
due to its Gouy phase, even cavities that are tuned for the
fundamental are generally detuned for a HOM. Almost any dynamics that
couple the fundamental mode with a higher order mode will thus cause
degradations of this sort. As they are sourced by the squeezed state,
their effects on the noise are squeezing-level-dependent and can be
mitigated by reducing the magnitude of the injected squeezed vacuum at
the expense of decreasing the quantum noise reduction provided by this
vacuum.

\paragraph{Squeezed state rotation}
The squeezed state rotates relative to the angle at which it is
injected as it propagates through the optomechanical system when the
upper and lower sidebands acquire differential phases as~\cite{McCuller2021,Kwee2014}
\begin{equation}
  \theta(\Omega) = \frac{\arg \rrse(+\Omega) - \arg \rrse^*(-\Omega)}{2}.
  \label{eq:sqz-metrics-rotation}
\end{equation}
When the signal is detected at a quadrature angle $\phi\neq -\theta(\Omega)$,
some of the anti-squeezing of the squeezed vacuum injected into the AS
port is observed along with the squeezing.  Radiation pressure
generates such a rotation, and detuned optical cavities, known as
filter cavities, are purposely used in gravitational-wave detectors to
counteract this
rotation~\cite{Kimble2001,McCuller2020,Zhao2020,Ganapathy2023};
however, such a rotation is often unwanted as in the case of that due
to mode mismatch.  The effects of such a frequency-dependent rotation
are shown as the dashed green line labeled ``AS Port anti-SQZ'' in
\cref{fig:CE-quantum-budget}.


\begin{table*}
  \begin{ruledtabular}
    \begin{tabular}{l l c c c c c}
      & & Quadratic Healed & Quadratic Harmed & HOA Healed & HOA Harmed & Fundamental \\
      & & ($\alpha=-1, \beta=-1$) & ($\alpha=-1, \beta=+1$) & ($\alpha=+1, \beta=-1$) & ($\alpha=+1, \beta=+1)$ & \\[0.5ex]
      \hline \\[-2ex]
      \multirow{5}{*}{\rotatebox[origin=c]{90}{\makebox[4.2cm][c]{Direct Coupling [$\unit{\W} \big/ \unit{\W}$]}}}
      & MM loss SN $\Omega \gg \wrse$
          & $\displaystyle\frac{\pi^2c^2\Upsilon}{L_\ta^2}\frac{1}{\Fa^2}\frac{1}{\Omega^2}$
          & $\displaystyle\frac{4c^2\Upsilon}{L_\ta^2}\left(\frac{\Fs}{\Fa}\right)^2\frac{1}{\Omega^2}$
          & $\displaystyle 4\Upsilon$
          & $\displaystyle\frac{16\Upsilon\Fs^2}{\pi^2}$
          & {---} \\[1.5ex]
      & MM loss SN $\Omega \ll \wrse$
          & $\displaystyle\frac{\pi^2\Upsilon}{\Fs^2}$
          & $4\Upsilon$
          & $\displaystyle \frac{4L_\ta^2\Upsilon}{c^2}\left(\frac{\Fa}{\Fs}\right)^2 \Omega^2$
          & $\displaystyle\frac{16L_\ta^2\Fa^2\Upsilon}{\pi^2 c^2}\Omega^2$
          & {---} \\[2.5ex]
      & MM loss QRPN
          & $\displaystyle\frac{\Upsilon\pi^2 \Fa^2\Krp^2}{4\Fs^4}$
          & $\displaystyle\frac{\Upsilon\Fa^2\Krp^2}{\Fs^2}$
          & $\displaystyle\frac{\Upsilon\Fa^2\Krp^2}{\Fs^2}$
          & $\displaystyle\frac{4\Upsilon\Fa^2\Krp^2}{\pi^2}$
          & {---} \\
      & SEC loss SN $\Omega\gg\wrse$
          & {---}
          & {---}
          & {---}
          & {---}
          & $\displaystyle\frac{2\Fs\varepsilon_\ts}{\pi}$ \\[2ex]
      & SEC loss SN $\Omega\ll\wa$
          & {---}
          & {---}
          & {---}
          & {---}
          & $\displaystyle\frac{\pi\varepsilon_\ts}{2\Fs}$ \\[2ex]
      & SEC loss QRPN
          & {---}
          & {---}
          & {---}
          & {---}
      & $\displaystyle\frac{\varepsilon_\ts\Fa^2\Krp^2}{2\pi\Fs}$ \\[1.5ex]
      \hline \\[-2ex]
      \multirow{5}{*}{\rotatebox[origin=c]{90}{\makebox[3.7cm][c]{Strain-Referred [$1\unit{\rtHz}$]}}}
      & MM loss SN
          & $\displaystyle\frac{\pi}{2kL_\ta}\sqrt{\frac{\Upsilon}{\Fa\Fs}\qfrac}$
          & $\displaystyle\frac{1}{kL_\ta}\sqrt{\frac{\Upsilon\Fs}{\Fa}\qfrac}$
          & $\displaystyle\frac{\Omega}{ck}\sqrt{\frac{\Upsilon\Fa}{\Fs}\qfrac}$
          & $\displaystyle\frac{2\Omega}{\pi ck}\sqrt{\Upsilon\Fa\Fs\qfrac}$
          & {---} \\[2ex]
      & MM loss QRPN
          & $\displaystyle\frac{4\pi Q}{L_\ta\Fs}\sqrt{\frac{\Upsilon\Fa}{\Fs}}$
          & $\displaystyle\frac{8Q}{L_\ta}\sqrt{\frac{\Upsilon\Fa}{\Fs}}$
          & $\displaystyle\frac{8Q}{L_\ta}\sqrt{\frac{\Upsilon\Fa}{\Fs}}$
          & $\displaystyle\frac{16Q}{\pi L_\ta} \sqrt{\Upsilon\Fa\Fs}$
          & {---} \\
      & SEC loss SN $\Omega\gg\wa$
          & {---}
          & {---}
          & {---}
          & {---}
          & $\displaystyle\frac{\Omega}{2ck} \sqrt{\frac{2\Fa\varepsilon_\ts}{\pi}\qfrac}$ \\[2ex]
      & SEC loss SN $\Omega\ll\wa$
          & {---}
          & {---}
          & {---}
          & {---}
          & $\displaystyle\frac{1}{2kL_\ta} \sqrt{\frac{\pi\varepsilon_\ts}{2\Fa}\qfrac}$ \\[2ex]
      & SEC loss QRPN
          & {---}
          & {---}
          & {---}
          & {---}
          & $\displaystyle\frac{8Q}{L_\ta}\sqrt{\frac{\Fa\varepsilon_\ts}{2\pi}}$ \\[1.5ex]
      \hline \\[-2ex]
      \multirow{5}{*}{\rotatebox[origin=c]{90}{\makebox[1cm][c]{HOM Res.}}}
      & SQZ rotation gain
          & $\displaystyle\frac{16\Fs^3\Fa}{\pi^4}\frac{L_\ta}{c}$
          & $\displaystyle\frac{\Fa}{\Fs}\frac{L_\ta}{c}$
          & $\displaystyle\frac{8\Fs^2\Fa}{\pi^3}\frac{L_\ta}{c}$
          & $\displaystyle\frac{2\Fa}{\pi}\frac{L_\ta}{c}$
          & {---} \\[1.5ex]
       & Loss gain
          & $\displaystyle\frac{8\Fs^2}{\pi^2}$
          & $2$
          & $\displaystyle\frac{4\Fs}{\pi}$
          & $\displaystyle\frac{4\Fs}{\pi}$
          & {---} \\[1.5ex]
       & Dephasing gain
          & $\displaystyle\frac{16\Fs^4}{\pi^4}$
          & $1$
          & $\displaystyle\frac{4\Fs^2}{\pi^2}$
          & $\displaystyle\frac{4\Fs^2}{\pi^2}$
          & {---}
    \end{tabular}
  \end{ruledtabular}
  \caption{Summary of important squeezing degradations and their dependence on detector
    parameters. MM loss is mode mismatch loss, SEC loss is optical loss in the signal
    extraction cavity, SN is shot noise, and QRPN is quantum radiation pressure noise.
    Strain-referred QRPN is multiplied by the factor
    $Q=\sqrt{P_\ta\hbar\omega_0}/Mc\Omega^2$ and $\Krp =-16kP_\ta/Mc\Omega^2$. The
    column headers also summarize the notation used throughout the paper where $\alpha$
    determines the type of mismatch (\cref{eq:phenom-mismatch-mapping}) and
    $\beta=\rme^{2\rmi\psi_\ts}$ determines the resonance condition of the HOM in the
    SEC with $\beta=-1$ corresponding to mode healing of an exactly anti-resonant HOM
    and $\beta=+1$ corresponding to mode harming of an exactly resonant HOM; $\psi_\ts$
    is the one-way Gouy phase of the HOM in the SEC. Note that the behavior of a general
    HOM will be mostly characteristic of exact anti-resonance $\psi_s=\pi/2$ ($\beta=-1$) unless the
    HOM is close to an SEC resonance. The direct loss couplings are described in
    \cref{subsec:direct-loss-coupling,subsec:mismatch-loss-qrpn} and the strain-referred
    loss is described in \cref{subsec:strain-referred-loss}. Note that the shot noise
    loss scalings below the arm cavity pole are not relevant in practice because the
    QRPN loss will be dominant here. \Cref{subsec:hom-degradations} describes the
    degradations around HOM arm resonances. The HOM resonance gains are the frequency
    independent factors in the McCuller metrics in
    \cref{eq:hom-resonance-rotation,eq:hom-resonance-loss,eq:hom-resonance-dephasing}:
    SQZ rotation is $g_\ts/\wcc$, loss is $2g_\ts$, and dephasing is $g_\ts^2$.
    }
  \label{tab:loss-scalings}
\end{table*}

\paragraph{Dephasing}
An intrinsic dephasing is generated when the magnitude of the optomechanical response
to the upper and lower sidebands differ; it is quantified
by~\cite{McCuller2021}
\begin{equation}
  \Xi(\Omega) =
  \frac{\left[|\rrse(+\Omega)| - |\rrse^*(-\Omega)|\right]^2}{4\Gamma(\Omega)\eta(\Omega)}.
  \label{eq:sqz-metrics-dephasing}
\end{equation}
This phase noise can be understood by the fact that the noise in the upper and lower
sidebands of a squeezed vacuum is increased relative to that of the unsqueezed vacuum,
but in a correlated way so that the noise in one quadrature is reduced while the noise
in the orthogonal quadrature is increased~\cite{Bachor2004,Barsotti2019}. A phase noise
is thus introduced when the sidebands experience different losses which preserve the
increased noise while degrading the correlations responsible for defining the angle of
the squeeze ellipse~\cite{McCuller2021}. The phase noise generated by this intrinsic
dephasing is shown as the dotted green line in \cref{fig:CE-quantum-budget}. The other
non-green dashed curves in \cref{fig:CE-quantum-budget} are technical sources of phase
noise, such as squeeze angle fluctuations due to OPA cavity length
fluctuations~\cite{Dwyer2013}, as is discussed further in \cref{sec:qn-budget}.

In the following sections we give simple approximations to these
metrics for the phenomenological model of \cref{subsec:rse-mismatched}
in the limits of exact HOM resonance $\psi_\ts=0$ and anti-resonance
$\psi_\ts=\pi/2$ in the SEC. There is a continuum of behavior between
these two cases, but the dynamics of a general field will mostly be
characteristic of anti-resonance unless the mode is close to an SEC
resonance. These expressions well explain many aspects of the exact
calculations shown in all of the figures and described in
\cref{sec:simulation} since the dominant degradation effects will
generally be due to one of many HOMs. The degradations can then be
qualitatively understood by considering this dominant HOM as the
single HOM in this simpler model keeping in mind that the specifics of
its coupling and interference with the other HOMs not considered alter
the exact details of its dynamics and thus the resulting squeezing
degradations.

\Cref{subsec:broadband-loss} discusses the broadband loss due both quadratic and higher
order internal mode mismatch and to optical loss in the SEC.
\Cref{subsec:hom-degradations} describes the squeezing degradations that arise from a
HOM which becomes resonant in the arm cavities, how the locations of these resonances
are determined by the detector design, and how they may be leveraged to diagnose the
thermal state of the detectors and to tune them to improve the detector performance.
\Cref{tab:loss-scalings} summarizes how the degradations described in these sections
depend on detector parameters. \Cref{subsec:broadband-rotation} describes a broadband
rotation of the squeezed state induced by a detuning of the SEC which can arise either
by a mode mismatch or an SEC length detuning. \Cref{subsec:external-mismatch} discusses
the interactions between internal and external mismatch.


As this work is primarily focused on the effects of mode mismatch, we
neglect optical loss in most of the approximations to an exact model
in order to arrive at tractable expressions. However, all sources of
loss detailed in \cref{tab:detector-parameters} are included in all of
the numerical results presented in the figures and are discussed where
appropriate in the text.

\subsection{Broadband loss}
\label{subsec:broadband-loss}


Out of all of the squeezing degradations due to mode mismatch, broadband loss generally
has the most significant impact to the overall sensitivity of a gravitational-wave
detector. Since radiation pressure obscures some characteristics of the purely optical
propagation of quantum vacuum throughout the optomechanical system, we first discuss the
direct coupling of quantum vacuum entering the optomechanical system to the readout in
\cref{subsec:direct-loss-coupling} in the absence of radiation pressure and then discuss
QRPN in \cref{subsec:mismatch-loss-qrpn}. The quantity that is ultimately relevant to
gravitational-wave detectors, and the one that is shown in noise budgets such as
\cref{fig:CE-quantum-budget}, is the effective loss that contaminates the strain
measurement itself, also known as the signal- or strain-referred loss, as is discussed
in \cref{subsec:strain-referred-loss}. Both of these losses are summarized in
\cref{tab:loss-scalings}.

The loss due to the mode mismatch between the arm cavities and the SEC
is sometimes treated as being a contributor to the optical loss
present in the SEC. They do have some similarities since the
unsqueezed quantum vacuum responsible for each of these losses couple
into the fundamental mode in similar ways; however, there are
important differences, even in addition to the coherent nature of the
former, and they must be treated separately. We therefore describe
both of these losses to highlight the connection between the two. In
both LIGO~\Asharp{} and Cosmic Explorer the total round-trip SEC loss
goal is \qty{500}{\ppm}. This should be taken as the total
\textit{equivalent} SEC loss properly budgeting the combined effects of
both mode mismatch and optical loss as explained in this section.

\subsubsection{Direct loss coupling in the absence of radiation pressure}
\label{subsec:direct-loss-coupling}

Since we do not consider the effects of radiation pressure in this
section, the quantum noise gain is $\Gamma(\Omega) = 1$ and all of the
equations of \cref{sec:cc-dynamics} directly apply. Furthermore, the
analytic expressions we give in this section are all for balanced
sidebands where $|\tfr_\mu(+\Omega)|=|\tfr_\mu(-\Omega)|$, and so the
loss in \cref{eq:sqz-metrics-loss} is just $\Lambda_\mu(\Omega) =
|\tfr_\mu(\Omega)|^2$.

\begin{figure}
  \centering
  \includegraphics[width=\columnwidth]{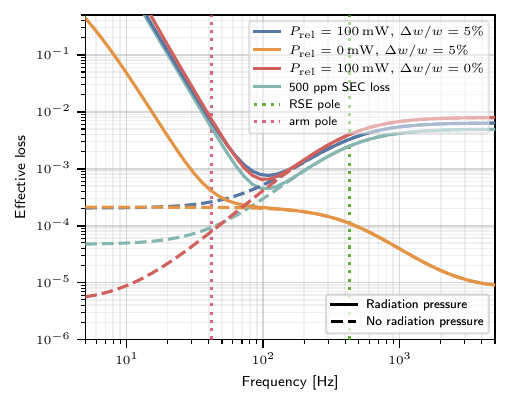}
  \caption{Direct mode mismatch loss for $\dww=\qty{5}{\%}$ and
    $\Pabs=\qty{100}{\mW}$ along with both types of aberrations
    separately for LIGO \Asharp{} using the parameters of
    \cref{tab:detector-parameters}. The dashed lines do not include
    radiation pressure and correspond to the discussion in
    \cref{subsec:direct-loss-coupling} while the solid lines include
    radiation pressure as discussed in
    \cref{subsec:mismatch-loss-qrpn}. The low-pass nature of quadratic
    mismatch and high-pass nature of higher order aberrations is
    evident, along with the fact that higher order modes can interfere
    destructively to produce less loss. \qty{500}{\ppm} of optical SEC loss is
    also shown for comparison.}
  \label{fig:mismatch-loss}
\end{figure}

Optical SEC loss could be due to, for example, the anti-reflective
coatings of the optics in that cavity. In the coupled cavity model,
this loss is due to the fundamental mode quantum vacuum entering the
system at the port $\mu_\text{a,r}$ of \cref{fig:coupled-cavity} or
$\mu_\text{a,r}^0$ of \cref{fig:phenom-cc-sflow}. The transmission of a
fundamental field entering the system here to the detected field read
out at $\mu_\text{as,r}^0$ is
\begin{equation}
  \tfr_\text{sec}(\Omega) = \frac{E_\text{as,r}^0}{E_\text{sec,i}^0} =
  \sqrt{\frac{\pi\varepsilon_\ts}{2\Fs}}\frac{1 + \rmi\Omega/\wa}{1 +
    \rmi\Omega/\wrse}.
  \label{eq:sec-loss}
\end{equation}
Such a field simply experiences the SEC RSE optical loop suppression
\cref{eq:fundamental-sec-closed-loop} followed by the transmission
$t_\ts\approx\sqrt{2\pi/\Fs}$ through the SEM. The resulting
$\Lambda_\text{sec}(\Omega) = |\tfr_\text{sec}(\Omega)|^2$ for LIGO
\Asharp{} is shown as the dashed teal curve in
\cref{fig:mismatch-loss} for $\varepsilon_\ts=\qty{500}{\ppm}$.

The unsqueezed quantum vacuum responsible for mode mismatch loss are
the higher order modes of the squeezed field $E_\text{as,i}$ injected
into the interferometer at $\mu_\text{as,i}$ shown in
\cref{fig:coupled-cavity} which scatter into the fundamental mode when
that field encounters the thermal aberrations described by the surface
$\mat{S}_{ij}$ and lens $\mat{L}_{ij}$ operators as discussed in
\cref{subsec:cc-aberrations} and summarized in \cref{tab:aberration-wide}. In the two mode model of
\cref{subsec:rse-mismatched,fig:phenom-cc-sflow}, it is possible to
calculate the mismatch loss directly by propagating the single HOM
$E_\text{as,i}^1$ entering the interferometer along with the
fundamental squeezed field $E_\text{as,i}^0$ to the fundamental mode
$E_\text{as,r}^0$ which is measured. This HOM couples into the
fundamental mode of the SEC at $\mu_\text{a,r}^0$ through
$t_\ts\tfr_\text{a,h}(\Omega)$ where $\tfr_\text{a,h}(\Omega)$ is the
transmission of the HOM from the HOM SEC into the fundamental SEC
given by \cref{eq:mismatched-hom-arm-transmission}; equivalently, this
is the reflection of the HOM off of the mismatched arm cavity and the
subsequent scattering into the fundamental. This field then
experiences the fundamental's SEC optical loop suppression, exactly as the
vacuum due to optical SEC loss does, to give a total transmission of
the HOM to the fundamental readout of
\begin{subequations}
\label{eq:broadband-hom-transmission}
\begin{align}
  \tfr_\text{rse,h}(\Omega) &= \frac{E_\text{as,r}^0}{E_\text{as,i}^1} =
  \sqrt{\Upsilon} b \frac{1 - r_\ts}{1 - \beta r_\ts} \frac{(\alpha - 1) +
    (1 + \alpha) \rmi\Omega/\wa}{1 + \rmi\Omega/\wrse} \raisetag{10ex} \\[1ex]
  &=
  \begin{cases}
    \displaystyle \frac{-\rmi\pi\sqrt{\Upsilon}}{\Fs} \frac{1}{1 + \rmi\Omega/\wrse} &
    \alpha = -1, \beta=-1 \\[1em]
    \displaystyle
    \frac{2\sqrt{\Upsilon}}{1 + \rmi\Omega/\wrse} &
    \alpha = -1, \beta=+1 \\[1em]
    \displaystyle -2\rmi\sqrt{\Upsilon}\frac{\rmi\Omega/\wrse}{1 + \rmi\Omega/\wrse} &
    \alpha=+1, \beta=-1 \\[1em]
    \displaystyle \frac{4\Fs\sqrt{\Upsilon}}{\pi} \frac{\rmi\Omega/\wrse}{1 + \rmi\Omega/\wrse}
    & \alpha=+1, \beta=+1
    \end{cases}
\end{align}
\end{subequations}
where $\beta=\rme^{2\rmi\psi_\ts}$ and is $+1$ for resonant HOMs and $-1$ for
anti-resonant HOMs; and $b=\rme^{-\rmi\pi(1 - \beta)/4}$ and is 1 for $\beta=+1$ and
$-\rmi$ for $\beta=-1$. The resulting loss due to mode mismatch
$\Lambda_\text{mm}(\Omega) = |\tfr_\text{rse,h}(\Omega)|^2$ is detailed in
\cref{tab:loss-scalings}. It is shown in \cref{fig:mismatch-loss} as the dashed blue
curve for \Asharp{} using the exact model of \cref{fig:coupled-cavity,sec:simulation}
with $\dww=\qty{5}{\%}$ and $\Pabs=\qty{100}{\mW}$---calculated now using
\cref{eq:general-mismatch-loss} since there are many HOMs which cannot be easily
propagated as in \cref{eq:broadband-hom-transmission}. The figure also shows the loss if
only the quadratic or only the higher order aberrations were present. The low-pass
dynamics of quadratic mismatch and high-pass dynamics of higher order aberrations are
evident in both \cref{eq:broadband-hom-transmission,fig:mismatch-loss}. This is
explained by \cref{eq:arm-reflection-interference} in the model of
\cref{subsec:rse-mismatched} by the different interference between the HOM and the fundamental
on reflection of the arm cavity leading to the HOM and fundamental SECs being AC coupled
through higher order aberrations while being DC coupled through quadratic
mismatch.\footnote{The low frequency deviation seen in \cref{fig:mismatch-loss} from the
high-pass dynamics of higher order aberrations predicted by
\cref{eq:arm-reflection-interference,eq:broadband-hom-transmission} is due to the fact
that $\rfr_\ta(\Omega)$ is not exactly $+1$ for a non-resonant field due both to the
small phase shift present in \cref{eq:arm-reflection-pade} even when $|\dwa|\gg1$ and
due to optical loss in the arm cavity.} Though the details are more complicated, the
same is true in the general case where the arm reflection is described by
\cref{eq:exact-cc-reflection,eq:cc-reflection-phenom-model}, rather than by
\cref{eq:approx-cc-reflection,eq:mismatch-matrices,eq:phenom-mismatch-mapping}, and
aberrations where $\Ur\approx\Ui^{-1}$ tend to have interference leading to low-pass
dynamics while aberrations where $\Ur\approx\Ui$ tend to have interference resulting in
high-pass dynamics.

Higher order aberrations will usually be the largest source of
mismatch loss for the frequencies where mismatch loss is significant
due to their high-pass nature. However, \cref{fig:mismatch-loss} also
underscores the fact that the dynamics responsible for mode mismatch
loss are coherent: even though quadratic mismatch is sub-dominant by
several orders of magnitude at high frequencies, its presence still
\textit{decreases} the noise relative to the case where only higher
order aberrations are present \textit{in the particular thermal state
  and optical configuration shown in the figure}. This is because in
the particular case of \cref{fig:mismatch-loss}, the mode content
interferes destructively in this particular thermo-optical system. This is a
generic, but not universal, occurrence and it is difficult to predict
what kind of interference will arise for a given thermal state and
optical configuration.

\begin{figure}
  \centering
  \includegraphics[width=\columnwidth]{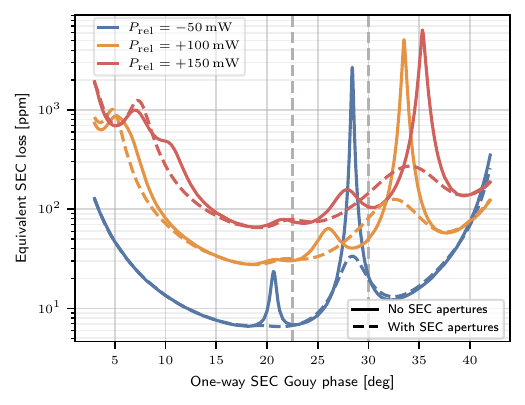}
  \caption{Mode mismatch loss for Cosmic Explorer varying
    $\Pabs$ while keeping $\dww=\qty{5}{\%}$ as a function of
    SEC Gouy phase $\Psi_\ts$.  The vertical dashed lines mark the 8th order
    resonance at \ang{22.5} and the 6th order resonance at \ang{30} in
    the absence of thermal aberrations and apertures. The Gouy phase
    is the one-way Gouy phase as computed from the round-trip ABCD
    matrix of the SEC; see \cref{subsec:cavity-eigenmodes}. Recall
    \cref{fig:thermal-lens} and the fact that $\Pabs$ can be
    negative in our parameterization.
  }
  \label{fig:loss-vs-SEC-gouy-phase}
\end{figure}

The Gouy phase and finesse of the SEC have a critical impact on the loss due to mode
mismatch within the coupled cavity.  Since mode mismatch loss and optical SEC loss have
the same spectral shape above the arm cavity pole in most cases of interest, one way of
quantifying this impact is to find the amount of optical SEC loss which would be
equivalent to the amount of mismatch loss given a particular thermal state \textit{and}
optical configuration. This parameterization gives a measure of how much that particular
mode mismatch loss would contribute to the optical SEC loss if it were, incorrectly,
included in that loss budget as is often done. This is shown in
\cref{fig:loss-vs-SEC-gouy-phase} for CE as a function of the SEC Gouy phase for three
different thermal states. As described in the last paragraph of
\cref{subsec:hom-couplings}, we imagine that the thermal actuators can perfectly correct
the thermal aberrations in some hot state with of order one watt of power absorbed in
the test mass coatings, and $\Pabs$ is the difference between the actual absorption and
this perfect compensation. The Gouy phase $\Psi_\ts$ shown in the figure is the one-way
Gouy phase of the SEC as calculated from the round-trip ABCD matrix of the cavity in
this perfectly matched state. In contrast, the Gouy phase referred to in most of our
discussion, denoted as $\psi_\ts$ and $\beta=\rme^{2\rmi\psi_s}$, is the true one-way
phase that a given mode accumulates. In the absence of thermal aberrations and
apertures, $\psi_\ts = N\Psi_\ts$ for a mode of order $N$; see
\cref{subsec:cavity-eigenmodes} for details about the general relationship between these
two notions of Gouy phase. It should be emphasized that figures like this should be
considered in the target hot state when being used in a detector design. 


Since the HOM vacuum experiences a different SEC than the fundamental,
it will be enhanced or suppressed to a varying degree depending on its
Gouy phase $\psi_\ts$ in the cavity. The round-trip phase for an
anti-resonant HOM ($\beta=-1$) is negative, and so it is suppressed by
the SEC dynamics. Such modes are said to be ``healed'' by the cavity. The
round-trip phase for a resonant HOM ($\beta=+1$) is positive so it is
enhanced in the SEC. Such modes are said to be ``harmed'' by the
cavity. \Cref{tab:loss-scalings} gives the mismatch loss for these two
cases for both quadratic and higher order aberrations. Mode harmed
loss is enhanced by a factor of $\Fs$ \textit{relative} to SEC loss
and mode healed loss is suppressed by a factor of $\Fs$
\textit{relative} to SEC loss. There will be a continuum of behavior
between exact anti-resonance $\beta=-1$ and exact resonance
$\beta=+1$, and thus between suppression by $\Fs$ or enhancement by
$\Fs$, but most HOMs will be more characteristic of anti-resonance and
thus suppression by the SEC to some degree.

Most of the peaks in the noise evident in
\cref{fig:loss-vs-SEC-gouy-phase} are due to a single HOM becoming
resonant in the SEC (all HOMs are resonant at \ang{0} and mode orders
which are integer multiples of four are resonant at \ang{45}). These
resonances become both larger and narrower as $\Fs$ is increased. In
the absence of higher order aberrations and apertures, the eighth
order modes are resonant at \ang{22.5} and the sixth order modes are
resonant at \ang{30}; these frequencies are marked with vertical
dashed lines in the figure. As higher order aberrations are
introduced, the magnitude of the loss increases and the locations of
the resonances shift in a direction determined by the sign of $\Pabs$. Note
that the Gouy phase of the cavity $\Psi_\ts$ is not changing here
because the quadratic part of the thermal lens is being held constant;
rather, the relation $\psi_\ts=N\Psi_\ts$ between the true phase that a
given HOM accumulates and the ABCD cavity Gouy phase is no longer
valid in the presence of higher order aberrations or
apertures. Varying the quadratic mismatch does change the cavity Gouy
phase $\Psi_\ts$, thus further shifting the resonances, in addition to
changing the overall magnitude of the loss due to the interference
described above. This underscores the necessity of controlling both
the quadratic and higher order aberrations in order to keep the SEC
Gouy phase constant, and thus prevent shifting HOMs from becoming
resonant.


The addition of apertures in the cavity clips the HOMs and therefore
introduces extra loss for the HOMs which are clipped. Furthermore, the
higher order HOMs (the ones mainly responsible for the higher order
aberrations) will be clipped more than the lower order ones due to
their larger spatial extent. This increased loss will drastically
reduce the extent to which the modes are enhanced, or mode harmed, in the SEC and will
slightly increase the mode mismatch loss away from the resonance
peaks. Note that a change in the quadratic mismatch will be
accompanied by a change in the beam size and thus also the extent
to which the HOMs are clipped by the apertures.



\Cref{fig:loss-vs-SEC-gouy-phase} does not illustrate the fact that
for some thermal states simply increasing the finesse may change the
high-pass vs.\ low-pass character of the mismatch loss. This may seem
counterintuitive since whether the mismatch has a low-pass or
high-pass character depends only on whether the mismatch is mainly due
to quadratic mismatch or to higher order aberrations and not on the
dynamics of the SEC
(see \cref{eq:arm-reflection-interference,eq:broadband-hom-transmission}). However,
when there are many HOMs present, the HOMs mainly responsible for one
type of mismatch may be mode harmed while the HOMs mainly responsible
for the other type are mode healed. Thus, whether the total mismatch
loss is mostly high-pass or mostly low-pass can change as the finesse
is varied and the loss due to the two types of mismatch is variably
enhanced or suppressed. Similarly, the frequency dependence of the
mismatch can change as the Gouy phase, and hence which modes are
resonant, changes.

The desired instrument bandwidth is the main driver for the SEC finesse with a higher
finesse resulting in a wider bandwidth (\cref{eq:cc-pole}). In order to keep a fixed
bandwidth, the finesse must scale as $\Fs\propto \Fa L_\ta$, and so longer and higher
finesse arm cavities generally result in higher finesse SECs. Nevertheless, there is
some choice in the bandwidth, and this discussion has highlighted the fact that higher
finesse SECs are generally favorable for reducing loss due to mode mismatch as long as a
reasonable SEC Gouy phase is chosen along with the ability to control it in the presence
of realistic thermal aberrations. The case of $\Pabs=\qty{100}{\mW}$ and
$\dww=\qty{5}{\%}$ shown in \cref{fig:loss-vs-SEC-gouy-phase} with a \ang{20} Gouy phase
results in \qty{30}{\ppm} equivalent SEC loss for Cosmic Explorer with the baseline SEM
transmission of $T_\ts=\qty{2}{\%}$ ($\Fs\approx 310$). For comparison, LIGO currently
has $T_\ts=\qty{32.5}{\%}$  ($\Fs\approx 18$) which results in \qty{650}{\ppm}
equivalent loss for the same thermal state and Gouy phase. A potential near-term change
to slightly broaden the bandwidth using $T_\ts=\qty{20}{\%}$ ($\Fs\approx 30$) results
in \qty{370}{\ppm} equivalent loss, and a more wideband tuning using $T_\ts=\qty{5}{\%}$
($\Fs\approx 125$), proposed for its increased sensitivity to high frequency
signals~\cite{Ganapathy2020}, would result in \qty{90}{\ppm} equivalent SEC loss.

We also stress that the relatively low mismatch loss shown in
\cref{fig:CE-quantum-budget} and in \cref{fig:loss-vs-SEC-gouy-phase}
for some ranges of Gouy phase---having an equivalent SEC loss that is a
small fraction of the \qty{500}{\ppm} target---should not be taken as
an indication that internal mismatch loss will not be important in a future
detector with a higher finesse SEC such as CE. In reality, the thermal
aberrations will not be as simple as the beam-heating of a test mass
optic shown in \cref{fig:thermal-lens} considered here, and
aberrations with more spatial structure may have significantly
more---or even occasionally less---loss than this analysis suggests
due to the details of the residual mode content, optical dynamics, and
the ensuing interference of the HOMs. Nevertheless, the impact of this loss for a given thermal aberration will generally be less due to the increased mode healing.

\subsubsection{Quantum radiation pressure noise due to mode mismatch}
\label{subsec:mismatch-loss-qrpn}


The direct loss coupling described above does not include quantum
radiation pressure noise (QRPN). Recall that amplitude quadrature
fluctuations are converted into phase quadrature fluctuations through
the optomechanical interaction between the light in the arm cavities
and the test mass mirrors. This correlates the upper and lower
sidebands so that $E^0_\text{as,r}(\Omega)/E^0_\text{sec,i}(\Omega)$
is no longer given by \cref{eq:sec-loss} and
$E^0_\text{as,r}(\Omega)/E^1_\text{as,i}(\Omega)$ is no longer given by
\cref{eq:broadband-hom-transmission} for the low frequencies where the
interferometer ponderomotively squeezes the vacuum and radiation
pressure is significant. Rather, the field $E^0_\text{as,r}(\Omega)$
in reflection of the coupled cavity is a function of both the upper
$E_\mu(+\Omega)$ and conjugate lower $E_\mu^*(-\Omega)$
sidebands. The necessary transfer functions are still computed as
described above except that the upper and lower sidebands must be
propagated together and the reflection from the ETM---in the exact
case illustrated in \cref{fig:coupled-cavity}---or from the arm
cavity---in the phenomenological case illustrated in
\cref{fig:phenom-cc-sflow}---must be replaced by matrices which mix
the sidebands.

In particular, in the analysis of \cref{subsec:rse-mismatched}, the
perfectly matched arm reflection given by
\cref{eq:arm-reflection-pade} can be evaluated at $\Omega=0$ when
analyzing QRPN, and the following substitutions should be made
\begin{subequations}
\begin{align}
  \rfr_{\ta0}(\Omega)&\to
  \begin{bmatrix}
    -1 + \rmi\Krp_\ta(\Omega) / 2 & \rmi\Krp_\ta(\Omega) / 2 \\
    -\rmi\Krp_\ta(\Omega) / 2 & -1 -\rmi\Krp_\ta(\Omega) / 2
  \end{bmatrix} \\
  \rfr_{\ta1}(\Omega) &\to
  \begin{bmatrix}
    1 & 0 \\
    0 & 1
  \end{bmatrix}
\end{align}
\end{subequations}
where the first column corresponds to the upper sideband and the
second to the conjugate lower. Here, $\Krp_\ta(\Omega) =
2\Fa/\pi\times2\Krp_\text{fm}(\Omega)$ is the optomechanical coupling
of the arm cavity where $\Krp_\text{fm}(\Omega)$ is the optomechanical
coupling of a free mass given by
\cref{eq:free-mass-rp-coupling}. Since there is essentially no HOM
power in the arm cavity for the HOM vacuum to beat with, the HOM is
not ponderomotively squeezed on direct reflection from the arm and its
sidebands are not correlated by $\rfr_{\ta1}(\Omega)$. However, the
HOM vacuum is still responsible for radiation pressure through the
process where it scatters into the fundamental mode before reflecting
off of the arm cavity. In the picture suggested by
\cref{fig:phenom-cc-sflow}, the HOM and fundamental SECs are thus no
longer AC coupled through higher order aberrations in the presence of
radiation pressure.

The resulting QRPN due to mode mismatch loss from
\cref{eq:sqz-metrics-loss} is
\begin{subequations}
\begin{align}
  \Gamma(\Omega)\Lambda_\text{mm}(\Omega) &=
  \Upsilon |\Krp_\text{rse}(\Omega)|^2
  \left(\frac{1 + \alpha r_\ts}{1 - \beta r_\ts}\right)^2 \\
  &= \Upsilon|\Krp_\text{rse}(\Omega)|^2\times
  \begin{cases}
    \displaystyle \left(\!\frac{\pi}{2\Fs}\!\right)^2 & \alpha=\beta=-1 \\
    1 & \alpha = -\beta \\
    \displaystyle \left(\!\frac{2\Fs}{\pi}\!\right)^2 & \alpha=\beta=+1
  \end{cases}
\end{align}
\end{subequations}
which is to be compared to the QRPN due to optical SEC loss
\begin{equation}
  \Gamma(\Omega)\Lambda_\text{sec}(\Omega) =
  \frac{\Fs\varepsilon_\ts}{2\pi}\,|\Krp_\text{rse}(\Omega)|^2.
\end{equation}
These losses are shown in \cref{fig:mismatch-loss} for LIGO \Asharp{}
as the solid lines. The dashed lines in \cref{fig:mismatch-loss}
correspond to the analysis of \cref{subsec:direct-loss-coupling} where
there is no radiation pressure. It is also useful to instead look at
$\Lambda_\text{mm}(\Omega)$ including QRPN but with the quantum noise
gain $\Gamma(\Omega)\propto |\Krp_\text{rse}(\Omega)|^2\propto
P_\ta^2/M^2\Omega^4$ factored out as is shown in
\cref{fig:Asharp-relgamma-budget}. The magnitude of the QRPN due to HOM vacuum
scattering into the fundamental mode due to mode mismatch relative to that due to all
sources of fundamental mode vacuum can be seen in the noise budgets of
\cref{fig:CE-quantum-budget,fig:Asharp-relgamma-budget}.

Note that while the shot noise and radiation pressure due to quantum
vacuum entering the system along the injection path are equal at the
RSE SQL frequency $\wsql^\text{rse}$ given by
\cref{eq:cc-sql-frequencies}, shot noise and radiation pressure due to
the vacuum entering inside the coupled cavity through either higher
order aberrations or optical SEC loss are equal at the generally
higher frequency $\Omega_\text{int}^6 = (\Fa\wa/\pi)^2\times
(\wsql^\text{fm})^4 = (c/4L_\ta)^2\times (\wsql^\text{fm})^4$ as is
evident in \cref{fig:CE-quantum-budget}.

\subsubsection{Strain-referred loss}
\label{subsec:strain-referred-loss}

We have so far discussed the direct coupling of optical SEC loss and
mode mismatch loss due to HOM vacuum scattering into the fundamental
mode of the field which is ultimately detected. This is shown in
\cref{fig:mismatch-loss} and is the transmission of the unsqueezed
vacuum fields entering the system at the nodes $\mu_\text{as,i}$ and
$\mu_\text{a,r}$ in \cref{fig:coupled-cavity} to the fundamental
$E_\text{as,r}^0$. However, the quantity that is directly relevant for
gravitational-wave detectors is the strain- or signal-referred loss
which is shown in \cref{fig:CE-quantum-budget}. This is the effective
loss as it would appear if it entered the system in the same way that
a gravitational wave strain signal enters; in
\cref{fig:coupled-cavity} a strain signal excites the optical field at
the node $\mu_x$. Since the optomechanical
plant~\cref{eq:optical-plant}, also known as the sensing
function~\cite{Abbott2016}, describes the propagation of a
gravitational wave signal to the readout, the strain-referred loss is
given by dividing the direct coupling by the optomechanical plant in
order to ``calibrate'' the noise into an equivalent displacement noise
and then dividing by the arm length $L_\ta$ to give an equivalent
strain noise.\footnote{This is not precisely strain for detectors with
long arms such as CE because displacement and strain do not differ
simply by a factor of arm length for frequencies comparable to or
larger than the FSR. Rather, they have an additional frequency and
source location dependent
correction~\cite{Rakhmanov2008,Essick2017}. We only show displacement
noise in this paper to avoid this complication. Simply dividing by
$L_\ta$, however, correctly captures the scaling of the noises with
arm length which is relevant for a strain measurement.}  The amplitude
spectral density of this strain-referred noise is then
\begin{equation}
  S_{hh}^{1/2}(\Omega) = \frac{1}{L_\ta}
  \sqrt{\frac{\Gamma(\Omega)\Lambda(\Omega)}{|C(\Omega)|^2}\frac{\hbar\omega_0}{2}},
  \label{eq:strain-referred}
\end{equation}
where $\Lambda(\Omega)$ is the appropriate direct coupling, either
$\Lambda_\text{mm}(\Omega)$ or $\Lambda_\text{sec}(\Omega)$ in this
case. The $\hbar\omega_0/2$ is the half-quanta of vacuum energy which
must be propagated through the direct coupling $\Lambda(\Omega)$ in
order to produce the quantum noise measured on a photodetector in
physical units.

Strain-referred SEC loss and mode mismatch loss are not affected by
the fundamental mode SEC because they experience the dynamics of this
cavity in exactly the same way as a strain signal does once the
relevant fields enter the cavity (at the node $\mu_\text{a,r}^0$ of
\cref{fig:phenom-cc-sflow}). Mathematically, the optomechanical plant
(\cref{eq:cc-transmission-zpk,eq:optical-plant}) and the relevant
vacuum transmissions to the readout
(\cref{eq:sec-loss,eq:broadband-hom-transmission}) are
\begin{align}
  \frac{|C(\Omega)|^2}{2k^2 P_\ta} = |\tfr_\text{rse}(\Omega)|^2
  &= |t_\ts H_\ts(\Omega) \!\times\! \sqrt{1 - \Upsilon}\,\tfr_\ta(\Omega)|^2 \\
  \left|\frac{E_\text{as,r}^0}{E_\text{as,i}^1}\right|^2 =  |\tfr_\text{rse,h}(\Omega)|^2
  &= |t_\ts H_\ts(\Omega) \!\times\! t_\ts\tfr_\text{a,h}(\Omega)|^2 \\
  \left|\frac{E_\text{as,r}^0}{E_\text{sec}^0}\right|^2 = |\tfr_\text{sec}(\Omega)|^2
    &= |t_\ts H_\ts(\Omega) \!\times\! \sqrt{\varepsilon_\ts}|^2.
\end{align}
Therefore, for these internal losses and mismatches, the SEC dynamics
encoded in the fundamental's optical loop suppression $H_\ts(\Omega)$ cancels,
the only difference is how they enter the SEC, and they therefore
differ only by the \textit{arm} transmission $\tfr_\ta(\Omega)$. This
is in contrast to all of the other losses generated by quantum vacuum
entering the system outside the coupled cavity shown in
\cref{fig:CE-quantum-budget} which are modified by the \textit{coupled
  cavity} transmission $\tfr_\text{cc}(\Omega)$ once calibrated into
an equivalent strain. The effects of the SEC are imprinted on the
mismatch loss only through the transmission of the HOMs from the HOM
SEC---where they experience the HOM SEC optical loop suppression---into the
fundamental SEC given by \cref{eq:mismatched-hom-arm-transmission}.

This strain-referred loss is given in \cref{tab:loss-scalings} for all
of the noises discussed above. Several facts are of particular
interest:
\begin{itemize}[leftmargin=0.05\columnwidth,rightmargin=0.02\columnwidth]
\item Shot noise due to HOA and SEC loss above the arm cavity pole
  rise like $\Omega$ and are independent of arm length. Therefore
  these noises will become relatively more important for detectors
  with long arms since most other noises fall with some power of arm
  length.
\item Shot noise due to HOA and SEC loss above the arm cavity pole are
  enhanced by a factor of $\sqrt{\Fa}$.
\item Shot noise due to HOMs that are mode healed is suppressed by a
  factor of $\sqrt{\Fs}$ and shot noise due to HOMs that are mode
  harmed is enhanced by a factor of $\sqrt{\Fs}$.
\item Mode harmed HOA QRPN is enhanced by a factor of $\sqrt{\Fs}$,
  quadratic healed QRPN is suppressed by a factor of $\Fs^{3/2}$, and
  both quadratic harmed and HOA healed QRPN are suppressed by a factor
  of $\sqrt{\Fs}$.
\item All sources of QRPN are enhanced by a factor of
  $\sqrt{\Fa}$ and are inversely proportional to arm length.
\item Optical SEC loss is independent of SEC finesse. It is more
  important for CE because of its independence with arm length. It is
  more important for other detectors which happen to have large SEC
  finesses because those detectors also have large arm finesses.
\end{itemize}
Note that while the shot noise behavior below the arm cavity pole is interesting in
understanding the optical dynamics, it is ultimately not relevant to a gravitational-wave detector because the QRPN contributions will be dominant at these low frequencies.

Finally, we note that the possibility of ``tuning'' the detector to
have extra sensitivity at frequencies relevant to post-merger neutron
star physics at the expense of broadband sensitivity is sometimes
discussed~\cite{Srivastava2022,Martynov2019,Ackley2020}. This is
achieved by making the SEC pole $\wsec$ sufficiently small that the
single pole approximation for the RSE transmission
\cref{eq:cc-transmission-zpk} breaks down. In this case, the pole at
$\wrse$ in the SEC loop suppression
\cref{eq:fundamental-sec-closed-loop} is replaced by a complex pair of
poles resulting in an optical plant with a resonant gain of bandwidth
$\wsec$ around these poles at a frequency of approximately
$\sqrt{\wrse\wsec}$.  This thus results in a resonant dip in several
noise sources when they are calibrated into an equivalent
strain. However, as described above, the internal SEC loss and mode
mismatch loss are not affected by the dynamics of the SEC encoded in
the loop suppression and thus do not gain the benefit of a resonant
dip when calibrated into strain. This is one of several reasons why
attempting such a tuning is of limited utility in the presence of SEC
loss or internal mode mismatch loss.

\subsection{Degradations around a higher order mode arm cavity resonance}
\label{subsec:hom-degradations}

When a higher order mode becomes resonant in the arm cavities, the
behavior of the HOM and the fundamental is swapped: the reflection off
of the arm for the fundamental is $\rfr_{\ta0}(\Omega)=+1$ since it is
non-resonant, and the reflection off of the arm for the HOM
$\rfr_{\ta1}(\Omega)$ is given by
\cref{eq:arm-reflection-pade}. Furthermore, the arm cavities are
detuned for the HOM due to the extra Gouy phase accumulated relative
to the fundamental so that $\delta\omega_\ta\neq0$ in
\cref{eq:arm-reflection-pade}.

The squeezing degradations around a HOM arm cavity resonance have many
parallels to those due to the filter cavity used to generate the
frequency-dependent squeezed state rotation needed for broadband
quantum noise reduction. The required rotation is imparted to a
frequency-independent squeezed state by reflecting it off of the
filter cavity which is itself a cavity detuned for the
fundamental~\cite{Kimble2001,McCuller2020,Zhao2020,Ganapathy2023}. The
arm cavities therefore act like filter cavities for the HOM due to the
same principle, and the HOM will experience all of the degradations
that the fundamental experiences due to the filter
cavity~\cite{McCuller2021,Kwee2014}---including an (unwanted)
rotation.\footnote{This is similar to the effect studied in
Ref.~\cite{Toyra2017} where HOMs become resonant in the filter cavity
itself. However, the effect described there should not be significant
in practice because it relies on optical parameters chosen to be
problematic.} These degradations will then couple into the fundamental
through the mode mismatch.

These dynamics around a higher order mode resonance at a frequency
$\delta\omega_\ta$ result in a reflection for the fundamental mode off
of the coupled cavity of
\begin{multline}
  \rfr_\text{rse}(\Omega) = -\frac{1 - \rmi \Omega/\wsec}{1 +
    \rmi\Omega/\wsec} \\ + \frac{g_\ts\Upsilon}{(1 +
    \rmi\Omega/\wsec)^2} \frac{(1 - \alpha) + (1 + \alpha)\rmi (\Omega
    - \delta\omega_\ta) / \wa}{1 + \rmi(\Omega -
    \delta\omega_\ta)/\wcc}
  \label{eq:hom-resonance-rse-reflection}
\end{multline}
where the relevant coupled cavity pole is
\begin{equation}
  \wcc = \frac{1 + \beta r_\ts}{1 - \beta r_\ts} \wa =
  \begin{cases}
    \wsr & \beta = -1 \\
    \wrse & \beta = +1
  \end{cases}
  \label{eq:hom-resonance-cc-pole}
\end{equation}
and where
\begin{equation}
  g_\ts = \frac{t_\ts^2}{(1 - r_\ts)^2} \frac{1 + \alpha\beta r_\ts}{1 +
    \beta r_\ts} =
  \begin{cases}
    1 & \alpha=-1, \beta=+1 \\[0.5em]
    \displaystyle\left(\!\frac{2\Fs}{\pi}\!\right)^2 & \alpha=-1, \beta=-1 \\[1em]
    \displaystyle\frac{2\Fs}{\pi} & \alpha=+1, \beta=\pm1
  \end{cases}
  \label{eq:hom-resonance-rse-gain}
\end{equation}
The SEC is tuned to operate in RSE for the fundamental. A HOM with a one-way Gouy phase
of $\psi_\ts=\pi/2$ ($\beta=-1$) experiences an extra round-trip $\pi$ phase shift in
the SEC relative to the fundamental, and therefore a total round-trip phase of zero near
an arm resonance. Such a HOM therefore has dynamics characteristic of SR rather than
RSE, being enhanced by a factor of $\Fs/\pi$ for frequencies $|\dwa|\ll\wa$,  and is
characterized by the SR coupled cavity pole. On the other hand, a HOM with a one-way
Gouy phase of $\psi_\ts=0$ ($\beta=+1$) experiences the same total round-trip phase as
the fundamental and therefore has dynamics characteristic of RSE, being suppressed by a
factor of 2 for frequencies $|\dwa|\ll\wa$, and is characterized by the same RSE coupled
cavity pole.
Since most HOMs will be more characteristic of a HOM with $\psi_\ts=\pi/2$,
these degradations will generally be narrow with a width characteristic of the SR
pole---thus getting narrower as the SEC finesse and RSE pole get larger (see \cref{eq:cc-pole}).

The resonance condition of the HOM in the SEC is only one effect which determines the
gain $g_\ts$ in \cref{eq:hom-resonance-rse-gain}, i.e.\ the overall degree to which the
dynamics of the HOM are amplified. The other is the interference between the fields
reflecting off of the mismatched arm cavity which is determined both by the type of
mismatch and by the round-trip phase of the HOM in the SEC. Recall that in the model of
\cref{fig:phenom-cc-sflow,subsec:rse-mismatched}, this reflection is given by
\cref{eq:mismatched-arm-reflection} and is made by the interference of the HOMs which
are promptly reflected from the arm (the first line) and those that enter the HOM
coupled cavity, experience the dynamics of the SEC, and leak back out (the second line).
Around an arm HOM resonance, this second term is proportional to $\alpha(1 +
\alpha\rfr_{\ta1})^2\,\rme^{2\rmi\psi_\ts}\propto\alpha\beta$, and the sign of
$\alpha\beta$ determines whether the interference is constructive or destructive. When
$\alpha = \beta$, the interference is such that the dynamics are enhanced by a factor of
2, and when $\alpha=-\beta$ the dynamics are suppressed by a factor of $\pi/\Fs$. Taken
together, this explains the gain in \cref{eq:hom-resonance-rse-gain}. The most
significant degradations will arise from a HOM with SR-like dynamics ($\beta=-1$)
enhanced by a factor of $\Fs/\pi$ which are excited by quadratic mismatch ($\alpha=-1$)
so that the interference is constructive. The least significant degradations will arise
from a HOM with RSE-like dynamics ($\beta=+1$) suppressed by a factor of 2 which are
excited by quadratic mismatch so that the interference is destructive.

In \cref{subsubsec:hom-metrics} we discuss the squeezing degradations
which result from the dynamics described by
\cref{eq:hom-resonance-rse-reflection}; in
\cref{subsubsec:hom-resonance-locations} we discuss how the locations
of these resonances are determined; and in \cref{subsubsec:adf} we
discuss how these degradations could be monitored in order to
characterize the detectors and as an aid in tuning them to minimize
mode mismatch.


\subsubsection{Rotation, loss, and dephasing}
\label{subsubsec:hom-metrics}

\begin{figure}
  \includegraphics[width=\columnwidth]{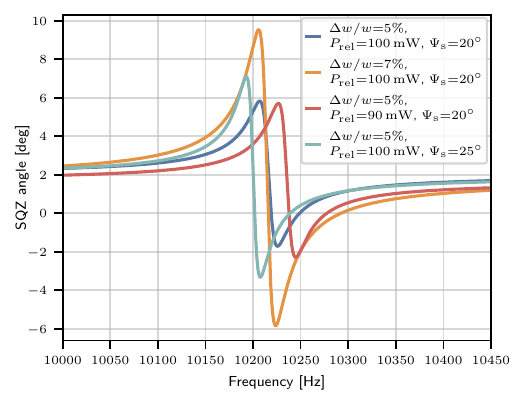}
  \caption{Squeezed state rotation around the first HOM arm cavity resonance in
    \Asharp{}\ with an SEM transmissivity of $T_\ts=\qty{20}{\%}$ for
    slightly different thermal states. This exact behavior is well
    accounted for by \cref{eq:hom-resonance-rotation}. The RSE pole in
    this case is \qty{750}{\Hz} and the SR pole is \qty{2}{\Hz}. Fits
    to these curves give widths of between 6 and \qty{10}{\Hz} showing
    that the dynamics are characteristic of those of a signal recycled
    interferometer for the fundamental mode. Measuring this rotation with
    an audio diagnostic field~\cite{Ganapathy2022} may prove useful in
    characterizing the detectors and optimizing their configurations
    as discussed in \cref{subsubsec:adf}.
  }
  \label{fig:ligo-HOM-resonance-rotation}
\end{figure}

As can be seen from \cref{fig:CE-quantum-budget}, the most prominent degradation around
a higher order mode resonance for reasonable values of mismatch for Cosmic Explorer is
due to the anti-squeezing caused by the squeezed state rotation. For more extreme values
of mismatch, the dephasing can become even more significant but can no longer be
described by the approximate expressions of this section.  Either the upper or the lower
HOM sideband will be resonant in the arm and experience the $\pi$ phase shift on
reflection of a resonant cavity, and the squeezed state will therefore rotate within the
band where this resonance occurs. This bandwidth will be characterized by the
appropriate coupled cavity pole \cref{eq:hom-resonance-cc-pole}. The rotation thus
generated by the reflection \cref{eq:hom-resonance-rse-reflection} is
\begin{equation}
  \theta(\Omega) = -g_\ts\Upsilon
  \frac{(\dwa) / \wcc}{1 + (\dwa)^2 / \wcc^2}.
  \label{eq:hom-resonance-rotation}
\end{equation}
This rotation is enhanced by a factor of $g_\ts/\wcc$ which is summarized in
\cref{tab:loss-scalings}. \Cref{fig:ligo-HOM-resonance-rotation} shows this rotation for
four slightly different thermal states in LIGO as is discussed further in
\cref{subsubsec:adf}.

Next consider the loss which is given by
\begin{equation}
  \Lambda(\Omega) =
  \begin{cases}
    \displaystyle \frac{2g_\ts\Upsilon}{1 + (\dwa)^2 / \wcc^2} & \alpha=-1 \\[2ex]
    \displaystyle 2g_\ts\Upsilon \left[1 + \frac{(\dwa)^2 /
      \wcc^2}{1 + (\dwa)^2 / \wcc^2}\right] & \alpha=+1
  \end{cases}
  \label{eq:hom-resonance-loss}
\end{equation}
Due to its low-pass dynamics, for quadratic mismatch only one of the sidebands will be
transmitted into the SEC and this transmission diminishes for frequencies above the
cavity pole. This leads to a Lorentzian loss. On the other hand, due to its high-pass
dynamics, for higher order aberrations one of the sidebands is always maximally
transmitted into the SEC and always contributes to the loss. When the other sideband is
resonant in the arm, it is not transmitted into the SEC and does not contribute. This
leads to an inverse Lorentzian-like loss with a dip reaching half the maximum on exact
resonance.

Finally, consider the dephasing. In the case of quadratic mismatch, the
single sideband which is transmitted into the SEC on resonance is
attenuated above the cavity pole. For higher order aberrations, one of
the sidebands is always maximally transmitted into the SEC and the
second only significantly couples into the SEC above the cavity
pole. In both cases the sidebands are maximally imbalanced on
resonance, and this leads to the following Lorentzian dephasing profile
\begin{equation}
  \Xi(\Omega) =
  \frac{g_\ts^2\Upsilon^2}{\left[1 + (\dwa)^2 / \wcc^2\right]^2}.
  \label{eq:hom-resonance-dephasing}
\end{equation}
For the particular thermal state shown in
\cref{fig:CE-quantum-budget}, this dephasing adds approximately
\qty{520}{\milli\radian} of phase noise around the most prominent HOM
resonance at about \qty{3}{\kHz}.

As we have seen from the fundamental dynamics of the HOMs summarized in
\cref{tab:loss-scalings}, the quadratic mismatch will produce the most significant
squeezing degradations around HOM arm cavity resonances. Since quadratic mismatch mostly
excites second order HOMs while the higher order aberrations excite mostly higher order
HOMs with larger spatial extent, the inclusion of realistic apertures in the arm cavity
introduces more clipping loss for the HOMs excited by higher order aberrations than for
those excited by quadratic mismatch thus making higher order aberrations even less
important than the scalings of \cref{tab:loss-scalings} suggest.
However, just as quadratic mismatch affects the broadband loss despite being
significantly subdominant to the effects of higher order aberrations as discussed in
\cref{subsec:broadband-loss}, the higher order aberrations affect the magnitude of
the degradations discussed here which are mainly due to quadratic mismatch due to the
interference between the HOMs which are responsible.

\subsubsection{Placement of higher order mode resonances}
\label{subsubsec:hom-resonance-locations}

In the absence of thermal aberrations and apertures, a higher order mode of order $N$
will become resonant in the arm cavities around frequencies~\cite{Siegman1986}
\begin{subequations}
    \label{eq:arm-res-freqs}
\begin{align}
  \delta\omega_\ta &= \left|pN \omega_\text{tms} - q \omega_\text{fsr}\right|
  = \omega_\text{fsr}\left|pN\frac{\Psi_\ta}{\pi} - q \right| \\
  \omega_\text{tms} &= \omega_\text{fsr} \frac{\Psi_\ta}{\pi}, \qquad
  \omega_\text{fsr} = \frac{\pi c}{L_\ta}
\end{align}
\end{subequations}
for integers $p$ and $q$ where $\omega_\text{fsr}$ is the free
spectral range (FSR), $\omega_\text{tms}$ is the transverse mode
spacing (TMS), and $\Psi_\ta$ is the one-way arm cavity Gouy
phase. The FSR is the difference in frequencies between resonances of
the fundamental mode, and the TMS is the frequency difference between
HOM resonances. In the context of gravitational-wave detectors it is
desirable to have a large TMS so that there are fewer HOM resonances
within the detection band. This is a more significant problem for
detectors with long arms, like CE, since the TMS is inversely
proportional to arm length.

\begin{figure}
  \centering
  \includegraphics{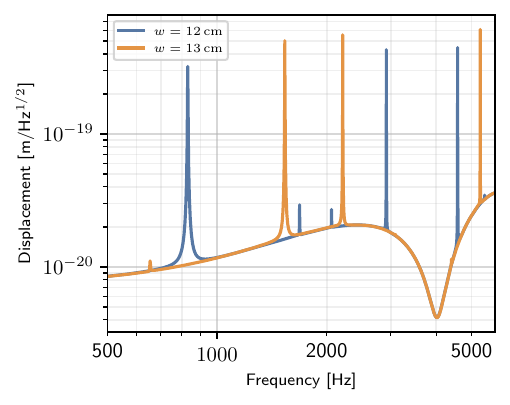}
  \caption{Higher order mode resonances in CE for two different arm
    cavity geometries with $\qty{70}{\cm}~\varnothing$ arm cavity
    apertures. Only second order mode resonances are significant after
    the addition of apertures. The free spectral range is
    \qty{3750}{\Hz} and the transverse mode spacing is \qty{1460}{\Hz}
    for the $w=\qty{12}{\cm}$ case and \qty{1110}{\Hz} for the
    $w=\qty{13}{\cm}$ case. In both cases, the first peak is due to
    the second order resonance in the first FSR, the second peak is
    due to the same resonance in the zeroth FSR, and the third peak is
    due to the resonance in the second FSR.}
  \label{fig:ce-hom-resonances}
\end{figure}

The arm cavity Gouy phase, and thus the HOM locations, is determined
by the geometry of the arm cavity, i.e.\ the cavity $g$-factors or,
equivalently, the radii of curvature of the input and end test masses
as
\begin{equation}
  \Psi_\ta = \arccos\left(\sgn g_\ti\, \sqrt{g_\ti g_\text{e}}\right), \qquad
  g_i = 1 - \frac{L_\ta}{R_i}.
  \label{eq:arm-rt-gouy}
\end{equation}
In reality, the presence of thermal aberrations and apertures in the arms will slightly
shift the locations of these resonances $\delta\omega_\ta$: the one-way Gouy phase
accumulated by a HOM of order $N$ is no longer precisely $\psi_\ta = N\Psi_\ta$ in the
presence of apertures or thermal aberrations as explained in
\cref{subsec:cavity-eigenmodes}. Furthermore, the dynamics of the SEC for HOMs not
exactly resonant or anti-resonant in that cavity will slightly shift the frequencies at
which the degradations will occur from the exact arm resonances. These details must be
accounted for when characterizing an existing detector or designing a new one, however
the dominant factor determining where the degradations will occur is, by far, the arm
cavity geometry. \Cref{fig:ce-hom-resonances} show how the arm cavity geometry
determines the resonances in CE.

The locations of these resonances cannot be arbitrarily placed
however. The arm cavity geometry also determines the size of the beams
on the test mass mirrors as~\cite{Siegman1986}
\begin{equation}
  w^2_{\ti(\text{e})} = \frac{\lambda L_\ta}{\pi}
  \sqrt{\frac{g_{\text{e}(\ti)}}{g_{\ti(\text{e})} (1 - g_\ti g_\text{e})}},
\end{equation}
and so the location of the HOM resonances can equivalently be thought of as being
determined by the size of the beams on the test masses. There are other constraints on
the $g$-factors and beam sizes which must be satisfied or implications to be considered
that are too numerous to discuss here but which therefore limit where these resonances
can be placed. Minimizing the impact of these squeezing degradations will therefore be
just one of many considerations that must be made when designing the arm cavities of a
new detector.


Finally we note that a real gravitational-wave detector has two arms
which do not have precisely the same geometry for a variety of
technical reasons and which will furthermore be slightly
astigmatic. Astigmatism in the SEC will also slightly alter their
behavior. There will therefore be four closely spaced peaks
responsible for these squeezing degradations near each HOM resonance
which will nonetheless largely behave like the single peak described
here.

\subsubsection{Use in tuning and characterizing detectors}
\label{subsubsec:adf}

It may be possible to take advantage of these squeezed state degradations
around a HOM arm cavity resonance by carefully measuring them in order to
characterize the detector state and tune the interferometer for better
mode matching. It is possible to measure the McCuller metrics directly
using an audio diagnostic field (ADF) injected into the system along
with the squeezed state~\cite{Ganapathy2022}. HOM resonances outside
the detection band provide as much information as those within the
band, and this technique can thus be useful in both LIGO and CE.
It is also important to note that this technique is a transfer function
measurement and does not require measuring noise spectra or
distinguishing quantum from classical noise as is often done to
characterize detectors and to tune them to maximize the broadband
quantum noise reduction.

\Cref{fig:ligo-HOM-resonance-rotation} shows the squeezed state rotation around the
first HOM arm cavity resonance in LIGO for four different thermal states and illustrates
the kind of information that could, in principle, be gained by monitoring this feature.
The blue curve shows a nominal case with $\dww = \qty{5}{\%}$ and $\Pabs=\qty{100}{\mW}$
relative absorption. The orange curve shows the same state with the quadratic mismatch
increased to $\dww=\qty{7}{\%}$. The amplitude of the rotation is simply increased
because this feature is predominantly due to the quadratic mismatch as discussed above.
The red curve shows the original state with the relative absorbed power decreased to
$\Pabs=\qty{90}{\mW}$ keeping the quadratic mismatch constant. In this case the
amplitude is constant because it is largely unaffected by higher order aberrations. The
frequency of the maximum rotation is slightly shifted, however, because the higher order
aberrations slightly shift the arm cavity Gouy phase away from the frequency determined
by the arm cavity geometry alone (\cref{eq:arm-res-freqs,eq:arm-rt-gouy}). Finally, the
teal curve shows the same thermal state but with an SEC Gouy phase of
$\Psi_\ts=\ang{25}$ rather than the $\Psi_\ts=\ang{20}$ for the other three cases. In
this case the amplitude is increased because the HOM dynamics are enhanced in the SEC to
a slightly greater degree at this Gouy phase. The peak is also shifted because the SEC
Gouy phase also slightly shifts the peak rotation away from the arm HOM resonance
determined by the arm cavity geometry alone. The measurements of
Ref.~\cite{Ganapathy2022} show that it is, in principle, possible to measure changes of
this magnitude with the ADF.

As in \cref{fig:loss-vs-SEC-gouy-phase}, the SEC Gouy phase has been artificially held
constant in \cref{fig:ligo-HOM-resonance-rotation}; see \cref{subsec:cavity-eigenmodes}
for details. In reality, the thermal lensing in the ITM substrates would induce both
higher order aberrations and change the Gouy phase, so the effects are not completely
orthogonal. (The actual detectors do have actuators capable of independently changing
only the SEC Gouy phase without affecting any of the test mass thermal aberrations,
however.) Nevertheless, it is an especially sensitive indicator of quadratic mismatch
and, especially when combined with other independent measurements, this illustrates how
monitoring the squeezing degradations around a HOM arm cavity resonance can be used as a
guide in how to tune the thermal actuators to improve the internal mode matching.

Finally, as is discussed in \cref{subsec:external-mismatch}, the
inevitable mode mismatch with the external optical cavities not
studied here will produce degradations that are in many ways
indistinguishable from the degradations due to the internal quadratic
mismatch and higher order aberrations discussed here. Crucially, the
degradations around a HOM resonance are only caused by internal
mismatch, and so monitoring them can be an important tool in
disentangling what mismatch is most responsible for the observed
degradations in order to know where to focus on improving them.

\subsection{Broadband rotation}
\label{subsec:broadband-rotation}

\begin{figure}
  \centering \includegraphics{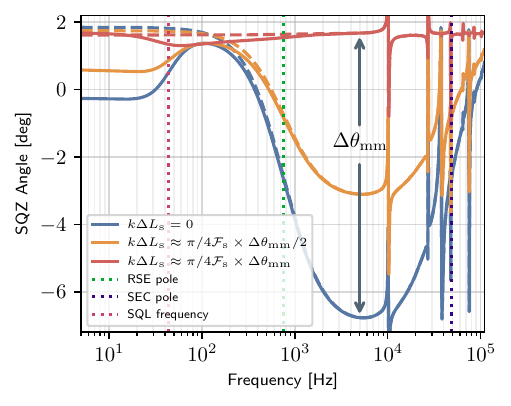}
  \caption{Broadband rotation of the squeezed state in \Asharp{} with
    an SEM transmissivity of $T_\ts=\qty{20}{\%}$. The peaks are due
    to the HOM arm cavity resonances discussed in
    \cref{subsec:hom-degradations} superimposed on the broadband
    rotation described by \cref{eq:BB-rotation}. The dashed lines do
    not include radiation pressure and thus only illustrate the purely
    optical rotation described by \cref{eq:BB-rotation}. The solid
    lines include radiation pressure and thus also include the
    rotation due to the shifted SQL frequency.
    The rotation cannot be completely flattened between the SQL and
    RSE pole by introducing a length detuning $\Delta L_\ts$ because,
    independent of any mismatch or length detuning, two filter
    cavities are required to perfectly compensate the rotation due to
    the ponderomotive squeezing of the
    interferometer~\cite{Kimble2001,Ding2025}.}
  \label{fig:BB-rotation}
\end{figure}

Detuning the SEC by an angle $\phi_\ts$ induces a broadband squeezed
state rotation of\footnote{The approximate expression of
\cref{eq:BB-rotation} describes the same rotation as does Eq.~(69) of
Ref.~\cite{McCuller2021} but is consistent with the exact model of
\cref{subsec:rse-mismatched} over a wider parameter regime. Note that
in the notation of Ref.~\cite{McCuller2021}, $\wrse$ is
$\gamma_\text{A}$, $\wsec$ is $\gamma_\text{S}$, and $\Fs=\pi/u_\ts$.}
\begin{equation}
  \theta(\Omega) = - \frac{4\Fs}{\pi}
  \frac{\phi_\ts\,(\Omega/\wrse)^2}{(\Omega/\wrse)^2 + (1 -
    \Omega^2/\wrse\wsec)^2}.
  \label{eq:BB-rotation}
\end{equation}
Such a rotation is illustrated in \cref{fig:BB-rotation}. The optical
response to the detuning $\phi_\ts$ produces no rotation at low
frequencies below the RSE pole $\wrse$, a rotation reaching
$\Delta\theta_\ts\simeq 4\Fs/\pi\times \phi_\ts$ in the mid-band
between the RSE pole and the SEC pole $\wsec$, and a rotation back to
zero at high frequencies above the SEC pole.

One way that such a rotation can arise is by detuning the length of
the SEC by $\Delta L_\ts$ to introduce an extra phase of
$\phi_\ts=k\Delta L_\ts$ to the
fundamental~\cite{McCuller2021}. However, even if there is no detuning
of the SEC length, the cavity will still be detuned for a HOM which is
not exactly resonant or anti-resonant due to its extra Gouy phase. The
cavity will then become detuned for the fundamental, not by detuning
the length, but through the coupling with the HOM introduced by a mode
mismatch. The $\phi_\ts$ introduced by a higher order aberration for a
HOM near resonance ($\psi_\ts = \delta\psi_\ts$) or anti-resonance
($\psi_\ts = \pi/2 + \delta\psi_\ts$) is
\begin{equation}
  \phi_\text{mm} \approx
  \begin{cases}
    \displaystyle\frac{\delta\psi_\ts \Upsilon}{\pi^2/4\Fs^2 + (\delta\psi_\ts)^2}
    & \beta \approx +1 + 2\rmi\delta\psi_\ts \\[1em]
    \displaystyle-\delta\psi_\ts\Upsilon
    & \beta \approx -1 - 2\rmi\delta\psi_\ts
  \end{cases}.
  \label{eq:HOA-detuning-angle}
\end{equation}
The detuning due to a nearly resonant HOM is therefore enhanced by the SEC finesse with
the maximum $\Fs\Upsilon/\pi$ occurring at a Gouy phase of $\pi/2\Fs$, while the
detuning due to a nearly anti-resonant HOM is independent of finesse.  Quadratic
mismatch does not produce any significant broadband rotation due to its low-pass nature.

The rotation described by \cref{eq:BB-rotation} is due solely to the optical response of
a field reflecting off of a detuned coupled cavity. There are two more related but
distinct effects due to such a detuning that arise under the influence of radiation
pressure. First, detuning the SEC by introducing a length offset $\Delta L_\ts$ will
also produce an optical spring whereby the dynamics of the mirrors themselves are
modified so that the light provides a restoring (or anti-restoring) force between the
mirrors~\cite{BnC2001,BnC2002}. This will modify the optomechanical plant $C(\Omega)$
\cref{eq:optical-plant}---as well as the test mass susceptibility. (There will also be
an associated optical resonance introduced in the plant at high frequencies because the
system is now not exactly operating in RSE and so the signal sidebands have become
slightly imbalanced.)  The same changes to the optomechanical plant will also occur for
a detuning caused by a mode mismatch such as \cref{eq:HOA-detuning-angle}. Since this is
a change only to the optical plant it will only affect the strain-referred noise; see
\cref{subsec:strain-referred-loss}.

The second effect is that the optomechanical coupling
$\Krp_\text{rse}$ will be modified by an SEC detuning introduced
through any mechanism. The SQL frequency will therefore be slightly
shifted and the squeezed state rotation due to the interferometer
$\theta_\text{ifo}(\Omega)$ will be slightly altered. The filter
cavity is tuned to produce a rotation of $\theta_\text{fc}(\Omega)$ to
cancel the rotation caused by the ponderomotive squeezing of the
interferometer $\theta_\text{ifo}(\Omega)$ by targeting a specific SQL
frequency. With a $\theta_\text{ifo}(\Omega)$ altered by a SEC
detuning, the rotation $\theta_\text{fc}(\Omega)$ provided by the
filter cavity no longer perfectly compensates the ponderomotive
squeezing of the interferometer and there will be a residual total
``misrotation'' of the squeezed state $\theta(\Omega) =
\theta_\text{fc}(\Omega) + \theta_\text{ifo}(\Omega)$ around the SQL
frequency. This only affects the squeezed state rotation
$\theta(\Omega)$ and does not modify the optomechanical plant
$C(\Omega)$ or the dynamics of the mirrors. This difference is
important when considering the interactions between the internal
mismatch discussed here and the external mismatch as discussed in
\cref{subsec:external-mismatch}.

Both the rotation of the squeezed state $\theta(\Omega)$---due both to the optical
rotation described by \cref{eq:BB-rotation} and to the shifted SQL frequency---and the
optical spring generated by a mode mismatch can be corrected by detuning the SEC length
by $k\Delta L_\ts\approx \pi/4\Fs\times\Delta\theta_\text{mm}\approx -\phi_\text{mm}$.
Note that a higher finesse SEC will produce a larger squeezed state rotation
$\Delta\theta_\text{mm}$ for a given mismatch; however, the rotation due to an SEC
length detuning is amplified by the same amount and so a similar $\Delta L_\ts$ will be
needed to correct the rotation from the same mode mismatch $\phi_\text{mm}$ regardless
of the SEC finesse. The magnitude of $\phi_\text{mm}$ will still be amplified by the SEC
finesse for nearly resonant HOMs though.
\Cref{fig:BB-rotation} shows the squeezed state rotation as an SEC detuning due to a
length offset $k\Delta L_\ts$ is introduced to compensate for the rotation caused by the
detuning due to a mode mismatch \cref{eq:HOA-detuning-angle}. Note that the length
offset needed to compensate the rotation in this way is generally microscopic and a
small fraction of the wavelength of the laser: the necessary $k\Delta L_\ts$ is
typically of order \qtyrange{0.01}{1}{\deg}. Since the current and future detectors use
lasers of wavelength between \qtyrange{1}{2}{\um}, this corresponds to a $\Delta L_\ts$
of at most a few \unit{nm}; this is easily realized by the length actuators used in the
detectors.

Note that since increasing the SEC finesse $\Fs$ both increases the RSE pole $\wrse$ and
decreases the SQL frequency $\wsql^\text{rse}$, increasing the finesse would broaden the
region over which minimal rotation occurs both by decreasing the start of the low
frequency rotation around $\wsql^\text{rse}$ and increasing the start of the high
frequency rotation around $\wrse$.

Finally we note that the detuning due to mode mismatch also induces
broadband dephasing due to the imbalanced sidebands which can
occasionally be significant. However, introducing the same SEC length
detuning needed to cancel the broadband rotation also cancels this
broadband dephasing, i.e.\ the same detuning needed to re-balance the
phase of the upper and lower sideband also re-balances their
magnitude.


\subsection{Interaction with external mode mismatch}
\label{subsec:external-mismatch}

While a detailed study of the interaction between the internal
mismatch discussed here and the external mismatch is left for future
work, we can get a rough idea of what those interactions will be and
how this will affect efforts to understand and tune the detectors. We
will find that the similarities between the internal and external
mismatch will confound efforts to understand them separately in
practice in order to know how to tune the detectors to reduce the
squeezing degradations. Crucially, however, there are key differences
in how the internal and external mismatch affect the rotation of the
squeezed state that can be used to disentangle them to some degree.

The phenomenological two mode model of the internal mismatch presented
here can be combined with the two mode model of the external mismatch
described in Ref.~\cite{McCuller2021}. The two external cavities that
are most significant in terms of mode matching are the output mode
cleaner (OMC), which filters the signal from the interferometer before
detection, and the optical parametric amplifier (OPA) in which the
squeezed state is generated. The mode matching with the filter cavity
is negligible compared with the matching between these two
cavities~\cite{Ganapathy2023}.

Recall that the reason that it is only possible to have a beam size
error in the mismatch between the mode of the arm cavity and the mode
of the SEC is because these cavities form a coupled cavity with
standing waves on either side. The OMC, OPA, and the interferometer do
not form a coupled cavity system and it is therefore possible to have
both beam size and curvature errors in the mismatches between these
three cavities. Ref.~\cite{McCuller2021} parametrizes where in this
phase space of beam size and curvature errors the mismatch takes place
by the angle $\psi_\text{R}$, and denotes the magnitude of the
mismatch between the OPA and the interferometer by $\Upsilon_\text{I}$
and between the OPA and the OMC by $\Upsilon_\text{O}$. Active
wavefront control (AWC) is provided by the telescopes which match the
OMC and OPA to the interferometer (and therefore to each other) and
which are made by mirrors which can change both the magnitude and
phasings of these mismatches by adjusting their radii of
curvature~\cite{Srivastava2021,Cao2020}. It is also important to note
that the external mismatch should be mostly quadratic and excite
mostly the second order modes.

Ref.~\cite{McCuller2021} showed that the rotation due to external mode mismatch in this
model is approximately\footnote{\Cref{eq:external-rotation} describes the same rotation
as does Eq.~(89) of Ref.~\cite{McCuller2021} over a wider parameter regime.}
\begin{equation}
  \theta_\text{ext}(\Omega) =
  -\frac{2\Upsilon_\text{I}\Upsilon_\text{O}\sin\psi_\text{R}\,(\Omega/\wrse)^2}
       {(\Omega/\wrse)^2 + (1 - \Omega^2/\wrse\wsec)^2}.
  \label{eq:external-rotation}
\end{equation}
We therefore expect external mismatch to cause a squeezed state rotation with the same
frequency dependence as the rotation described by \cref{eq:BB-rotation} caused by an SEC
detuning $\phi_\ts$, due either to internal mismatch or a length detuning, does except
that the maximum rotation in the mid-band in this model is
$\Delta\theta_\text{ext}\simeq
2\Upsilon_\text{I}\Upsilon_\text{O}\sin\psi_\text{R}$ rather than
$\Delta \theta_\ts\simeq 4\Fs/\pi \times \phi_\ts$.
The important point is that in reality, regardless of the
details of the mismatch, this rotation will be indistinguishable from
one generated by an SEC detuning and that its magnitude will be a
function of both the magnitude and phasings of the external
mismatch---and therefore controllable to some extent by the AWC. The
external mismatch will also cause a rotation around the SQL frequency
due to radiation pressure in the same way as the internal
mismatch. However, unlike internal mismatch, the external mismatch
does not detune the SEC and it does not create an optical spring and
thus does not modify the optomechanical plant $C(\Omega)$ or the
dynamics of the mirrors.

As with the internal mismatch where a SEC length detuning $\Delta
L_\ts$ could be introduced to cancel the rotation caused by internal
mode mismatch---both that described by \cref{eq:BB-rotation} and that
due to the modified SQL frequency---a length detuning can counteract
the rotation due to external mismatch to some degree. However, the
mechanism of this rotation is not through a SEC detuning, and it will
therefore not in general be possible to find a $\Delta L_\ts$ that
simultaneously cancels the rotation due to both effects. Furthermore,
attempting to cancel the rotation due to external mismatch in this way
will introduce an optical spring since the external mismatch does not
produce an optical spring in the first place. Thus, attempting to
measure the rotation $\Delta\theta$ as suggested in
\cref{subsec:broadband-rotation}, either with the ADF or through some
other means, in order to determine the correct length detuning to
introduce to cancel the rotation ($k\Delta L_\ts\approx \pi/4\Fs
\times\Delta\theta$) is not likely to be successful because
$\Delta\theta$ will be a mix of the $\Delta\theta_\text{mm}$ and
$\Delta\theta_\text{ext}$. A more successful strategy may be to try to
introduce an SEC detuning to reduce the optical spring followed by
adjusting the AWC to reduce the remaining squeezed state rotation
$\theta(\Omega)$ due to the external mismatch. Though this will run
into difficulties if the optical spring is generated partially through
other technical means not discussed here.

Finally, the external mismatch will not cause any of the degradations
around higher order mode arm cavity resonances discussed in
\cref{subsec:hom-degradations}. In particular, while external and
internal mismatch affect the broadband loss and rotation in the same
way, the magnitude of the rotation around a HOM resonance will be a
clean error signal for the internal mismatch, especially the quadratic
part.

Ref.~\cite{McCuller2021} further showed that the loss due to external
mode mismatch in this model is approximately
\begin{equation}
  \Lambda_\text{ext}(\Omega) = \frac{\Upsilon_\text{O}}{1 + \Omega^2/\wrse^2}
  + \left(\Upsilon_\text{O} + \Upsilon_\text{R}\right) \frac{\Omega^2/\wrse^2}{1 + \Omega^2/\wrse^2}
  \label{eq:external-loss}
\end{equation}
where
\begin{equation}
  \Upsilon_\text{R} \approx 4\Upsilon_\text{I} - 4 \sqrt{\Upsilon_\text{I}\Upsilon_\text{O}} \cos\psi_\text{R}.
\end{equation}
We thus expect external output mismatch to coherently add to the
quadratic mismatch to produce low-pass losses and for a combination of
the output and input mismatches to coherently add to the higher order
aberrations to produce high-pass losses with the same RSE pole
frequency. The external mismatch will also be responsible for
radiation pressure noise. The phasing of the external mismatch will
also affect the frequency dependence of this loss, and can thus be
influenced by the AWC, both because it changes the relative amount of
high-pass and low-pass loss in \cref{eq:external-loss} and because it
will change the interference with the HOMs generated with the internal
mismatch between the arms and SEC.

Taken together, some aspects of a detector's behavior due to thermal
changes could plausibly be explained by something like the following.
As the thermal state of the interferometer drifts, the frequency-dependent losses due to the internal mismatch will change due to the sensitivity of these losses to the details of the thermal
aberrations as discussed in \cref{subsec:broadband-loss}. The detuning
of the SEC due to a mismatch such as \cref{eq:HOA-detuning-angle} will
also change requiring a different SEC length offset $\Delta L_\ts$ to
be introduced to correct the broadband rotation \cref{eq:BB-rotation}
and the optical spring as discussed in
\cref{subsec:broadband-rotation}. In practice, it may not be possible
to continue introducing a larger length offset if needed since this
may saturate photodetector electronics. The HOM peaks will shift
predominantly by quadratic changes to the arm cavity geometry but also
due to changes in other interferometer parameters such as the Gouy
phase of the SEC; the magnitude will also change predominantly due to
changes in the quadratic matching but also due to changes in other
detector parameters to a lesser degree as described in
\cref{subsubsec:adf}.  Even though the modes of the external cavities
are unaffected by the thermal changes in the interferometer, the
external mismatch will be affected since the matching of those
cavities to the altered interferometer mode will have changed; this is
described by $\Upsilon_\text{I}$, $\psi_\text{R}$, and
$\Upsilon_\text{R}$ in the simple model described here.  The
internal thermal changes will thus also be accompanied by changes to the
frequency-dependent losses (\cref{eq:external-loss}) and broadband
rotation (\cref{eq:external-rotation}) due to external mismatch which
will need to be corrected with the AWC. These external effects
will be largely indistinguishable from the effects of the internal
mismatch studied throughout the rest of this work.  The use of thermal
actuators to correct the thermal aberrations will likewise produce all
of these effects needing further adjustment of the AWC and SEC length
offset.

\section{Summary of effects and implications for detector design}
\label{subsec:detector-implications}

In this section we summarize how detector design choices impact the squeezing
degradations discussed above. First we note that
\begin{itemize}[leftmargin=0.05\columnwidth,rightmargin=0.02\columnwidth]
\item Higher order aberrations are the most significant source of broadband mismatch
  loss for the frequencies where mismatch loss is significant due to their high-pass
  dynamics (\cref{subsec:broadband-loss}).
\item Quadratic mismatch generates the most significant degradations around frequencies
  that a HOM is resonant in the arms (\cref{subsec:hom-degradations}) since 1) the
  degradations caused by quadratic mismatch are amplified to a greater extant than those
  caused by HOA are due both to their interference with the fundamental mode and to their
  resonance conditions in the SEC; and 2) since the HOMs mainly responsible for
  quadratic mismatch are not clipped by the arm apertures as much as the HOMs mainly
  responsible for HOA are.
\item Higher order aberrations are the most significant source of broadband rotation
  (\cref{subsec:broadband-rotation}). While their effects may be mitigated by properly
  tuning the detector to a large degree, doing so is not independent of external
  mismatch (\cref{subsec:external-mismatch}) and may furthermore be limited by other
  technical challenges, so it is still important to minimize these effects.
\end{itemize}

The location of several frequency scales are determined by a
combination of other detector parameters and affect the squeezing
degradations in the following ways
\begin{description}[font=\normalfont,leftmargin=0.08\columnwidth,rightmargin=0.02\columnwidth]
\item[$\wrse$ RSE pole \cref{eq:cc-pole}] Sets the instrument
  bandwidth. The pole frequency of the full coupled cavity low-pass dynamics of
  quadratic mismatch and the high-pass dynamics of HOA is also $\wrse$ (see
  \cref{eq:broadband-hom-transmission,fig:mismatch-loss}). The frequency dependence of
  the external mismatch loss is similarly around $\wrse$ (see \cref{eq:external-loss}).
  Finally, the broadband rotation also occurs around $\wrse$ (see
  \cref{eq:BB-rotation,fig:BB-rotation,eq:external-rotation}).

\item[$\wsr$ SR pole \cref{eq:cc-pole}] Generally sets the bandwidth
  of the degradations around HOM arm cavity resonances; see
  \cref{eq:hom-resonance-cc-pole} and surrounding
  discussion. Typically, broader instrument bandwidths thus lead to
  narrower HOM resonance degradations.

 \item[$\wsql^\text{rse}$ SQL frequency \cref{eq:cc-sql-frequencies}]
   The frequency at which QRPN and shot noise coming from all external vacuum are equal;
   QRPN and shot noise coming from SEC loss and higher order aberrations are equal at a
   higher frequency (\cref{subsec:mismatch-loss-qrpn}). The squeezed state also rotates
   around $\wsql^\text{rse}$ due to SEC detunings (caused by both mode mismatch or
   length detunings) as well as external mismatch (see \cref{fig:BB-rotation}).
   Furthermore, the larger is the ratio $\wrse/\wsql^\text{rse}$ the better a single
   filter cavity can properly compensate the ponderomotive squeezing from the
   interferometer and the larger is the region with minimal broadband rotation. Finally,
   a larger filter cavity finesse is required for a lower SQL frequency, and the
   squeezing degradations due to the filter cavity are enhanced by the filter cavity
   finesse~\cite{Kwee2014,McCuller2021}; however, these degradations are pushed towards
   lower frequencies as $\wsql^\text{rse}$ is decreased.

\item[$\wsec$ SEC pole \cref{eq:cav-finesse-pole}] The broadband
  rotation starts reversing direction around $\wsec$ (see
  \cref{eq:BB-rotation,fig:BB-rotation,eq:external-rotation}). The bandwidth of the
  instrument will also start to be limited by $\wsec$ if it gets close enough to
  $\wrse$. It is generally desirable to keep $\wsec$ sufficiently large that its effects
  are kept out of the detection band. If $\wsec$ is intentionally made sufficiently
  small to target post-merger neutron star physics as is sometimes discussed, a resonant
  dip with a width and frequency set by $\wsec$ will appear in several noise sources;
  however, SEC loss and internal mode mismatch loss do not experience this benefit
  (\cref{subsec:strain-referred-loss}).
\end{description}

In addition to determining the frequency scales described above, some
important consequences of detector parameters to the squeezing
degradations are the following
\begin{description}[font=\normalfont,leftmargin=0.08\columnwidth,rightmargin=0.02\columnwidth]
\item[$\Fa$ Arm cavity finesse] The arm cavity finesse enhances every
  relevant degradation, as summarized in \cref{tab:loss-scalings}, and
  should be kept as small as possible from a squeezing degradation
  perspective.

\item[$\Fs$ SEC finesse] The SEC finesse determines the extent to
  which HOMs will be suppressed (mode healed) or enhanced (mode harmed) in the
  SEC---both for the mismatch loss (see
  \cref{fig:loss-vs-SEC-gouy-phase,eq:broadband-hom-transmission,tab:loss-scalings}),
  and for the degradations around a HOM arm resonance  (see
  \cref{eq:hom-resonance-rotation,eq:hom-resonance-loss,eq:hom-resonance-dephasing,tab:loss-scalings}).
  It also amplifies the broadband rotation due to an SEC detuning (see
  \cref{eq:BB-rotation}). Notably, it does not affect the optical SEC loss. The SEC
  finesse will be chosen primarily to set the desired instrument bandwidth. In order to
  keep a fixed bandwidth, it is necessary to choose $\Fs \propto \Fa L_\ta$
  (cf.~\cref{eq:cc-pole,eq:cav-finesse-pole}). Longer and higher finesse arm cavities
  therefore generally require higher SEC finesses.

\item[$\Psi_\ts$ SEC Gouy phase] Determines the resonance conditions
  of the HOMs in the SEC and thus which HOMs will be enhanced or
  suppressed in that cavity; see \cref{fig:loss-vs-SEC-gouy-phase}.

\item[$\Psi_\ta$ Arm cavity Gouy phase] Predominately sets the
  location of the degradations around arm cavity HOM resonances; see
  \cref{subsubsec:hom-resonance-locations,fig:ce-hom-resonances}.

\item[$L_\ts$ SEC length] The main effect of $L_\ts$ is through its
  determination of the SEC pole $\wsec$, and this favors a cavity
  short enough to keep the effects of $\wsec$ out of the detection
  band.\footnote{The proposed NEMO detector favors a long
  SEC~\cite{Ackley2020} because that detector aims to move the effects
  of $\wsec$ into the detection band to target neutron star
  physics. But as described in \cref{subsec:strain-referred-loss},
  $\wsec$ determines the width of these resonances while their
  location is approximately $\sqrt{\wrse\wsec}=\sqrt{\wa c/L_\ts}$. So
  if a detector with longer arms like CE, with a
  correspondingly lower arm cavity pole $\wa$, were to try to target
  the same signals, it would still need a short SEC length in order to
  keep this resonance above the roughly \qty{2}{\kHz}
  required~\cite{Srivastava2022}. This is another reason why trying to
  target neutron star physics in this way is of limited utility in
  CE.} However, shorter cavities require stronger
  telescopes to realize and thus may make it difficult to robustly
  achieve the required levels of mode matching in practice.

\item[$L_\ta$ Arm cavity length] Since most noises decrease as some
  power of the arm length, the biggest effect of longer arms is the
  increased sensitivity~\cite{Evans2021,Evans2017}. Notably, however,
  SEC loss and the loss due to higher order aberrations are
  independent of arm length and therefore become relatively more
  significant as the arm length is increased
  (\cref{subsec:strain-referred-loss,tab:loss-scalings}). Longer arms
  also have lower free spectral ranges---thus limiting the instrument
  bandwidth---and lower transverse mode spacings---thus having more
  degradations due to HOM arm cavity resonances in the detection band; see
  \cref{subsec:hom-degradations}.

\item[SEC apertures] Smaller apertures in the SEC can drastically
  reduce the degree to which HOMs resonant in the SEC are mode harmed
  while at the same time introduce extra loss away from these
  resonances; see \cref{fig:loss-vs-SEC-gouy-phase}.

\item[Arm cavity apertures] Apertures in the arm cavities reduce the
  degradations around HOM arm cavity resonances. In practice, these
  degradations due to higher order aberrations can be largely
  eliminated by moderate apertures leaving only the effects of
  quadratic mismatch and second order modes.

\end{description}

\section{Outlook}
\label{sec:conclusion}

We have identified two types of internal mode mismatch between the arm
cavities and the signal extraction cavity in a gravitational-wave
detector: those due to the quadratic mismatch between the wavefront of
two optical modes and those due to all residual higher order
aberrations. These two types of mismatch are predominantly responsible
for different frequency-dependent squeezed state degradations which
have been detailed here for the first time. While we have focused on
the degradations due to thermal effects in the test mass optics, the
main issue in gravitational-wave detectors, the degradations to the
squeezed states due to these two types of mismatch generated through
other means will have similar behavior.


We have studied these degradations theoretically using a modal model
of a coupled cavity system which includes the exact couplings between
the higher order modes produced by the thermal aberrations generated
by the absorption of a small fraction of the power circulating in the
interferometer arm cavities by the test mass optics.
However, given our lack of detailed knowledge of the exact thermal
aberrations in and optical parameters of the current detectors and
given the state-of-the-art modeling tools existing today, it is
unlikely that such models will quantitatively describe these complex
thermo-optomechanical systems by exactly predicting the squeezing
degradations.  We believe that experiments on simpler optical systems
are needed in order to validate the effects described in this work, to
quantitatively understand the detailed behavior resulting from
complications not discussed, and to investigate and develop the
techniques suggested for diagnosing and mitigating the effects of the
squeezing degradations including the effects of external mode
mismatch. On the modelling tool perspective, further verification is
needed to determine the model accuracy when dealing with apertured
HOMs and large thermal aberrations, which should also be verified in
small scale experiments.


We have also described a simpler phenomenological model---the full
details of which are given in \cref{sec:full-gwinc-model}---which
makes no attempt to predict the mode couplings which result from a
given thermal state, but which better elucidates the physics of the
squeezing degradations. Even so, it is often possible to find
phenomenological parameters of this model which produce degradations
that agree with those of the more complicated but exact model for
small mismatch. For some applications, this type of model is therefore
sufficient for characterizing, improving, and designing the detectors
in practice.

Building on the work of Ref.~\cite{McCuller2021} in particular, this
work extends the understanding of mode mismatch in gravitational-wave
detectors which is necessary to continue to improve the astrophysical
sensitivity of the current observatories and to design the next
generation ones. This analysis has also shown that the optical
configuration and thermal state of these detectors are inextricably
linked and that the optical system and thermal actuators should
therefore be designed simultaneously informed by an understanding of
the effects of mode mismatch on squeezing degradations as well as its
impacts on other technical challenges.

\begin{acknowledgments}
We thank Lee McCuller, Huy-Tuong Cao, and Aidan Brooks for early
discussions about internal mode mismatch which helped to inspire this
work, and thank Sheila Dwyer and Evan Hall for detailed comments on
the manuscript.
KK thanks Sheila Dwyer and Vicky Xu for assistance in studying
squeezing degradations at the LIGO Hanford observatory and for further
discussions.
KK was supported by NSF PHY--2309200, PHY--2309064, and PHY--2309267.
DB was supported from the Australian Research Council (ARC) on Grant
DE230101035.
DB and KK would like to thank OzGrav (ARC Grant CE170100004 and
CE230100016) for their support for this research with various travel
funded over the years.  This material is based upon work supported by
NSF's LIGO Laboratory, which is a major facility fully funded by the
US National Science Foundation.  LIGO was constructed by the
California Institute of Technology and Massachusetts Institute of
Technology with funding from the NSF and operates under NSF
Cooperative Agreement PHY--2309200. Advanced LIGO was built under NSF
PHY--0823459. The LIGO~A+ Upgrade to Advanced LIGO is supported by NSF
PHY--1834382
\end{acknowledgments}

\appendix

\section{Quantum noise budget factorization in terms of the M\lowercase{c}Culler squeezing metrics}
\label[appendix]{sec:qn-budget}

The noise budget factorization in terms of the McCuller metrics,
developed in Ref.~\cite{McCuller2021}, used throughout this paper is
not novel and has been used in, for example,
Refs.~\cite{Jia2024,Capote2024}. We here give a more explicit
description of this factorization and the meaning of all of the traces
shown in the noise budgets than has been given elsewhere.

In the simplest case where a squeezed state is detected without
encountering any optomechanical system that would either introduce a
frequency dependence, due to cavity dispersion for example, or source
radiation pressure, due to acting on a suspended mirror for example,
the quantum noise relative to the $\hbar\omega_0/2$ of unsqueezed
vacuum is (cf.~Eq.~(6) of Ref.~\cite{McCuller2021})
\begin{subequations}
  \label{eq:findep-qn-factorization}
  \begin{align}
    N &= \eta\left(S_- \cos^2\phi + S_+ \sin^2\phi\right) + (1 - \eta) \\
    S_\pm &= \left(1 - \phi_\text{rms}^2\right) \rme^{\pm 2r} + \phi_\text{rms}^2 \rme^{\mp 2r}
  \end{align}
\end{subequations}
where $r$ is the amplitude of the squeezed state injected into the
system, $\eta$ is the efficiency, $\phi$ is the relative angle between
the injected squeezed state and the measurement quadrature, and
$\phi_\text{rms}$ is the RMS fluctuations in that angle. The quantum
noise in this case is due to three frequency independent effects, or
degradations: 1) loss $\Lambda = 1 - \eta$ due to squeezed vacuum
being replaced by unsqueezed vacuum; 2) phase noise
$\phi_\text{rms}^2$ due to mixing the squeezed and anti-squeezed
quadratures; and 3) simply observing more noise by detecting a
quadrature other than the one which was squeezed, i.e.\ observing
$\phi\neq 0$.

Ref.~\cite{McCuller2021} defines four frequency dependent metrics so
that the noise of a general optomechanical system can be factored in
the same way as \cref{eq:findep-qn-factorization}. Using these
metrics, the noise $N(\Omega)$ relative to the $\hbar\omega_0/2$ of
unsqueezed vacuum is (Eqs.~(7) to (9) of Ref.~\cite{McCuller2021})
\begin{subequations}
  \label{eq:qn-factorization}
\begin{align}
  N(\Omega) &= \Gamma(\Omega) \left[ \eta(\Omega) S(\Omega) +
    \Lambda(\Omega)\right] \label{eq:factorization-N} \\ S(\Omega) &=
  S_-(\Omega) \cos^2\left[\phi + \theta(\Omega)\right] +
  S_+(\Omega)\sin^2\left[\phi + \theta(\Omega)\right] \\ S_\pm(\Omega)
  &= \left[1 - \Xi'(\Omega)\right]\rme^{\pm 2r} + \Xi'(\Omega)
  \rme^{\mp 2r}.
\end{align}
\end{subequations}
In a general system, the rotation of the squeezed state
$\theta(\Omega)$ will be frequency dependent and so $\phi +
\theta(\Omega)$ replaces the angle $\phi$ in
\cref{eq:findep-qn-factorization}. In the context of gravitational
wave detectors, a filter cavity is employed to attempt to keep $\phi +
\theta(\Omega) = 0$ at all frequencies. The noise gain
$\Gamma(\Omega)$ describes the radiation pressure responsible for the
ponderomotive squeezing in a general optomechanical
system. In a general system, both the loss $\Lambda(\Omega)$ and
efficiency $\eta(\Omega)$ are frequency-dependent; where
$\Gamma(\Omega)\approx1$, they approximately satisfy the usual
frequency-independent relationship $\Lambda(\Omega) \approx 1 -
\eta(\Omega)$. Phase noise is quantified by an effective dephasing
$\Xi'(\Omega)$ which includes an intrinsic dephasing $\Xi(\Omega)$ as
well as other sources of phase noise such as RMS phase noise
$\phi_\text{rms}^2$.

These metrics are given explicitly
by~\cite{McCuller2021,Ganapathy2022}
\begin{align}
  \theta(\Omega) &=
  \frac{1}{2} \arg \left( \frac{m_p + \rmi m_q}{m_p
    - \rmi m_q} \right) \label{eq:general-theta} \\
  \Xi(\Omega) &= \frac{1}{2} - \sqrt{
    \frac{(|m_p|^2 - |m_q|^2)^2 + 4[\re(m_q m_p^*)]^2} {4(|m_p|^2 +
      |m_q|^2)^2} } \label{eq:general-intrinsic-dephasing} \\
  \eta(\Omega)\Gamma(\Omega) &= |m_p|^2 + |m_q|^2,
  \label{eq:general-etaGamma}
\end{align}
where $m_p(\Omega)$ and $m_q(\Omega)$ are the observed noise due to a
squeezed state injected in the phase and amplitude quadratures,
respectively.\footnote{If the phase quadrature is measured, then $m_p
=[\rrse(+\Omega) + \rrse^*(-\Omega)]/2$ and $m_q=[\rrse(+\Omega) -
  \rrse^*(-\Omega)]/2\rmi$. In this case,
\cref{eq:general-theta,eq:general-intrinsic-dephasing,eq:general-etaGamma}
simplify to
\cref{eq:sqz-metrics-efficiency,eq:sqz-metrics-rotation,eq:sqz-metrics-dephasing}
in the sideband picture.} They can be measured experimentally by
measuring the response to a diagnostic field injected into the system
as described in Ref.~\cite{Ganapathy2022}. They are calculated by
\cref{eq:full-simulation-quadratures} for the noise budgets in this
paper.

\begin{figure}
  \centering
  \includegraphics[width=\columnwidth]{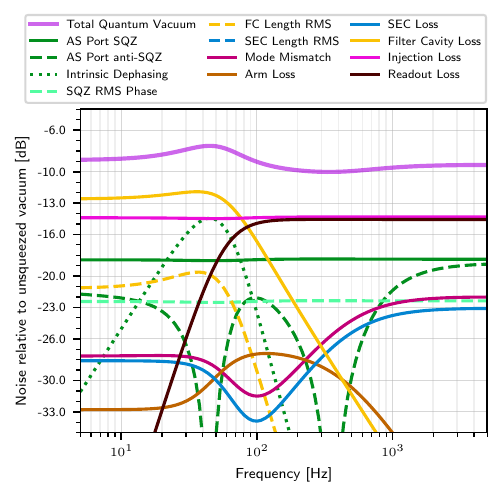}
  \caption{\Asharp{} quantum noise budget plotted as
    $10\log_{10}[N(\Omega)/\Gamma(\Omega)]$ for the thermal state of
    \cref{fig:mismatch-loss}.}
  \label{fig:Asharp-relgamma-budget}
\end{figure}

Rather than plotting the noise relative to shot noise $N(\Omega)$,
noise budgets are usually shown in terms of the equivalent
displacement noise. The amplitude spectral density of this
displacement-referred noise is related to $N(\Omega)$ by
\begin{equation}
  S^{1/2}_{xx}(\Omega) = \sqrt{\frac{N(\Omega)}{|C(\Omega)|^2} \frac{\hbar\omega_0}{2}}
\end{equation}
where $C(\Omega)$ is the interferometer response to differential arm
motion, known as the optomechanical plant, with units of
\unit{\sqrt{\W}/\m} (cf.~\cref{eq:strain-referred}). Since $N(\Omega)$
is relative to shot noise, the $\hbar\omega_0/2$ converts to an
amplitude spectral density in physical
units. \Cref{fig:CE-quantum-budget} is in terms of $S_{xx}^{1/2}$.

It is also common to look at the noise relative to unsqueezed vacuum
in decibels. Rather than directly comparing to the frequency
independent $\hbar\omega_0/2$ of unsqueezed shot noise using
$N(\Omega)$, it is more useful to compare to the unsqueezed vacuum
that travels through the optomechanical system along the same path
that the squeezed state takes and which therefore experiences the same
ponderomotive squeezing,
i.e.\ $10\log_{10}[N(\Omega)/\Gamma(\Omega)]$. \Cref{fig:Asharp-relgamma-budget}
shows the \Asharp{} noise budget for the thermal state of
\cref{fig:mismatch-loss} in this way.

The noises in these budgets are explicitly the following:

\textbf{AS Port SQZ:} The noise arising from the squeezed state injected
  into the anti-symmetric (AS) port of the interferometer. Its
  contribution to $N(\Omega)/\Gamma(\Omega)$ is
\begin{equation}
  \eta(\Omega) \left[1 - \Xi'(\Omega)\right]\rme^{-2r}
  \cos^2\left[\phi + \theta(\Omega)\right].
\end{equation}
In an ideal system with no squeezing degradations and a filter cavity
that perfectly cancels the rotation due to the ponderomotive
squeezing of the interferometer so that $\phi + \theta(\Omega) = 0$,
this would be the only source of quantum noise. With no filter cavity
and no detuning, this is the shot noise due to the field injected into
the AS port.

\textbf{AS port Anti-SQZ:} The noise due to measuring the anti-squeezing rather than the
squeezing. Its contribution to $N(\Omega)/\Gamma(\Omega)$ is
\begin{equation}
  \eta(\Omega)S_+(\Omega) \sin^2\left[\phi + \theta(\Omega)\right].
\end{equation}
If there is a filter cavity, this is the noise due to the filter cavity not perfectly
compensating the squeezing angle rotation throughout the optical system in order to keep
the total $\theta(\Omega) = \phi + \theta_\text{ifo}(\Omega) + \theta_\text{fc}(\Omega)
= 0$ at all frequencies, i.e.\ a ``misrotation'' relative to the optimal angle. With no
filter cavity and no detuning, it is just the standard radiation pressure contribution
due to the ponderomotive anti-squeezing.

\textbf{Dephasing:} There are several sources of dephasing. The total effective dephasing
degrades the squeezed state with a contribution to $N(\Omega)/\Gamma(\Omega)$ of
\begin{equation}
  \eta(\Omega)\Xi'(\Omega) \rme^{+2r} \cos^2\left[\phi +
    \theta(\Omega)\right].
  \label{eq:general-dephasing}
\end{equation}
The effective dephasing $\Xi'(\Omega)$ includes an intrinsic dephasing inherent to the
system and several sources of technical noise:

\textbf{Intrinsic Dephasing:} The fundamental phase noise $\Xi(\Omega)$
  given by \cref{eq:general-intrinsic-dephasing}. It is due to the
  upper and lower sidebands experiencing different loss or by the
  interaction with a lossy mechanical system.

\textbf{SQZ RMS Phase:} The usual frequency independent
RMS phase noise $\phi_\text{rms}$ in the angle $\phi$; it is simply
$\Xi_{\phi_\text{rms}} = \phi_\text{rms}^2$.

\textbf{Length RMS:} Since the rotation of the squeezed state depends on
  the detunings of the cavities that it encounters as it propagates
  throughout the optical system, any fluctuation in the length of
  a cavity will also be converted into a phase fluctuation as
\begin{equation}
  \Xi_{L_\text{rms}}(\Omega) = \left|\frac{\partial
    \theta(\Omega)}{\partial L}\right|^2 L_\text{rms}^2,
  \label{eq:rms-length-dephasing}
\end{equation}
where $L$ is the length of the cavity and $L_\text{rms}$ is the RMS
fluctuation in that length.  The two ``Length RMS'' traces shown in
the budgets of this paper are the dephasings due to RMS fluctuations
in the lengths of either the SEC or the filter cavity.

The individual sources of dephasing are combined into the total
effective dephasing as described in Appendix~B of
Ref.~\cite{McCuller2021}. Namely, the effective dephasing due to the
individual dephasings $\Xi_1$ and $\Xi_2$ is\footnote{The individual
dephasing contributions shown in the noise budgets of this paper are
not precisely the individual $\Xi_i(\Omega)$ due to the way in which
they must be combined according to \cref{eq:dephasing-sum}. Rather,
they are $\Xi_i - 2\Xi_i \Xi_{i-1}'$ where $\Xi_{i-1}'$ is the
effective dephasing computed by combining the previous $i-1$ dephasing
sources. The dephasing sources in these budgets are combined in the
order in which they appear. So the intrinsic dephasing is just $\Xi$
and the SQZ RMS phase is $\phi_\text{rms}^2 - 2\Xi\phi_\text{rms}$,
etc. The relative difference $2\Xi'_{i-1}$, however, is generally
small; only \qty{2}{\%} for \qty{100}{\milli\radian} of dephasing.}
\begin{equation}
  \Xi'(\Omega) = \Xi_1(\Omega) + \Xi_2(\Omega) -
  2\,\Xi_1(\Omega)\Xi_2(\Omega).
  \label{eq:dephasing-sum}
\end{equation}

\textbf{Loss:} The remaining traces in the noise budgets shown in this
paper are due to the various sources of loss where squeezed photons
are lost and replaced with unsqueezed vacuum as discussed in
\cref{subsec:broadband-loss}. They are calculated using
\cref{eq:general-mismatch-loss,eq:general-total-loss} for the noise
budgets in this paper. The optical loss for a given source is the
transfer function from that source to the readout. The mode mismatch
loss is the fraction of total power in the higher order modes rather
than in the fundamental mode.

\section{Cavity eigenmodes}
\label[appendix]{subsec:cavity-eigenmodes}

The Laguerre-Gauss (LG) modes \cref{eq:LG-modes}---or the equivalent
Hermite-Gauss (HG) modes---are not the true eigenmodes of an optical
cavity. They are a very good approximation in a cavity with no
aberrations and large apertures, but the differences between the true
eigenmodes can become significant in the presence of realistic thermal
aberrations or apertures. Of particular significance is the round-trip
Gouy phase of each eigenmode in the cavity which determines the
resonance conditions of the cavity and thus the extent to which
dynamics are enhanced or suppressed.


Colloquially, the Gouy phase can refer to one of four quantities. It
could be the excess phase that the fundamental mode accumulates over a
plane wave \textit{relative to its beam waist} as determined by the
$q$ parameter at a point, as in \cref{eq:LG-expansion}, which we
denote as $\Xi$; or it could be the excess phase that the fundamental
accumulates \textit{between two spatial locations}, which we denote as
$\Psi$. The Gouy phase can also refer to the same quantities for a
particular higher order mode relative to the fundamental. We denote
these with the lower case $\xi$ and $\psi$. When the true eigenmodes
are described by the LG or HG modes, $\psi = N\Psi$ for a mode of
order $N$, but this is not always the case.

When only the quadratic effects of an optical cavity are accounted for, the eigenmodes
of that cavity are given by \cref{eq:LG-modes} (or the equivalent HG modes). To describe
an optical field at a spatial point $\mu$, as in \cref{eq:LG-expansion}, it is then
necessary to specify the complex beam parameter $q_\mu$ at that point, i.e.\ specifying
the beam size $w_\mu$ and defocus $S_\mu$ at that point. It is possible to use any beam
parameter, but the natural choice and the one that requires the fewest modes to describe
that field is the one describing the quadratic fundamental eigenmode of the cavity. That
mode is the eigenmode of the round-trip ABCD matrix for the cavity.\footnote{If the true
eigenmodes differ significantly from the LG or HG modes, the natural choice of $q$
parameters---in the sense of requiring the fewest terms in an expansion
\cref{eq:LG-expansion} to well describe a true eigenmode---is not always obvious or the
one given by the eigenmode of \cref{eq:sec-round-trip-abcd}.} For example, in the case
of the signal extraction cavity shown in \cref{fig:coupled-cavity}, starting from the
node $n_{\text{a,i}}$, this is the matrix
\begin{equation}
  \matQ{S}(L_\ts)\, \matQ{F}(2/R_\ts)\, \matQ{S}(L_\ts)\, \matQ{F}(1/f_\text{th})\,
  \matQ{F}(-2n/R_\ti)\, \matQ{F}(1/f_\text{th})
  \label{eq:sec-round-trip-abcd}
\end{equation}
where $\matQ{F}(D)$ is the ABCD matrix for a focusing element (a lens or a mirror) with
defocus $D$ and $\matQ{S}(L)$ is the ABCD matrix for a space of length $L$. We write the
ABCD matrices in a different font to emphasize that they are $2\times 2$ matrices
transforming the complex $q$ parameters, while the matrices in
\cref{eq:cc-operators,fig:coupled-cavity} are high dimensional matrices transforming the
HOMs themselves, i.e.\ acting on the vector of $c_{p\ell}(q_\mu)$ coefficients in
\cref{eq:LG-expansion}.

Once the $q$ parameters have been determined from the cavity eigenmode, thus defining
the HOMs, the Gouy phase accumulated between two spatially separated points is the
difference between the Gouy phases $\Xi_\mu$ in \cref{eq:LG-expansion} at each point.
For example, the Gouy phase accumulated between the SEM and the ITM AR surface is
$\Psi=\Xi_\text{a,i} - \Xi_\text{s,r}$. The total phase accumulated for a HOM of order
$N = 2p + |\ell|$ relative to the fundamental is then $\psi = N\Psi$. The total
round-trip Gouy phase in the cavity is~\cite{KojiRTGouy}
\begin{equation}
  2\Psi = \sgn B\, \arccos\left(\frac{A + D}{2}\right)
  \label{eq:rt-gouy-abcd}
\end{equation}
where $A,B,$ and $D$ are the elements of the round-trip ABCD matrix, given by
\cref{eq:sec-round-trip-abcd} in the example of the SEC. For the arm cavity with only
the ITM and ETM, \cref{eq:rt-gouy-abcd} is just \cref{eq:arm-rt-gouy}.

In reality, the eigenmodes of the cavity are the eigenmodes of the
round-trip operator of that cavity.\footnote{Unlike the LG and HG
modes, the true cavity eigenmodes are bi-orthogonal rather than
orthogonal and are not guaranteed to be complete~\cite{Siegman1986}.}
In the case of the SEC starting from the same node $n_{\text{a,i}}$,
this is
\begin{equation}
  \mat{P}_\ts(-r_\ts\mat{1})\mat{P}_\ts \Las (r_\ti \Mss)\Lsa.
  \label{eq:sec-round-trip}
\end{equation}
The phase of the eigenvalue of each mode is the true round-trip Gouy
phase $2\psi$ for that mode. Each eigenmode can be expanded as in
\cref{eq:LG-expansion} and, while the modes $u_{p\ell}$ have Gouy
phases $\xi=N\Xi$ determined from \cref{eq:rt-gouy-abcd}, the summation of
all of these LG modes results in a phase determined by
\cref{eq:sec-round-trip}, and this phase $\psi$ is not necessarily an
integer multiple of the phase $\Psi$ of the first eigenmode.

Now consider how thermal aberrations affect the mode couplings and the
Gouy phase as calculated by
\cref{eq:sec-round-trip,eq:rt-gouy-abcd}. First we note that in
addition to the phase evolution of an optical field described by the
OPD \cref{eq:opd-decomposition}, there will be apertures $A(r,\phi)$
due to the finite spatial extent of the optics for example. The
reasoning leading to the general coupling between modes
\cref{eq:general-opd-coupling} being broken up into a quadratic and
higher order part is still valid in the presence of apertures and so
\begin{equation}
  \braket{q_2}{A(r,\phi)\,\rme^{-\rmi kZ(r,\phi)}}{q_1} =
  \braket{q_2}{A(r,\phi)\,\rme^{-\rmi k z_\text{hoa}(r,\phi)}}{\hat{q}_1}
\end{equation}
in general. All of the lens $\mat{L}_{ij}$ and surface $\mat{S}_{ij}$
operators are thus the same as in \cref{eq:cc-operators} with the
addition of $A(r,\phi)$ multiplying the exponentials.

Several effects will modify the couplings and round-trip Gouy
phase. First, any quadratic change due to the quadratic term $ar^2$ in
\cref{eq:opd-decomposition} directly changes the eigenmode $q$
parameter and thus changes the Gouy phase as calculated by
\cref{eq:rt-gouy-abcd}. Second, higher order aberrations
$z_\text{hoa}(r)$, and thus the operators in \cref{eq:cc-operators},
do not change the eigenmode or the round-trip Gouy phase as computed
by the quadratic effects and \cref{eq:rt-gouy-abcd}, but do change
the Gouy phase of the true eigenmodes as calculated by
\cref{eq:sec-round-trip}. Since quadratic effects vary the beam size
$w$, they also affect the extent to which modes are clipped by
apertures. This is another effect that alters the operators in
\cref{eq:cc-operators} and thus the Gouy phases of the true
eigenmodes. Furthermore, any quadratic change to the beam parameters
$q$ changes the modal basis of the HOMs and thus the matrix
elements of the operators \cref{eq:cc-operators} irrespective of any
higher order aberration effects.

Finally, there are several methods of removing the quadratic term
$ar^2$ from an OPD as in \cref{eq:opd-decomposition}, and some of
these depend on the beam size~\cite{Bond2016}. Therefore different
methods will yield varying fractions of quadratic or higher order
aberrations for the same original OPD. The important point, however,
is that regardless of the details of how an OPD is broken up or how
the $q$ parameters are chosen, there will always be a quadratic part
that behaves like \cref{eq:cc-reflection-phenom-model} with $\Ur \approx
\Ui^{-1}$, thus having low-pass dynamics, and the remaining higher
order aberrations which behave like
\cref{eq:cc-reflection-phenom-model} with $\Ur \approx \Ui$, thus having
high-pass dynamics. Furthermore, the ensuing squeezing degradations
will be identical no matter how the coefficient $a$ in
\cref{eq:opd-decomposition} is determined and subtracted as long as
the equivalent $f_\text{th}=-1/2a$ is used for the substrate thermal
lens focal length instead of that giving an arbitrary $\dww$, and as
long as enough HOMs are used in the calculation. Indeed, it is not
even necessary to break the OPD up into a quadratic and higher order
part in a calculation in order to get the same numerical
results.\footnote{It is, however, computationally more efficient to
attribute some quadratic part of an OPD to the focal length of a thin
lens because the resulting $q$ parameters as determined by
\cref{eq:sec-round-trip-abcd} will then more closely resemble the true
eigenmodes, thus requiring fewer terms in an expansion
\cref{eq:LG-expansion} to accurately describe the true eigenmodes.}
Nevertheless, the squeezing degradations are still highly sensitive to
the full details of $Z(r)$ regardless of the details of the
calculation.

It is also important to understand how we study the effects of changing cavity Gouy
phases as presented in
\cref{fig:loss-vs-SEC-gouy-phase,fig:ligo-HOM-resonance-rotation,fig:ce-hom-resonances}.
In these analyses, the Gouy phase displayed on the $x$-axis or noted in the legends, is
the \textit{one-way} Gouy phase $\Psi$ of the fundamental mode in that cavity as
computed with \cref{eq:rt-gouy-abcd}. The question of which Gouy phase to design a
cavity for in the absence of thermal aberrations is a critical design choice and is
determined by the cavity geometry: lengths between optics, radii of curvatures of
mirrors, lenses purposely polished into optic substrates, etc. It is thus difficult to
change the Gouy phase in an analysis as that requires redesigning the cavities and
ensuring that they are well mode-matched. Furthermore, quadratic $\dww$ changes are
unavoidably accompanied by SEC Gouy phase changes. To avoid these complications we thus
adjust the Gouy phases $\Xi_\mu$ in \cref{eq:LG-expansion} ad hoc, independent of their
$q$ parameters determined by \cref{eq:sec-round-trip-abcd}, so that all of the phases
$\Psi_\mu$ in the propagators $\mat{P}_\ts$ produce the desired Gouy phase for the
fundamental mode. However, given these adjusted propagators, the exact eigenmodes and
Gouy phases as computed by \cref{eq:sec-round-trip} are still used and thus most of the
effects of higher order aberrations are captured. The one exception is that the beam
size change which must accompany a change in Gouy phase is not accounted for and thus
the change in the degree to which HOMs are clipped by apertures as would really happen
is not captured. This is not significant at the level of detail and for the goals of
this work, but it does need to be accounted for when characterizing an existing detector
or designing a new one. This treatment can also be justified by imagining that as the
Gouy phase and beam size change, the apertures are adjusted to keep the same aperture
ratios given in \cref{tab:detector-parameters} fixed.

Finally we note that the details discussed above could, in principle,
offer a mundane un-physical explanation for why in
\cref{fig:mismatch-loss} the total loss with the addition of quadratic
mismatch appears to be less than that with the higher order
aberrations alone as parameterized by our artificial separation of the
two effects. However, as described in \cref{sec:simulation}, our
calculation ensures that identical higher order aberrations are used
regardless of the amount of quadratic mismatch. Furthermore, we have
used enough HOMs in the simulation that the physical observables do
not depend on the modal basis of the HOMs which is indirectly
affected by $\dww$. Therefore, the explanation for this behavior is
the destructive interference of the HOMs.


\section{Phenomenological model}
\label[appendix]{sec:full-gwinc-model}

When generalized to include radiation pressure and external mode
mismatch, the phenomonelogical model used throughout the body of the
paper is a slight extension of the model presented in Appendix~E of
Ref.~\cite{McCuller2021} to include both quadratic mismatch and higher
order aberrations. Ref.~\cite{McCuller2021} considers the fundamental
and a single HOM with output mismatch $\Upsilon_\text{O}$ between the
OPA and the OMC, input mismatch $\Upsilon_\text{I}$ between the OPA
and the interferometer, and quadratic mismatch $\Upsilon_\text{A}$
between the arms and the SEC. Each of these mismatches also have a
phasing $\psi$ and are described by the $4\times4$ matrices
\begin{equation}
  \Dmat{U}(\Upsilon, \psi) =
  \begin{bmatrix}
    \sqrt{1 \!-\! \Upsilon}\,\Tmat{1} & -\sqrt{\Upsilon}\,\Tmat{R}(\psi)\\
    \sqrt{\Upsilon}\,\Tmat{R}(-\psi) & \sqrt{1 \!-\! \Upsilon}\,\Tmat{1}
  \end{bmatrix},
\end{equation}
where $\Tmat{1}$ and $\Tmat{R}(\psi)$ are the $2\times2$ identity and rotation matrices, respectively.
That model can be extended by replacing the mismatch matrix
$\Dmat{U}_\text{A} = \Dmat{U}(\Upsilon_\text{A}, \psi_\text{A})$ and
its inverse with ones including both types of internal mismatch as described
by \cref{eq:approx-cc-reflection,eq:mismatch-matrices,eq:phenom-mismatch-mapping}
\begin{subequations}
\begin{align}
  \Dmat{U}_\text{A} &\rightarrow
  \Dmat{U}_\text{i} =
  \Dmat{U}_\text{quad}(\Upsilon_\text{quad}, \psi_\text{quad})
  \Dmat{U}_\text{hoa}(\Upsilon_\text{hoa},\psi_\text{hoa})\\
  \Dmat{U}_\text{A}^{-1} &\rightarrow
  \Dmat{U}_\text{r} =
  \Dmat{U}_\text{hoa}(\Upsilon_\text{hoa}, \psi_\text{hoa})
  \Dmat{U}_\text{quad}^{-1}(\Upsilon_\text{quad}, \psi_\text{quad})
\end{align}
\end{subequations}
In particular, these substitutions should be made in Eqs.~(E13) to
(E18).

This is sufficient for many purposes. Since second order modes are
generally responsible for quadratic mismatch and higher order HOMs are
generally responsible for higher order aberrations, it can
occasionally be useful to include two HOMs in order to more carefully
investigate resonance effects. In this case, there are three couplings
and phases, and the mismatch matrices are of the form
\begin{align}
  & \Dmat{U}(\{\Upsilon\}, \{\psi\}) = \nonumber \\[1ex]
  &  \begin{bmatrix}
    \sqrt{1 \!-\! \Upsilon_{10} \!-\! \Upsilon_{20}}\, \Tmat{1} &
    -\sqrt{\Upsilon_{10}}\, \Tmat{R}(\psi_{10}) &
    -\sqrt{\Upsilon_{20}}\, \Tmat{R}(\psi_{20}) \\
    \sqrt{\Upsilon_{10}}\, \Tmat{R}(-\psi_{10}) &
    \sqrt{1 \!-\! \Upsilon_{10} \!-\! \Upsilon_{21}}\,\Tmat{1} &
    -\sqrt{\Upsilon_{21}}\, \Tmat{R}(\psi_{21}) \\
    \sqrt{\Upsilon_{20}}\, \Tmat{R}(-\psi_{20}) &
    \sqrt{\Upsilon_{21}}\, \Tmat{R}(-\psi_{21}) &
    \sqrt{1 \!-\! \Upsilon_{20} \!-\! \Upsilon_{21}}\,\Tmat{1}
    \end{bmatrix}
\end{align}
Extending this further is of limited utility, and a model such as
\cref{sec:simulation} should be used if more detail is needed. It
should also be noted that, even in the single HOM case, many
parameters are degenerate in such a model and some parameters can thus
not be easily fit to data. The utility of such a model is in getting a
feel for what combinations of mismatch are consistent with some data,
or in understanding what range of behavior could be expected in some
optical design.

\section{Simulation details}
\label[appendix]{sec:simulation}

\begin{table}
  \begin{ruledtabular}
    \begin{tabular}{l l l S S}
      Parameter & {Symbol} & {Units} & {LIGO \Asharp} & {CE} \\
      \hline
      Arm power  & $P_\ta$ & \unit{\mega\W} & 1.5 & 1.5 \\
      Arm length  & $L_\ta$ & \unit{\km} & 4 & 40 \\
      SEC length & $L_\ts$ & \unit{\m} & 55 & 120 \\
      Test mass mass & $M$ & \unit{\kg} & 100 & 320 \\
      ITM transmission & $T_\ti$ & \unit{\%} & 1.4 & 1.4 \\
      SEM transmission & $T_\ts$ & \unit{\%} & 32.5 & 2 \\
      Arm Gouy phase & $\Psi_\ta$ & \unit{\deg} & 310 & 220 \\
      SEC Gouy phase & $\Psi_\ts$ & \unit{\deg} & 20 & 20 \\
      SEC loss & $\varepsilon_\ts$ & \unit{\ppm} & 500 & 500 \\
      Arm loss & $\varepsilon_\ta$ & \unit{\ppm} & 75 & 40 \\
      Readout loss & $\varepsilon_\text{ro}$ & \unit{\%} & 3.5 & 3.5 \\
      Injection loss & $\varepsilon_\text{inj}$ & \unit{\%} & 4 & 3 \\
      Filter cavity loss & $\varepsilon_\text{fc}$ & \unit{\%} & 30 & 80 \\
      Filter cavity length & $L_\text{fc}$ & \unit{\m} & 300 & 4000 \\
      Filter cavity transmission & $T_\text{fc}$ & \unit{\ppm} & 1000 & 1700 \\
      Injected squeezing & $\rme^{2r}$ & \unit{\decibel} & 18 & 18 \\
      RMS phase noise & $\phi_\text{rms}$ & \unit{\milli\radian} & 10 & 10 \\
      Arm aperture ratio & {---} & {---} & 3.1 & 2.9 \\
      SEC aperture ratio & {---} & {---} & 2.6 & 2.5 \\
    \end{tabular}
  \end{ruledtabular}
  \caption{Baseline parameters used for LIGO \Asharp{} and Cosmic
    Explorer unless otherwise stated. The arm and SEC aperture ratios
    are the ratio between the diameter of the apertures and the
    diameter of the beams in the arm cavity and SEC, respectively. The
    Gouy phases are one-way and the losses are round-trip. Note that
    the injected squeezing is the idealized and lossless squeezing
    level generated at the source before encountering any losses.}
  \label{tab:detector-parameters}
\end{table}

The main numerical results of this paper shown in all of the figures
are obtained using the \textsc{finesse} simulation
package~\cite{finesse} to simulate the DARM coupled cavity system
shown in \cref{fig:coupled-cavity} using the parameters given in
\cref{tab:detector-parameters}. In order to simulate thermal lensing
in the ITM substrate, a thin lens is placed next to the AR surface of
the ITM. Finally, the squeezed state is injected into the coupled
cavity system through the AR surface of the SEM after reflecting off
of an external Fabry-Perot cavity acting as the filter cavity which is
not shown in the figure. When the SEC finesse $\Fs$ is varied, the
filter cavity bandwidth and detuning are reoptimized according to
Ref.~\cite{Kwee2014}.

The ITM lens initially has an infinite focal length and the radii of curvature of all
optics are adjusted to give perfect mode matching between all three cavities of the
system. Future work will expand on the discussion in \cref{subsec:external-mismatch} by
introducing mismatch between all optical cavities while at the same time adding an
additional two cavities to serve as the optical parametric amplifier (OPA), in which the
squeezed state is generated, and the output mode cleaner (OMC), which filters the signal
before detection.

Our treatment of the thermal aberrations is described in
\cref{sec:thermal-aberrations,fig:thermal-lens}. In particular, the thermal aberrations
due to the beam-heating of the laser are computed using the Hello-Vinet
model~\cite{Hello_Vinet_1990,Hello_Vinet_1990b,Vinet2009}. The total optical path length
is decomposed into the quadratic terms and the higher order aberrations (HOA) as shown
in \cref{fig:thermal-lens}. First the piston and then the quadratic terms are removed by
weighting the optical path length by a Gaussian with a radius equal to the beam size of
the laser on the relevant optic. The remaining terms are the HOA and are added as an OPD
map to the lens~\cite{Bond2016}. Normally the quadratic terms which were removed would
then be added to the focal length of the lens; however, in this work we study the
quadratic and higher order aberrations separately and thus set the focal length of the
lens to produce a given quadratic $\dww$ average beam size error. The thermoelastic
deformations of the surface of the mirrors are treated similarly except that the HOA due
to the thermoelastic deformations are added as surface maps to the HR surfaces of the
mirrors. Circular apertures are also added to the lens and mirror HR surfaces. Future
work will include the effects of thermal
actuators~\cite{Brooks2016,Rocchi2012,Jones2024} and other imperfections such as coating
defects~\cite{Brooks2021} or mis-centered laser beams on the test mass optics.

It is important to note that changing $\dww$ in this way by changing
the focal length of the thin lens changes the beam size on the back of
the ITM in the extraction cavity and therefore the eigenmodes of the
SEC. However, it does not change the beam size on the front of the ITM
in the arm cavity and thus does not change the eigenmodes of the arm
cavities; see discussion around \cref{eq:itm-defocus}. Since we use
the beam size of the beam on the HR surface of the mirrors when
removing the piston and quadratic terms of the OPD or surface
deformation, the higher order aberrations used for different $\dww$
are identical. See \cref{subsec:cavity-eigenmodes} for a detailed
discussion of how this parameterization of thermal aberrations affects
cavity eigenmodes.

Since this is a simple coupled cavity model, the DC fields are added ad hoc to produce a
given arm power in the fundamental mode of the arm cavity. The ITM and ETM are simulated
as free masses. We then simulate all even Hermite-Gauss modes up to order 10 and collect
transfer functions for the optical fields between several locations in the
optomechanical system into $42\times 42$ matrices corresponding to the phase and
amplitude quadratures for the fundamental and each of the 20 HOMs. The total path the
squeezed state takes from its injection into the filter cavity to the readout, the path
from $\mu_{\text{as,i}}$ to $\mu_{\text{as,r}}$ in \cref{fig:coupled-cavity}, is
$\Dmat{H}(\Omega)$. The transfer functions from each loss location to the readout are
collected in the matrices $\Dmat{T}_\mu(\Omega)$. For SEC loss, this is the path from
$\mu_\text{a,r}$ to $\mu_\text{as,r}$ in \cref{fig:coupled-cavity}, for example.
Finally, the transfer functions of ETM motion to the readout are collected into a
$42\times 1$ vector $\Dvec{T}_\text{rse}(\Omega)$ The units of $\Dmat{H}(\Omega)$ and
$\Dmat{T}_\mu(\Omega)$ are \unit{\sqrt{\W}/\sqrt{\W}} and the units of
$\Dvec{T}_\text{rse}(\Omega)$ are \unit{\sqrt{\W}/\m}.

The quantum noise budgets described in \cref{sec:qn-budget} are then
computed using a procedure similar to the one outlined in Appendix~E
of Ref.~\cite{McCuller2021}. First the local oscillator
$\Dvec{v}^\dag$ is defined by \cref{eq:lo} as described below and then
the McCuller metrics are calculated by
\cref{eq:general-theta,eq:general-intrinsic-dephasing,eq:general-etaGamma}
using the noise quadratures
\begin{equation}
  m_p(\Omega) = \Dvec{v}^\dag \Dmat{H}(\Omega)\Dvec{e}_{p0}, \qquad
  m_q = \Dvec{v}^\dag \Dmat{H}(\Omega) \Dvec{e}_{q0},
  \label{eq:full-simulation-quadratures}
\end{equation}
where $\Dvec{e}_{q0}$ and $\Dvec{e}_{p0}$ are the basis vectors for
the amplitude and phase quadratures of the fundamental mode,
respectively.  The loss due to mode mismatch is the fraction of the
total power in the higher order modes rather than in the fundamental
\begin{equation}
  \Gamma(\Omega)\Lambda_\text{mm}(\Omega) = \left|\Dvec{v}^\dag
  \Dmat{H}(\Omega)\right|^2 - \eta(\Omega)\Gamma(\Omega).
  \label{eq:general-mismatch-loss}
\end{equation}
The total loss $\Lambda(\Omega)$ in \cref{eq:factorization-N} is
obtained by additionally propagating the $\varepsilon_\mu$ of unsqueezed
vacuum entering at each location $\mu$ to the readout
\begin{equation}
  \Gamma(\Omega)\Lambda(\Omega) = \Gamma(\Omega)\Lambda_\text{mm}(\Omega) + \sum_\mu
  \left|\Dvec{v}^\dag \Dmat{T}_\mu(\Omega)\right|^2.
  \label{eq:general-total-loss}
\end{equation}
For vacuum entering the SEC, $\left|\Dvec{v}^\dag
\Dmat{T}_\mu(\Omega)\right|^2$ is the
$\Lambda_\text{sec}(\omega)$ of \cref{subsec:broadband-loss}. The
quantum noise gain is calculated as\footnote{As discussed in
Ref.~\cite{McCuller2021}, there is freedom in defining $\Gamma$ since
only the combination $\eta\Gamma$ can be measured. \Cref{eq:Gamma-sum}
sums over all sources of loss and corresponds to the choice that
$\Gamma = N|_{S=1}$, i.e.\ the gain is just the noise in the absence
of an injected squeezed state $r=0$. Equations~(40) and (E20) of
Ref.~\cite{McCuller2021} only sum over the internal losses as that
work is primarily focused on external mismatch.}
\begin{equation}
  \Gamma(\Omega) = \left|\Dvec{v}^\dag \Dmat{H}(\Omega)\right|^2 +
  \sum_\mu \left|\Dvec{v}^\dag \Dmat{T}_\mu(\Omega)\right|^2.
  \label{eq:Gamma-sum}
\end{equation}
The efficiency $\eta(\Omega)$ is then calculated by dividing the
$\eta\Gamma$ as computed by \cref{eq:general-etaGamma} by the $\Gamma$
computed by \cref{eq:Gamma-sum}.

The optomechanical plant is calculated from $\Dvec{T}_\text{rse}$ as
\begin{equation}
  C(\Omega) = \frac{1}{\sqrt{2}} \Dvec{v}^\dag
  \Dvec{T}_\text{rse}(\Omega)
\end{equation}
where the factor of $1/\sqrt{2}$ accounts for the presence of the
beamsplitter when mapping the dynamics of the coupled cavity onto
those of an interferometric gravitational-wave detector.

Finally, the dephasing due to RMS length fluctuations of an optical
cavity is computed by repeating the calculation of $\Dmat{H}$ after
detuning that cavity by a small amount. The squeezing angle
$\theta(\Omega)$ of this detuned system is then calculated using the
new $m_p$ and $m_q$ which is then used to find the derivative
$\partial\theta(\Omega)/\partial L$ needed for the calculation of the
dephasing given by \cref{eq:rms-length-dephasing}.

As discussed around Eq.~(E21) of Ref.~\cite{McCuller2021}, the local
oscillator should be defined taking the optical DC response of the coupled
cavity into account. To this end, the transfer functions
$\Dmat{T}_\text{lo}$ are computed with infinite mass mirrors at
$\Omega=0$ from the fields leaving the ITM HR surface at the point
with Gaussian $q$ parameter $-\qhr^*$ in \cref{fig:coupled-cavity} to
the readout. The local oscillator is then defined as
\begin{equation}
  \Dvec{v}^\dag =
  \left(
  \frac{\Dmat{P} \Dmat{T}_\text{lo} \Dvec{e}_{p0}}
       {\left|\Dmat{P} \Dmat{T}_\text{lo} \Dvec{e}_{p0}\right|}
  \right)^\dag
       \Dmat{R}(\zeta)
  \label{eq:lo}
\end{equation}
where $\zeta$ is the homodyne angle and $\Dmat{P} =
\Dvec{e}_{q0}\Dvec{e}_{q0}^\dag + \Dvec{e}_{p0}\Dvec{e}_{p0}^\dag$
represents the OMC by rejecting all HOMs. Even though when using
balanced homodyne readout, as we assume will be done for both
\Asharp{} and CE, the local oscillator can be defined simply as
$\Dvec{v}^\dag = \Dvec{e}_{p0}^\dag\Dmat{R}(\zeta)$, it is easier to
define it as \cref{eq:lo} because this best represents a pure phase
signal by taking into account the extra phases accumulated by the mode
scattering and propagation through the SEC. In doing so, we always
measure the phase quadrature $\zeta=0$ throughout this work without
further optimizing the homodyne angle to produce the best sensitivity.

While the procedure for finding the operating point of the system is
fairly clear---as described in the main text (especially
\cref{subsec:broadband-rotation})---it is difficult to write down an
algorithm that will reliably find the correct operating points in
practice. Therefore, the simulation for each thermal state is tuned by
hand using the optomechanical plant $C(\Omega)$, as would be obtained
through detector calibration, and the rotation of the squeezed state
$\theta(\Omega)$, as would be obtained from ADF
injections~\cite{Ganapathy2022} or some other means, to ensure that a
reasonable operating point is found. In particular, an SEC length
detuning $\Delta L_\ts$ is introduced both to cancel the broadband
squeezed state rotation as illustrated in \cref{fig:BB-rotation} and
ensure that the resulting $C(\Omega)$ is the proper RSE plant without
an optical spring.

\bibliography{references}

\end{document}